\documentclass[sigplan,10pt,nonacm=true]{acmart}

\acmConference[]{}{}{}{}
\acmISBN{}
\acmDOI{}
\startPage{1}

\setcopyright{none}
\usepackage[iso,american,cleanlook]{isodate}
\usepackage{url} 
\hypersetup{hidelinks=false,colorlinks=true,linkcolor=blue,urlcolor=blue,citecolor=blue,anchorcolor=blue}
\usepackage{booktabs}   %% For formal tables:
\usepackage{listings}
\usepackage{subfig}
\usepackage[export]{adjustbox}

\usepackage{wrapfig}
\usepackage{float}
\usepackage{url}

\newenvironment{proofsketch}{%
  \renewcommand{\proofname}{Proof Sketch}\proof}{\endproof}

\newcommand{\var}[1]{\lstinline+#1+}
\newcommand{\cSetP}[1]{\lstinline+Set<#1>+}

\newcommand{\figlabel}[1]{\label{figure:#1}}
\newcommand{\nakedfigref}[1]{\ref{figure:#1}}
\newcommand{\figref}[1]{Figure~\nakedfigref{#1}}
\newcommand{\figrangeref}[2]{Figures~\nakedfigref{#1}--\nakedfigref{#2}}

\newcommand{\linelabel}[1]{\label{line:#1}}
\newcommand{\nakedlineref}[1]{\ref{line:#1}}
\newcommand{\lineref}[1]{Line~\nakedlineref{#1}}
\newcommand{\linerangeref}[2]{Lines~\nakedlineref{#1}--\nakedlineref{#2}}

\newcommand{\remove}[1] {}
\newcommand{\extabstract}[1] {}
\newcommand{\code}[1] {\texttt{#1}}

\newtheorem{theorem}{Theorem}[subsection] % theorem counter resets every \subsection
\lstdefinestyle{numbers}
{numbers=left, stepnumber=1, numberstyle=\tiny, numbersep=10pt}
\lstdefinestyle{nonumbers}
{numbers=none}

\def\ContinueLineNumber{\lstset{firstnumber=last}}

\makeatletter
\lst@Key{countblanklines}{true}[t]%
    {\lstKV@SetIf{#1}\lst@ifcountblanklines}

\lst@AddToHook{OnEmptyLine}{%
    \lst@ifnumberblanklines\else%
       \lst@ifcountblanklines\else%
         \advance\c@lstnumber-\@ne\relax%
       \fi%
    \fi}
\makeatother

\begin{document}

%% Title information
\title[]{Avoiding Scalability Collapse by \\Restricting Concurrency}         %% [Short Title] is optional;
                                        %% when present, will be used in
                                        %% header instead of Full Title.

%% Author information
%% Contents and number of authors suppressed with 'anonymous'.
%% Each author should be introduced by \author, followed by
%% \authornote (optional), \orcid (optional), \affiliation, and
%% \email.
%% An author may have multiple affiliations and/or emails; repeat the
%% appropriate command.
%% Many elements are not rendered, but should be provided for metadata
%% extraction tools.

\newcommand{\orcidicon}[1]{\href{https://orcid.org/#1}{\includegraphics[scale=0.06]{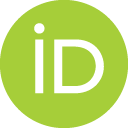}}}

\author{Dave Dice \orcidicon{0000-0001-9164-7747}} 
\orcid{0000-0001-9164-7747}             %% \orcid is optional
\affiliation{Oracle Labs}
\email{first.last@oracle.com}  

\author{Alex Kogan \orcidicon{0000-0002-4419-4340}}           
\orcid{0000-0002-4419-4340}             %% \orcid is optional
\affiliation{Oracle Labs}
\email{first.last@oracle.com}  

\begin{abstract}
Saturated locks often degrade the performance of a multithreaded application, leading 
to a so-called scalability collapse problem.
This problem arises when a growing number of threads circulating through a saturated lock 
causes the overall application performance to fade or even drop abruptly.
This problem is particularly (but not solely) acute on oversubscribed systems (systems with more threads
than available hardware cores).
 
In this paper, we introduce GCR (generic concurrency restriction), a mechanism that aims to avoid the scalability collapse.
GCR, designed as a generic, lock-agnostic wrapper, intercepts lock acquisition calls, and decides
when threads would be allowed to proceed with the acquisition of the underlying lock.
%Threads denied the ability to call the underlying lock by GCR are blocked, preventing them from
%wasting system resources while waiting for the lock and keeping the set of threads circulating through the lock small.
%In this paper, we show how a CR mechanism can be introduced into any lock by means of a generic, lock-agnostic wrapper.
%The wrapper, called GCR, intercepts lock acquisition and release calls, and decides, 
%based on the number of threads trying to acquire the underlying lock,
%when the calling thread would be allowed to do so.
Furthermore, we present GCR-NUMA, a non-uniform memory access (NUMA)-aware extension of GCR, 
that strives to ensure that threads allowed to acquire the lock are those that run on the same socket.
 
The extensive evaluation that includes more than two dozen locks, 
three machines and three benchmarks shows that GCR brings substantial speedup
(in many cases, up to three orders of magnitude) in case of contention and growing thread counts,
while introducing nearly negligible slowdown when the underlying lock is not contended.
%it brings substantial performance benefit (up to tremendous speedups of over 6700x) in case of contention and 
%growing thread counts.
%Moreover, with GCR, the performance of an application does not depend much on the underlying lock.
GCR-NUMA brings even larger performance gains starting at even lighter lock contention.
%Moreover, with GCR, the performance of an application does not depend much on the underlying lock.
%Interestingly, GCR-NUMA applied to a non-NUMA-aware lock beats state-of-the-art NUMA-aware locks by a wide margin.

\end{abstract}


\maketitle

\thispagestyle{fancy}

\section{Introduction}

% 1. Why CR is important
% 1.1 Abundant resources
% 1.2 Adding more resources throttles performance
The performance of applications on multi-core systems is often harmed by {\it saturated} locks, 
where at least one thread is waiting for the lock.
Prior work has observed that as the number of threads circulating through a saturated lock grows,
the overall application performance often fades or even drops abruptly~\cite{JSA10,Dice17,damon09-johnson,BKM12}, 
a behavior called \emph{scalability collapse}~\cite{Dice17}.
This happens because threads compete over shared system resources, such as computing cores and last-level cache (LLC).
For instance, the increase in the number of distinct threads circulating through the lock 
typically leads to increased cache pressure, resulting in cache misses.
At the same time, threads waiting for the lock consume valuable resources and might preempt the lock holder from
making progress with its execution under lock, exacerbating the contention on the lock even further.

\begin{figure}[t]
%\begin{center}
\centering
\includegraphics[width=1\linewidth]{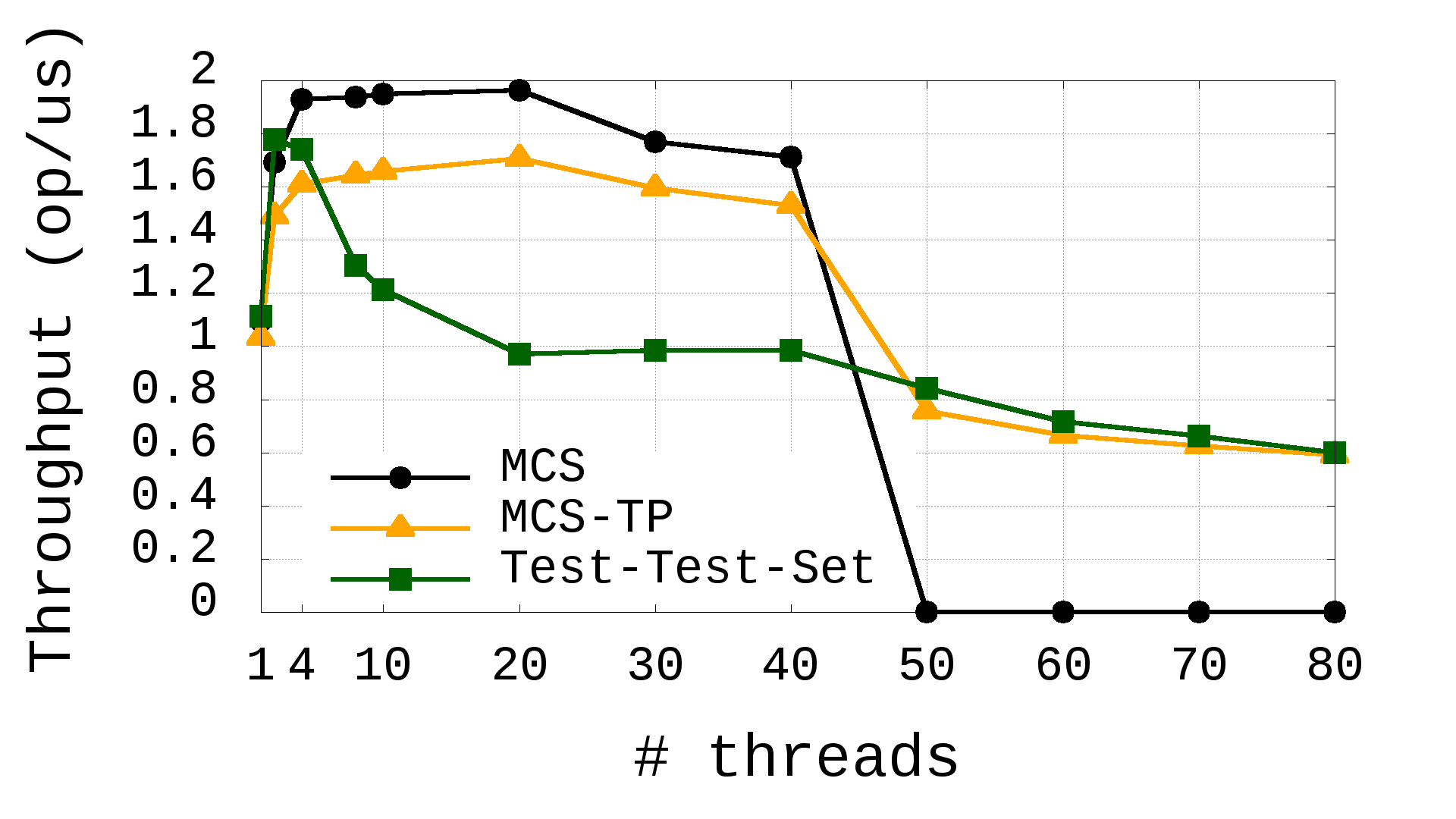}
\caption{Microbenchmark performance with different locks on a 2-socket machine with 20 hyper-threads per socket.}
\figlabel{fig:X6-2=avl-tree-with-delay-absolute-perf-example}
%\end{center}
\end{figure}

An example for scalability collapse can be seen in \figref{fig:X6-2=avl-tree-with-delay-absolute-perf-example} 
that depicts the performance of a key-value map microbenchmark with three popular locks 
on a 2-socket x86 machine featuring $40$ logical CPUs in total (full details of the microbenchmark and the machine are provided later).
The shape and the exact point of the performance decline differ between the locks, 
yet all of them are unable to sustain peak throughput.
With the Test-Test-Set lock, for instance, the performance drops abruptly when more than just a few threads are used, 
while with the MCS lock~\cite{MS91} the performance is relatively stable up to the capacity of the machine and 
collapses once the system gets oversubscribed (i.e., has more threads available than the number of cores).
Note that one of the locks, MCS-TP, was designed specifically to handle oversubscription~\cite{hipc05-he}, 
yet its performance falls short of the peak.

It might be tempting to argue that one should never create a workload where the underlying 
machine is oversubscribed, pre-tuning the maximum number of threads and using a lock, 
such as MCS, to keep the performance stable.
We note that in modern component-based software, 
the total number of threads is often out of the hands of the developer.
A good example would be applications that use thread pools, or even have multiple mutually unaware thread pools.
Furthermore, in multi-tenant and/or cloud-based deployments, where the resources of a 
physical machine (including cores and caches) are often shared between applications running inside 
virtual machines or containers, applications can run concurrently with one
another without even being aware that they share the same machine.
Thus, limiting the maximum number of threads by the number of cores does not help much.
Finally, even when a saturated lock delivers a seemingly stable performance, threads spinning and waiting
for the lock consume energy and take resources (such as CPU time) from other, unrelated tasks\footnote{We
also discuss other waiting policies and their limitations later in the paper.}.

In this paper we introduce {\it generic concurrency restriction} (GCR) to deal with the scalability collapse. 
GCR operates as a wrapper around any existing lock (including POSIX pthread mutexes, and specialized locks provided by an application). 
GCR intercepts calls for a lock acquisition and decides which threads would proceed with
 the acquisition of the underlying lock (those threads are called \emph{active})
and which threads would be blocked (those threads are called \emph{passive}).
%To achieve efficiency, the number of the active threads is kept low 
%(two, in the common case, where one thread holds the lock and another waits for the handoff).
Reducing the number of threads circulating through the locks improves cache performance,
while blocking passive threads reduces competition over CPU time, leading to better 
system performance and energy efficiency.
To avoid starvation and achieve long-term fairness, active and passive threads are shuffled periodically.
We note that the admission policy remains fully work conserving with GCR. 
That is, when a lock holder exits, one of the waiting threads will be able to acquire the lock immediately and enter its critical section.

In this paper we also show how GCR can be extended into a non-uniform access memory (NUMA) setting
of multi-socket machines.
In those settings, accessing data residing in a local cache is far cheaper than accessing  data in a cache located on a remote socket.
Previous research on locks tackled this issue by trying to keep the lock ownership on the same socket~\cite{RH03,topc15-dice,ppopp15-chabbi,DK19}, 
thus increasing the chance that the data accessed by a thread
holding the lock (and the lock data as well) would be cached locally to that thread.
The NUMA extension of GCR, called simply GCR-NUMA, takes advantage of that same idea 
by trying to keep the set of active threads composed of threads running on the same socket.
As a by-product of this construction, GCR-NUMA can convert any lock into a NUMA-aware one.

We have implemented GCR (and GCR-NUMA) in the context of the LiTL library~\cite{GLQ16,LiTL}, 
which provides the implementation of over two dozen various locks.
We have evaluated GCR with all those locks using 
a microbenchmark as well as two well-known database systems (namely, Kyoto Cabinet~\cite{kyotocabinet} and LevelDB~\cite{leveldb}),
on three different systems (two x86 machines and one SPARC). %, each running a different OS (Solaris and two flavors of Linux).
The results show that GCR avoids the scalability collapse, which translates to substantial speedup (up to three orders of magnitude)
in case of high lock contention for virtually
every evaluated lock, workload and machine.
Furthermore, we show empirically that GCR does not harm the fairness of underlying locks
(in fact, in many cases GCR makes the fairness better).
%and maintains overall performance close to the peak even
%when the number of threads is twice the number of available cores.
%This translates to substantial speedup in case of high lock contention.
%The speedup grows higher when the set of active threads is stable, that is the same threads access the lock
%over and over again.
GCR-NUMA brings even larger performance gains starting at even lighter lock contention.
%(Some of those features are exemplified in \figref{fig:X6-2=avl-tree-with-delay-absolute-perf-example}, 
%where we plot the performance of GCR and GCR-NUMA on top of one of the tested locks.)
%Due to lack of space, this paper includes only a small portion of performance evaluation results;
%a longer version of this paper is available at \url{https://arxiv.org/abs/1905.10818}.
We also prove that GCR keeps the progress guarantees of the underlying lock, 
i.e., it does not introduce starvation if the underlying lock is starvation-free.
%In fact, the performance of applications using GCR-NUMA remains close to the peak even
%when the number of threads is twice the number of available cores.

\section{Related Work}
\label{sec:related}
%For a through discussion on those policies, refer to~\cite{}.
%Johnson et al. \cite{} and Lim et al. \cite{} explored the trade-offs between 
%spinning and blocking.  

Prior work has explored adapting the number of active threads based on lock contention~\cite{JSA10,Dice17}.  
However, that work customized certain types of locks, exploiting their specific features, such as the fact that waiting 
threads are organized in a queue~\cite{Dice17}, or that lock acquisition can be aborted~\cite{JSA10}.
Those requirements limit the ability to adapt those techniques into other locks and use them in practice.
For instance, very few locks allow waiting threads to abandon an acquisition attempt, and many spin locks, such 
as a simple Test-Test-Set lock, do not maintain a queue of waiting threads.
Furthermore, the lock implementation is often opaque to the application, e.g., when POSIX pthread mutexes are used.
At the same time, prior research has shown that every lock has its own ``15 minutes of fame",
i.e., there is no lock that always outperforms others 
and the choice of the optimal lock depends on the given application, platform and workload~\cite{TGT13,GLQ16}.
Thus, in order to be practical, a mechanism to control the number of active threads 
has to be lock-agnostic, like the one provided by GCR.

Other work in different, but related contexts 
has observed that controlling the number of threads used by an application
is an effective approach for meeting certain performance goals.
For instance, Raman et al. \cite{pldi11-raman} demonstrate that with a run-time system that
monitors application execution to dynamically adapt the number of worker threads executing 
parallel loop nests.
In another example, Pusukuri et al. \cite{iiswc11-pusukuri} propose a system
that runs an application multiple times for short durations while varying the number of threads, 
and determines the optimal 
number of threads to create based on the observed performance.  
Chadha et al. \cite{cases12-chadha} identified cache-level thrashing as a 
scalability impediment and proposed system-wide concurrency throttling.  
Heirman et al. \cite{hpca14-heirman} suggested intentional undersubscription of threads
as a response to competition for shared caches. 
Hardware and software transactional memory systems use contention managers to 
throttle concurrency in order to optimize throughput \cite{spaa08-yoo}.  
The issue is particularly acute in the context of transactional memory as failed optimistic transactions
are wasteful of resources.

Trading off between throughput and short-term fairness has been extensively explored
in the context of NUMA-aware locks~\cite{RH03,topc15-dice,ppopp15-chabbi,DK19}.
\extabstract{
The main idea behind those locks is to restrict the set of active threads
circulating through the lock to a preferred socket over the short term, 
while enforcing long-term fairness by periodically handing-off the lock to threads on another socket.
}
Those locks do not feature a concurrency restriction mechanism, and in particular, do not avoid contention
on the intra-socket level and the issues resulting from that.
%Our evaluation shows that GCR-NUMA used on top of a non-NUMA-aware lock such as MCS~\cite{MS91}
%beats state-of-the-art NUMA-aware locks way before the number of threads exceeds the capacity of the system.

\remove{
\emph{Cohort locks} \cite{topc15-dice} explored the trade-off 
between throughput and short-term fairness.  Cohort locks restrict the active 
circulating set to a preferred NUMA node over the short term.  They sacrifice
short-term fairness for aggregate throughput, but still enforce long-term fairness.
NUMA-aware locks exploit the inter-socket topology, while our approach focuses 
on intra-socket resources.  
%% OPTIONAL ...
The NUMA-aware HCLH lock \cite{HCLH} edits the nodes of a queue-based lock in a fashion 
similar to that of MCSCR, but does not provide CR and was subsequently discovered 
to have an algorithmic flaw.

Johnson et al. \cite{damon09-johnson} and Lim et al. \cite{tocs93-lim} explored the trade-offs between 
spinning and blocking.  

Ebrahimi et al. \cite{micro11-Ebrahimi} proposed changes to the system scheduler,
informed in part by lock contention and mutual inter-thread DRAM interference,
to shuffle thread priorities in order to improve overall throughput.

Various hardware schemes have been proposed to mitigate LLC thrashing,
but none are available in commonly available processors \cite{js04-suh}.  
Intel \cite{Intel-CAT} allows static partitioning of the LLC in certain
models designed for real-time environments. 
}

\section{Background}
\label{sec:background}
\extabstract{
Locks continue to remain a key synchronization construct for most applications as well as the the topic of active research.
The following list includes 
only a subset of papers published just in the last couple of years and offering new lock constructs or investigating related areas:
\cite{Eastep-smartlocks,locks-ols,cacm15-bueso,ieeetocs15-cui,ppopp15-chabbi,apsys15-Kashyap,ppopp16-chabbi,ppopp16-wang,transact16-dice,GLQ16,sac16-gustedt,usenixatc16-falsafi,ppopp16-ramalhete,KMK17}.
}

Contending threads must wait for the lock when it is not available.
There are several common waiting policies.
The most simple one is unbounded spinning, also known as busy-waiting or polling.
There, the waiting threads spin on a global or local memory location and wait until the value in that location changes.
%The change may indicate that the lock became available, or that the lock has been handed off to another thread.
Spinning consumes resources and contributes to preemption when the system is oversubscribed, i.e., has more ready threads
than the number of available logical CPUs.
Yet, absent preemption, it is simple and provides fast lock handover times, and for those reasons used by many popular locks,
e.g., Test-Test-Set.

An alternative waiting policy is parking, where a waiting thread voluntarily releases its CPU 
and passively waits (by blocking) for another thread to unpark it when the lock becomes available.
%There are various ways to implement park-unpark facility, e.g., using futexes in Linux or special system calls in Solaris~\cite{Dice17}.
Parking is attractive when the system is oversubscribed, as it releases CPU resources for threads ready to run, including 
the lock holder.
However, the cost of the voluntary context switching imposed by parking is high, which translates to longer lock handover times
when the next owner of the lock has to be unparked.

To mitigate the overhead of parking and unparking on the one hand, and limit the shortcomings of unlimited spinning on the other hand, 
lock designers proposed a hybrid spin-then-park policy.
There, threads spin for a brief period, and park if the lock is still not available by the end 
of that time.
While tuning the optimal time for spinning is challenging~\cite{damon09-johnson,tocs93-lim}, 
it is typically set to the length of the context-switch round trip~\cite{Dice17}.

\section{Generic Concurrency Restriction}
\label{sec:GCR}
\subsection{Overview}
\label{sec:GCR-overview}

GCR wraps a lock API, that is, calls to, e.g., \code{Lock}/\code{Unlock} methods go through the corresponding methods of GCR.
In our implementation, we interpose on the standard POSIX \code{pthreads\_mutex\_lock} and \code{pthreads\_mutex\_unlock} methods.
Thus, using the standard LD\_PRELOAD mechanism in Linux and Unix, GCR can be made immediately available to any 
application that uses the standard POSIX API, even without recompiling the application or its locks.

In the following description, we distinguish between \emph{active} threads, that is, threads allowed by GCR to invoke the API of the underlying lock,
and \emph{passive} threads, which are not allowed to do so.
Note that this distinction is unrelated to the running state of the corresponding threads.
That is, active threads may actually be blocked (parked) if the underlying lock decides doing so, while passive threads 
may be spinning, waiting for their turn to join the set of active threads.
In addition, given that GCR by itself does not provide lock semantics (even though it implements the lock API), 
we will refer to the underlying lock simply as \emph{the lock}.

GCR keeps track of the number of active threads.
When a thread invokes the \code{Lock} method wrapped by GCR, GCR 
checks whether the number of active threads is larger than a preconfigured threshold.
If not, a thread proceeds by calling the lock's \code{Lock} method.
This constitutes the fast path of the lock acquisition.
Otherwise, GCR detects that the lock is saturated, and places the (passive) thread into a (lock-specific) queue.
This queue is based on a linked list; each node in the list is associated with a different thread.
Every thread in the queue but the first can choose whether 
to spin on a local variable in its respective node, yield the CPU and park, or any combination of thereof.
(The thread at the head of the queue has to spin as it monitors the number of active threads.)
In practice, we choose the spin-then-park policy for all passive threads in the queue but the first,
to limit the use of system resources by those threads.
Once the thread at the head of the queue detects that there are no active threads, it leaves the queue, 
signals\footnote{We use
the word \emph{signal} throughout the paper in its abstract form, unrelated to the OS inter-process communication mechanism of signals.} the next thread (if exists) that 
the head of the queue has changed (unparking that thread if necessarily), and proceeds by calling the lock's \code{Lock} method.

\extabstract{
\begin{figure}
\centering
\subfloat[][A typical spin-then-park lock]{\includegraphics[width=0.49\linewidth]{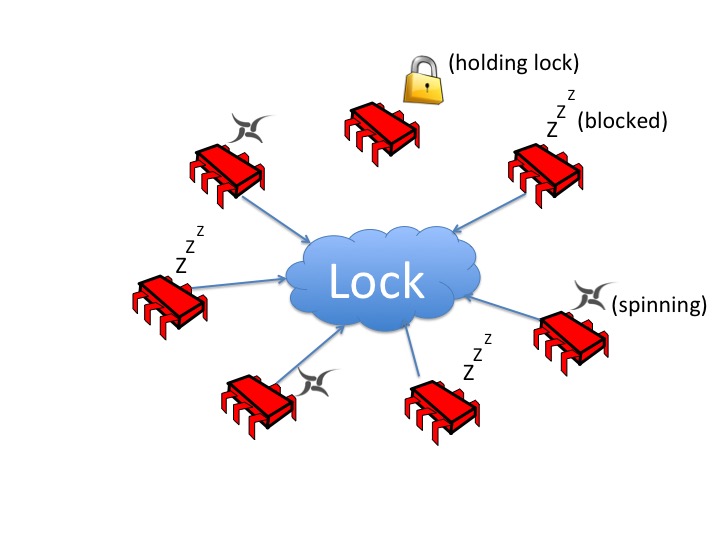}}
\subfloat[][The same lock with GCR]{\includegraphics[width=0.49\linewidth]{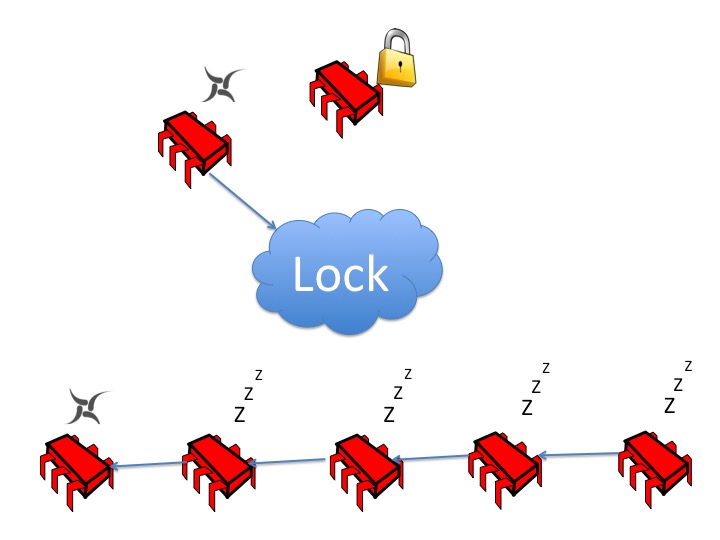}}
\caption{Schematic view of the GCR mechanism}
\label{fig:scheme}
\end{figure}

The schematic view of a typical lock implementation versus the one enabled with GCR is depicted in Figure~\ref{fig:scheme}.
Note that without GCR, it is up to the lock designer to manage the waiting policy for the lock.
As discussed in Section~\ref{sec:background}, various policies exist, but for the sake of visualization, we 
assume a hybrid spin-then-park policy where some threads waiting for the lock may be spinning while others may be parked.
When GCR is used, however, since the active set of threads is typically small (specifically, consisting of a lock holder and another thread
waiting to get the lock next), we expect the handoff of the lock to be frequent enough to avoid the need to park the waiting thread (or a few threads).
Thus, the handoff is expected to be fast.
At the same time, most passive threads, except for one, will be parked, even if the lock itself is not designed to do that, freeing system resources from excessive usage.
}

When a thread invokes GCR's \code{Unlock} method, 
it checks whether it is time to signal the (passive) thread at the head of the queue to join the set of active threads.
This is done to achieve a long-term fairness, preventing starvation of passive threads.
To this end, GCR keeps a simple counter for the number of lock acquisitions.
(Other alternatives, such as timer-based approaches, are possible.)
Following that, GCR calls the lock's \code{Unlock} method.

\subsection{Technical Details}
\label{sec:GCR-details}
The auxiliary data structures used by GCR are given in \figref{fig:structures}\footnote{For the clarity of exposition, 
we assume sequential consistency. 
Our actual implementation uses memory fences as well as \code{volatile} keywords and padding (to avoid false sharing) where necessarily.}.
The \code{Node} structure represents a node in the queue of passive threads.
In addition to the successor and predecessor nodes in the list, the \code{Node} structure
contains the \code{event} flag. 
This flag is used to signal a thread when its node moves to the head in the queue.

The \code{LockType} structure contains the internal lock metadata  (passed to the \code{Lock} and \code{Unlock} functions of that lock)
and a number of additional fields:
\begin{itemize}
\item \code{top} and \code{tail} are the pointers to the first and the last nodes in the queue of passive threads, respectively.
\item \code{topApproved} is a flag used to signal the passive thread at the top of the queue that it can join the set of active threads.
\item \code{numActive} is the counter for the number of active threads.
\item \code{numAcqs} is the counter for the number of lock acquisitions. It is used to move threads from the passive set to the active set, as explained below.
\end{itemize}

In addition to the \code{LockType} structure, GCR uses \code{nextLock} (\code{nextUnlock}) function pointer,
which is initialized to the \code{Lock} (\code{Unlock}, respectively) function of the underlying lock. 
The initialization code is straightforward (on Linux it can use the \code{dlsym} system call), and thus is not shown.

\begin{figure}
\begin{lstlisting}[style=nonumbers]
typedef struct _Node {
  struct _Node * next; 
  struct _Node * prev; 
  int event;
} Node;

typedef struct {
  lock_t internalMutex;

  Node * top;
  Node * tail;
  int topApproved;

  int numActive;

  int numAcqs;
} LockType ;

static int (*nextLock)(lock_t *);
static int (*nextUnlock)(lock_t *);
\end{lstlisting}
\caption{Auxiliary structures.}
\figlabel{fig:structures}
\end{figure}

\begin{figure}
\begin{lstlisting}[style=numbers]
int Lock(LockType *m) {   
    /* if there is at most one active thread */
    if (m->numActive <= 1) 			 										@\linelabel{lock:3}@
    	/* go to the fast path */				
        FAA(&m->numActive, 1);					 							@\linelabel{lock:5}@
        goto FastPath;						 								@\linelabel{lock:6}@
    }																	@\linelabel{lock:7}@
    
SlowPath:    															@\linelabel{lock:8}@
    /* enter the MCS-like queue of passive threads */								
    Node *myNode = pushSelfToQueue(m);									@\linelabel{lock:10}@
    
    /* wait (by spin-then-park) for my @\newline@       node  to get to the top */						
    if (!myNode->event) Wait(myNode->event); 									@\linelabel{lock:12}@
    
    /* wait (by spinning) for a signal to join the @\newline@       set of active threads */
    while (!m->topApproved) {    												@\linelabel{lock:14}@
    	Pause();   															@\linelabel{lock:15}@
	/* stop waiting if no active threads left */
	if (m->numActive == 0) break;											@\linelabel{lock:17}@
    }																	@\linelabel{lock:18}@
    
    if (m->topApproved != 0) m->topApproved = 0;								@\linelabel{lock:19}@
    FAA(&m->numActive, 1);												@\linelabel{lock:20}@

    popSelfFromQueue(m, myNode);											@\linelabel{lock:21}@    
    
FastPath:																@\linelabel{lock:22}@
    return nextLock(&m->internalMutex);										@\linelabel{lock:23}@
}
\end{lstlisting}
\caption{Lock procedure.}
\figlabel{fig:lock}
\end{figure}

The implementation of the GCR's \code{Lock} function is given in \figref{fig:lock}.
When executing this function, a thread first checks the current number of active threads 
by reading the \code{numActive} counter (\lineref{lock:3}).
If this number is at most one, 
it atomically (using a fetch-and-add (FAA) instruction if available) increments it (\lineref{lock:5}) 
and continues to the fast path (\lineref{lock:21}).
Note that the comparison to $1$ in \lineref{lock:3} effectively controls when the concurrency restriction is enabled.
That is, if we would want to enable concurrency restriction only after detecting that, say, $X$ threads are waiting for the lock,
we could use $X+1$ instead (+1 to account for the lock holder).
Also, we note that the check in \lineref{lock:3} and the increment in \lineref{lock:5} are not mutually atomic, 
that is, multiple threads can reach \lineref{lock:5} and thus increment the counter stored in \code{numActive} concurrently.
However, the lack of atomicity may only impact performance (as the underlying lock will become more contended), and not correctness.
Besides, this should be rare when the system is in the steady state.
Finally, note that the FAA operation in \lineref{lock:5} is performed by active threads only (rather than all threads), 
limiting the overhead of this atomic operation on a shared memory location.

In the fast path, the thread simply invokes the \code{Lock} function of the underlying lock (\lineref{lock:23}).

The slow path is given in \linerangeref{lock:8}{lock:21}.
There, the thread joins the queue of passive threads (\lineref{lock:10}); 
the code of the \code{pushSelfToQueue} function is given in \figref{fig:queue} and described below.
Next, the thread waits until it reaches the top of the queue (\lineref{lock:12}).
This waiting is implemented by spin-then-park waiting policy in the \code{Wait} function\footnote{The technical details of the parking/unparking mechanism 
are irrelevant for the presentation, but briefly,
we used futexes on Linux and a mutex with a condition variable on Solaris.},
and we assume that when \code{Wait} returns, the value of \code{event} is non-zero.
We note that we could use pure spinning (on the \code{event} field) in the \code{Wait} function as well.

Once the thread reaches the top of the queue, it starts monitoring the signal from active threads to join the active set.
It does so by spinning on the \code{topApproved} flag (\lineref{lock:14}).
In addition, this thread monitors the number of active threads by reading 
the \code{numActive} counter (\lineref{lock:17}).
Note that unlike the \code{topApproved} flag, this counter changes on every lock acquisition and release.
Thus, reading it on every iteration of the spinning loop would create unnecessary coherence traffic and slow down active threads 
when they attempt to modify this counter.
In Section~\ref{sec:optimization} we describe a simple optimization that allows to read this counter less frequently while still 
monitoring the active set effectively.

Once the passive thread at the top of the queue breaks out of the spinning loop, 
it resets the \code{topApproved} flag if needed (\lineref{lock:19}) and atomically increments the \code{numActive} counter (\lineref{lock:20}).
Then it removes itself from the queue of passive threads (\lineref{lock:21}) and continues with the code of the fast path.
The pseudo-code of the \code{popSelfFromQueue} function is given in \figref{fig:queue} and described below.

\ContinueLineNumber
\begin{figure}[t]
\begin{lstlisting}[style=numbers]
int Unlock (LockType * m) {											
  /* check if it is time to bring someone from @\newline@     the passive to active set */
  if (((m->numAcqs++ % THRESHOLD) == 0) && m->top != NULL) {				@\linelabel{unlock:3}@
    /* signal the selected thread that it can go */
    m->topApproved = 1;												@\linelabel{unlock:5}@
  }																@\linelabel{unlock:6}@

  FAA(&m->numActive, -1);											@\linelabel{unlock:7}@

  /* call underlying lock */
  return nextUnlock(&m->internalMutex);									@\linelabel{unlock:9}@
}																@\linelabel{unlock:end}@
\end{lstlisting}
\caption{Unlock procedure.}
\figlabel{fig:unlock}
\end{figure}

The \code{Unlock} function is straight-forward (see \figref{fig:unlock}).
The thread increments the \code{numAcqs} counter and checks whether it is time to bring a passive thread 
into the set of active threads (\lineref{unlock:3}).
Notice that in our implementation we decide to do so based solely 
on the number of lock acquisitions, while other, more sophisticated approaches are possible.
For our evaluation, \code{THRESHOLD} is set to 0x4000.
Afterwards, the thread atomically decrements the \code{numActive} counter (\lineref{unlock:7}).
Finally, it calls the \code{Unlock} function of the underlying lock (\lineref{unlock:9}).

\ContinueLineNumber
\begin{figure}[t]
\begin{lstlisting}[style=numbers]
Node *pushSelfToQueue(LockType * m) {									
  Node * n = (Node *)malloc(sizeof(Node));								@\linelabel{push:2}@
  n->next  = NULL;													@\linelabel{push:3}@
  n->event = 0;														@\linelabel{push:4}@
  Node * prv = SWAP (&m->tail, n);										@\linelabel{push:5}@
  if (prv != NULL) {													@\linelabel{push:6}@
    prv->next = n;													@\linelabel{push:7}@
  } else {															@\linelabel{push:8}@
    m->top = n;														@\linelabel{push:9}@
    n->event = 1;														@\linelabel{push:10}@
  }																@\linelabel{push:11}@	

  return n;															@\linelabel{push:end}@
}

void popSelfFromQueue(LockType * m, Node * n) {							
  Node * succ = n->next;												@\linelabel{pop:2}@
  if (succ == NULL) {													@\linelabel{pop:3}@
    /* my node is the last in the queue	*/						
    if (CAS (&m->tail, n, NULL)) {										@\linelabel{pop:5}@
      CAS (&m->top, n, NULL);											@\linelabel{pop:6}@
      free(n);														@\linelabel{pop:7}@
      return;															@\linelabel{pop:8}@
    }																@\linelabel{pop:9}@
    for (;;) {															@\linelabel{pop:11}@
      succ = n->next ;													@\linelabel{pop:12}@
      if (succ != NULL) break;											@\linelabel{pop:13}@				
      Pause();														@\linelabel{pop:14}@
    }																@\linelabel{pop:15}@
  }																@\linelabel{pop:16}@

  m->top = succ;													@\linelabel{pop:17}@
  /* unpark successor if it is parked in @\code{Wait}@ */
  succ->event = 1;													@\linelabel{pop:18}@

  free(n);															@\linelabel{pop:19}@	
}																@\linelabel{pop:end}@
\end{lstlisting}
\caption{Queue management procedures.}
\figlabel{fig:queue}
\end{figure}

The procedures for inserting and removing a thread to/from the queue of passive threads are fairly simple, yet not trivial.
Thus, we opted to show them in \figref{fig:queue}.
(Readers familiar with the MCS lock~\cite{MS91} will recognize close similarity to procedures used by that lock to manage waiting threads.)
In order to insert itself into the queue, a thread allocates and initializes a new node (\linerangeref{push:2}{push:4})\footnote{In our implementation, 
we use a preallocated array of \code{Node} objects, one per thread per lock, as part of the infrastructure provided by the LiTL library~\cite{LiTL}.
It is also possible to allocate a \code{Node} object statically in \code{Lock()} and pass it the the \code{push} and \code{pop} functions, 
avoiding the dynamic allocation of \code{Node} objects altogether.}.
Then it atomically swaps the tail of the queue with the newly created node (\lineref{push:5}) using an atomic swap instruction.
If the result of the swap is non-NULL, then the thread's node is not the only node in the queue; 
thus, the thread updates the \code{next} pointer of its predecessor (\lineref{push:7}).
Otherwise, the thread sets the \code{top} pointer to its newly created node (\lineref{push:9}) and sets the \code{event} flag (\lineref{push:10}).
The latter is done to avoid the call to \code{Wait} in \lineref{lock:12}.

The code for removing the thread from the queue is slightly more complicated.
Specifically, the thread checks first whether its node is the last in the queue (\lineref{pop:3}).
If so, it attempts to update the \code{tail} pointer to NULL using an atomic compare-and-swap (CAS) instruction (\lineref{pop:5}).
If the CAS succeeds, the thread attempts to set the \code{top} pointer to NULL as well (\lineref{pop:6}).
Note that we need CAS (rather than a simple store) for that as the \code{top} pointer may have been already updated concurrently in \lineref{push:9}.
This CAS, however, should not be retried if failed, since a failure means that the queue is not empty anymore and
thus we should not try to set \code{top} to NULL again.
The removal operation is completed by deallocating (or releasing for future reuse) the thread's node (\lineref{pop:7}). 

If the CAS in \lineref{pop:5} is unsuccessful, the thread realizes that its node is no longer the last in the queue, 
that is, the queue has been concurrently updated in \lineref{push:5}.
As a result, it waits (in the for-loop in \linerangeref{pop:11}{pop:15}) until the \code{next} pointer of its node is updated in \lineref{push:7} by a new successor.
Finally, after finding that its node is not the last in the queue (whether immediately in \lineref{pop:3} or after the failed CAS in \lineref{pop:5}), 
the thread updates the \code{top} pointer to its successor in the queue (\lineref{pop:17}) and 
signals the successor (\lineref{pop:18}) to stop waiting in the \code{Wait} function (cf.~\lineref{lock:12}).

\subsection{Correctness}
A lock is \emph{starvation-free} if every attempt for its acquisition eventually succeeds.
In this section, we argue that the GCR algorithm does not introduce starvation as long as the 
underlying lock is starvation-free, the OS scheduler does not starve any thread 
and the underling architecture supports starvation-free atomic increment and swap operations.
On a high level, our argument is built on top of two observations, namely that once a thread
enters the queue of waiting threads, it eventually reaches the top of the queue,
and that a thread at the top of the queue eventually calls the \code{Lock} function of the underlying lock.

\begin{lemma}
\label{lemma:1}
The \code{tail} pointer always either holds a \code{NULL} value or points to a node whose \code{next} pointer is \code{NULL}.
\end{lemma}

\begin{proofsketch}
The \code{tail} pointer initially holds \code{NULL}.
From inspecting the code, the \code{tail} pointer may change only in \lineref{push:5} or \lineref{pop:5}.
Consider the change in \lineref{push:5}.
The value of the \code{tail} pointer is set to a node whose \code{next} field was initialized to NULL (cf.~\lineref{push:3}).
Thus, the lemma holds when the change in \lineref{push:5} takes place.

The \code{next} pointer of a node gets modified only in \lineref{push:7}.
The node whose \code{next} pointer gets modified is the one pointed by \code{tail} before 
the change of \code{tail} in \lineref{push:3} took place.
As a result, when \lineref{push:7} is executed, the \code{tail} pointer does not point anymore to the node whose
\code{next} pointer gets updated, and the lemma holds.

Next, consider the change to \code{tail} in \lineref{pop:5} done with a CAS instruction.
If CAS is successful, the \code{tail} pointer gets a \code{NULL} value.
Otherwise, the \code{tail} pointer is not updated (and thus 
remains pointing to a node whose \code{next} pointer is \code{NULL}).
Thus, the lemma holds in both cases.
\end{proofsketch}

We define the state of the queue at time $T$ to be all nodes that are reachable from \code{top} and \code{tail} at time $T$.
We say that a passive thread enters the queue when it finishes executing \lineref{push:5} 
and leaves the queue when it finishes executing CAS in \lineref{pop:6} or an assignment in \lineref{pop:17}.

\begin{lemma}
\label{lemma:2}
The \code{event} field of any node in the queue, except for, perhaps, the node pointed by \code{top}, is $0$.
\end{lemma}

\begin{proofsketch}
First, we observe that only a thread whose node has a non-zero \code{event} value 
can call \code{popSelfFromQueue} (cf.~\lineref{lock:12}).

Next, we show that at most one node in the queue has a non-zero \code{event} value.
The \code{event} field is initialized to $0$ (\lineref{push:4}), and is set to $1$ either in \lineref{push:10}
or in \lineref{pop:18}.
In the former case, this happens when the corresponding thread finds the queue empty (\lineref{push:5}).
Thus, when it sets the \code{event} field of its node to $1$, the claim holds.
In the latter case, a thread $t$ sets the \code{event} field in the node of its successor in the queue.
However, it does so after removing its node from the queue (by updating the \code{top} pointer to its successor in \lineref{pop:17}).
Based on the observation above, $t$'s node contains a non-zero \code{event} field.
By removing its node from the queue and setting the \code{event} field in the successor node, $t$ maintains the claim.

Finally, we argue that the node with a non-zero \code{event} value is the one pointed by \code{top}.
Consider, again, the two cases where the \code{event} field gets set.
In the first case, it is set by a thread that just entered the queue and found the queue empty (and thus updated \code{top} to point to its node in \lineref{push:9}).
In the second case, it is set by a thread $t$ that just updated \code{top} to point to the node of its successor in the queue.
At this point (i.e., after executing \lineref{pop:17} and before executing \lineref{pop:18}), no node in the queue contains a non-zero \code{event} value.
Thus, based on the observation above, no thread can call \code{popSelfFromQueue}.
At the same time, any new thread entering the queue will find at least $t$'s successor there and thus, will not change the \code{top} pointer.
Hence, when $t$ executes \lineref{pop:18}, it sets the \code{event} field of the node pointed by \code{top}.
\end{proofsketch}

We refer to a thread whose node is pointed by \code{top} as a \emph{thread at the top of the queue}.
\begin{lemma}
\label{lemma:3}
Only a thread at the top of the queue can call \code{popSelfFromQueue}.
\end{lemma}

\begin{proofsketch}
After entering the queue, a thread may leave it (by calling \code{popSelfFromQueue}) 
only after it finds the \code{event} field in its node holding a non-zero value (cf.~\lineref{lock:12}).
According to Lemma~\ref{lemma:2}, this can only be the thread at the top of the queue.
\end{proofsketch}

We order the nodes in the queue according to their rank, which is the number of links (\code{next} pointers) to be traversed
from $top$ to reach the node.
The rank of the node pointed by \code{top} is $0$, the rank of the next node is $1$ and so on.
The rank of a (passive) thread is simply a rank of its corresponding node in the queue.

\begin{lemma}
\label{lemma:4}
The queue preserves the FIFO order, that is, a thread $t$ that enters the queue will leave the queue after 
all threads that entered the queue before $t$ and before all threads that enter the queue after $t$.
\end{lemma}

\begin{proofsketch}
By inspecting the code, the only place where the \code{next} pointer of a node may change is in \lineref{push:7} 
(apart from the initialization in~\lineref{push:3}).
If a thread executes \lineref{push:7}, then according to Lemma~\ref{lemma:1}, it changes the \code{next} pointer
of a node of its predecessor in the queue from \code{NULL} to the thread's node.
Thus, once a \code{next} pointer is set to a non-\code{NULL} value, it would never change again
(until the node is deleted, which will happen only after the respective thread leaves the queue).
Hence, threads that enter the queue after $t$ will have a higher rank than $t$.

Lemma~\ref{lemma:3} implies that only a thread with rank $0$ can leave the queue.
Thus, any thread that joins the queue after $t$ will leave the queue after $t$, as $t$'s rank will reach $0$ first.
Also, any thread that enters the queue before $t$ will have a lower rank than $t$.
Therefore, $t$ cannot leave the queue before all those threads do.
\end{proofsketch}

\begin{lemma}
\label{lemma:5}
A thread at the top of the queue eventually calls the \code{Lock} function of the underlying lock.
\end{lemma}

\begin{proofsketch}
By inspecting the code, when a thread $t$ reaches the top of the queue, its \code{event} field changes to $1$ (\lineref{push:10} and \lineref{pop:18}).
Thus, it reaches the \code{while} loop in \lineref{lock:14}.
It would stay in this loop as long as the \code{topApproved} field does not change or when all active threads have left.
If the latter does not happen, however, assuming that the lock is starvation-free and no thread holds it indefinitely long, active
threads will circulate through the lock, incrementing the \code{numAcqs} counter.
Once this counter reaches the threshold, the \code{topApproved} field will be set, releasing $t$ from the \code{while} loop.

Following that, the only other place where $t$ may spin before calling the \code{Lock} function of the underlying lock
is in the \code{for} loop in \lineref{pop:11}.
This would occur in a rare case where $t$ may not find a successor in \lineref{pop:2}, but the successor will appear and update
\code{tail} (in \lineref{push:5}) right before $t$ does (in \lineref{pop:5}).
However, assuming a scheduler that does not starve any thread, the successor will eventually update the \code{next}
pointer in $t$'s node, allowing $t$ to break from the \code{for} loop.
\end{proofsketch}

\begin{lemma}
\label{lemma:6}
A thread that enters the queue eventually becomes a thread at the top of the queue.
\end{lemma}

\begin{proofsketch}
Consider a thread $t$ entering the queue by executing \lineref{push:5}.
If it finds no nodes in the queue (i.e., \code{tail} holds a \code{NULL} value), $t$ sets \code{top}
to point to its node (\lineref{push:9}) and the Lemma trivially holds.
Otherwise, some other thread $t_1$ is at the top of the queue.
By Lemma~\ref{lemma:5}, that thread will eventually call the \code{Lock} function of the underlying lock.
However, before doing so, it will remove itself from the queue (\lineref{lock:21}).
According to Lemma~\ref{lemma:4}, the queue preserves the FIFO order.
Thus, the next thread after $t_1$ will become the thread at the top of the queue, and 
eventually call the \code{Lock} function of the underlying lock.
By applying the same argument recursively, we can deduct that $t$ will reach the top of the queue (after all threads
with a lower rank do so).
\end{proofsketch}

\begin{theorem}
When GCR is applied to a starvation-free lock L, the resulting lock is starvation-free.
\end{theorem}

\begin{proofsketch}
Given that the underlying lock L is starvation-free, we need to show that every thread calling GCR's \code{Lock} function would
eventually call L's \code{Lock} function.
In case a thread finds at most one active thread in \lineref{lock:3}, it proceeds by calling to the L's \code{Lock} function
(after executing an atomic FAA instruction, which we assume is starvation-free).
%\footnote{We note that this assumption 
%does not hold on platforms that do not support atomic FAA and require emulation via CAS or Load-Linked/Store-Conditional instructions.}).
Otherwise, it proceeds on the slow path by entering the queue (\lineref{lock:10}).
Assuming the atomic swap operation is starvation-free, the thread would eventually execute \lineref{push:5} and enter the queue.
By Lemma~\ref{lemma:6}, it will eventually reach the top of the queue, and by Lemma~\ref{lemma:5}, 
it will eventually call L's \code{Lock} function. 
\end{proofsketch}

\subsection{Optimizations}
\label{sec:optimization}
\textbf{Reducing overhead at low contention:}
When the contention is low, GCR may introduce overhead by repeatedly sending threads to the slow path where they would discover 
immediately that no more active threads exist and they can join the set of active threads.
This behavior can be mitigated by tuning the thresholds for joining the passive and active sets (i.e., constants in \lineref{lock:3} and~\lineref{lock:17}, respectively).
%Note that toning down the concurrency restriction, however, also means that threads can spin idly and waste system resources at lower contention levels.
In our experiments, we found that setting the threshold for joining the passive set at $4$ (\lineref{lock:3}) and the threshold for joining 
the active set at half of that number (\lineref{lock:17}) was a reasonable compromise between reducing the overhead of GCR and letting threads spin idly
at low contention (if the underlying lock allows so).

\textbf{Reducing overhead on the fast path:}
Even with the above optimizations, employing atomic instructions on the fast path to maintain the counter of active threads 
degrades performance when the underlying lock is not contended, or contended lightly.
To cope with that, one would like to be able to disable GCR (including counting the number of active threads)
dynamically when the lock is not contended, and turn it on only when the contention is back.
However, the counter of active threads was introduced exactly for the purpose of identifying the contention on the underlying lock.

We solve this "chicken and egg problem" as following: we introduce one auxiliary array, which is shared by all threads acquiring any lock.
Each thread writes down in its dedicated slot the address of the underlying lock it is about to acquire, and clears the slot after releasing that lock.
After releasing a lock, a thread scans periodically the array, and counts the number of threads trying to acquire the lock it just released.
If it finds that number at or above a certain threshold (e.g., $4$), it enables GCR by setting a flag in the \code{LockType}
structure.
Scanning the array after every lock acquisition would be prohibitively expensive.
Instead, each thread counts its number of lock acquisitions (in a thread-local variable) and scans the array
after an exponentially increasing number of lock acquisitions.
Disabling GCR is easier -- when it is time to signal a passive thread (cf.~\lineref{unlock:3}), and the queue of passive threads is empty while
the number of active threads is small (e.g., 2 or less), GCR is disabled until the next time
contention is detected.

\textbf{Splitting the counter of active threads:}
The counter of active threads, \code{numActive}, is modified twice by each thread for every critical section.
The contention over this counter can be slightly reduced by employing a known technique of splitting the 
counter into two, \code{ingress} and \code{egress}.
The former is incremented (using atomic FAA instruction) in the entry to the critical section (i.e., \code{Lock}), 
while the latter is incremented in the exit (\code{Unlock}) with a simple store as the increment is done under the lock.
The difference of those two counters gives the estimate for the number of currently active threads.
(The estimate is because those counters are not read atomically).

\textbf{Spinning loop optimization:}
As mentioned above, repeated reading of the \code{numActive} counter 
(or the \code{ingress} and \code{egress} counters, as described above) 
inside the spinning loop of the thread at the top of the queue (cf.~\linerangeref{lock:14}{lock:18} in \figref{fig:lock})
creates contention over the cache lines storing those variables.
We note that monitoring the size of the active set is required 
to avoid a livelock that may be caused when the 
last active lock holder leaves without setting the \code{topApproved} flag.
This is unlikely to happen, however, when the lock is heavily contended and acquired repeatedly by the
same threads.

Based on this observation, we employ a simple deterministic back-off scheme that increases the time interval 
between subsequent reads of the \code{numActive} counter by the spinning 
thread as long as it finds the active set not empty (or, more precisely, as long as it finds the size of the active set being larger than
the threshold for joining that set when the optimization for reducing overhead at low contention described above is used).
To support this scheme, we add the \code{nextCheckActive} field into the \code{LockType} structure, initially set to $1$, 
and also use a local counter initialized to $0$ before the \code{while} loop in \lineref{lock:14} of \figref{fig:lock}.
In every iteration of the loop, we increment the local counter and check if its value modulo \code{nextCheckActive} is equal $0$.
If so, we check the \code{numActive} counter.
As before, we break out of the loop if this counter has a zero value, after resetting \code{nextCheckActive} to 1 so the next thread
at the top of the queue will start monitoring the active set closely.
Otherwise, if the \code{numActive} is non-zero, we double the value of \code{nextCheckActive},
up to a preset boundary (1M in our case).

It should be noted that like many synchronization mechanisms, GCR contains several knobs that can be used to tune its performance.
In the above, we specify all the default values that we have used for our experiments.
While evaluating the sensitivity of GCR to each configuration parameter is in the future work, 
we note that our extensive experiments across multiple platforms and applications provide empirical evidence
that the default parameter values represent a reasonable choice.

\remove{
We have employed those three optimizations in our implementation of GCR.
The implementation of the three optimizations described so far is shown in~\figref{fig:spin-loop}.

\begin{figure}
\begin{lstlisting}[style=nonumbers]
int cnt = 0;
while (!m->topApproved) {    												
  Pause();   															
  if ((++cnt % m->nextCheckActive) == 0) {
    /* stop waiting if only a few active threads left */
    if (m->egress - m->ingress >= -2) {
      m->nextCheckActive = 1;
      break;
    }
    if (m->nextCheckActive < MAX_VALUE) {
      m->nextCheckActive *= 2;
    }
}    
\end{lstlisting}
\caption{Optimized spinning loop with a split counter of active threads.}
\figlabel{fig:spin-loop}
\end{figure}
}

\section{NUMA-aware GCR}
As GCR controls which threads would join the active set, it may well do so in a NUMA-aware way.
In practice, this means that it should strive to maintain the active set composed of threads running on the same
socket (or, more precisely, on the same NUMA node).
Note that this does not place any additional restrictions on the underlying lock, which might be a NUMA-aware lock by itself or not.
Naturally, if the underlying lock is NUMA-oblivious, the benefit of such an optimization would be higher.

Introducing NUMA-awareness into GCR requires relatively few changes.%, which are summarized next.
On a high level, instead of keeping just one queue of passive threads per lock, we keep a number of queues, one per socket.
Thus, a passive thread joins the queue corresponding to the socket it is running on.
In addition, we introduce a notion of a preferred socket, which is a socket that gets preference in decisions which threads
should join the active set.
In our case, we set the preferred socket solely based on the number of lock acquisitions 
(i.e., the preferred socket is changed in a round-robin fashion every certain number of lock acquisitions), 
but other refined (e.g., time-based) schemes are possible.

We say that a (passive) thread is eligible (to check whether it can join the active set) if it is running on 
the preferred socket \emph{or} the queue (of passive threads) of the preferred socket is empty.
When a thread calls the \code{Lock} function, we check whether it is eligible and let it proceed 
with examining the size of the active set (i.e., read the \code{numActive} counter) only if it is.
Otherwise, it immediately goes into the slow path, joining the queue according to its socket.
This means that once the designation of the preferred socket changes (when threads running
on that socket acquire and release the lock ``enough'' times),
active threads from the now not-preferred socket will become passive when they attempt to acquire the lock again.

Having only eligible threads monitor the size of the active set
has two desired consequences.
First, only the passive thread at the top of the queue corresponding to the preferred socket will 
be the next thread (out of all passive threads) to join the set of active threads.
This keeps the set of active threads composed of threads running on the same (preferred) socket and ensures
long-term fairness.
Second, non-eligible threads (running on other, non-preferred sockets) do not access the counter of 
active threads (but rather wait until they become eligible), reducing contention on that counter.
%Going forward, as active threads from the de-prioritized socket become passive, passive threads from the currently
%prioritized socket will replace them, keeping the lock between the threads running on the same socket.

\remove{
The code for the NUMA-aware GCR, which also includes the optimization described in Section~\ref{sec:optimization}, 
is given in the Appendix.
}

\section{Evaluation}
\label{sec:evaluation} 

\remove{
\begin{itemize}
\item CR in action (on various machines)
	\begin{itemize}
	\item{benchmark1}
	\item{benchmark2}
	\item{benchmark3}
	\item{Multiple tenants}
	\end{itemize}
\item Fairness
\item Overhead (static linkage)
\item The cost of generality (CR vs. Malthusian)
\item Other variants (NUMA awareness, ultra-fast)
\item The benefit of MWAIT?
\end{itemize}
}

\remove {
\begin{table*}
\begin{tabular}{clp{9cm}ccc}\\
\# & Lock Name & Brief Description & \makecell{Waiting \\ policy} &  \makecell{NUMA-\\aware?} & Source\\
1 & alockepfl (orig) & Uses a fixed-size array with one slot per thread; threads spin locally on the slot they acquire via atomic increment & Spin & N & \cite{} \\
2 & backoff (orig) & A simple test-and-set spinlock with exponential backoff & Spin & N & \cite{} \\
3 & cbomcs (spin) & A cohort lock with the MCS lock as a local (per-socket) lock and the backoff lock as a global lock  & Spin & Y & \cite{} \\
4 & cbomcs (spin+park) & Same as above but the local MCS lock uses spin-then-park waiting  & spin+park & Y & \cite{} \\
5 & clhepfl (orig) & A variant of the CLH lock (see below) & Spin & N & \cite{} \\
6 & clh (spin) & TBD & spin & N & \cite{} \\
7 & clh (spin+park) & TBD & spin+park & N & \cite{} \\
8 & concurrency (orig) & A simple wrapper over pthread lock with some stats & pthread & ? & \cite{} \\
\end{tabular}
\end{table*}
}

We implemented GCR as a stand-alone library conforming to the pthread mutex lock API defined by the POSIX standard.
%Thus, any software that uses this standard API is able to utilize GCR without any code change or even recompilation.
We integrated GCR into LiTL~\cite{LiTL}, an open-source project providing an implementation of dozens of various locks, including
well-known established locks, such as MCS~\cite{MS91} and CLH~\cite{CLH}, 
as well as more recent ones, such as NUMA-aware Cohort~\cite{topc15-dice} and HMCS locks~\cite{ppopp15-chabbi}.
The LiTL library also includes the implementation of a related Malthusian lock~\cite{Dice17}, which introduces a concurrency restriction mechanism into the MCS lock.
Furthermore, the LiTL library allows specifying various waiting policies (e.g., spin or spin-then-park) for locks that support that (such as MCS, CLH or Cohort locks).
Overall, we experimented with $24$ different lock+waiting policy combinations in LiTL (for brevity, we will refer to each lock+waiting policy combination simply as a lock).
For our work, we enhanced the LiTL library to support the Solaris OS (mainly, by supplying functionality to park and
unpark threads) as well as the SPARC architecture (mainly, by abstracting synchronization primitives).
We note that for polite spinning, we use the MWAIT instruction made available on the SPARC M7 processors~\cite{Dice17}.

We run experiments on three different platforms:
\begin{itemize}
\setlength\itemsep{0em}
\item Oracle X6-2 server with 2 Intel Xeon E5-2630 v4 processors featuring 10 hyper-threaded cores each (40 logical CPUs in total) and running Fedora 25.
\item Oracle T7-2 server with 2 SPARC M7 processors featuring 32 cores each, each supporting 8 logical CPUs (512 logical CPUs in total) and running Solaris 11.3.
\item Oracle X5-4 server with 4 Intel Xeon E7-8895 v3 processors featuring 18 hyper-threaded cores each (144 logical CPUs in total) and running Oracle Linux 7.3.
\end{itemize}
As most high-level conclusions hold across all evaluated platforms, 
we focus our presentation on X6-2; unless said otherwise, the reported numbers are from this platform.

\extabstract{
We run experiments on three different platforms.
The first one is Oracle X6-2 server with 2 Intel Xeon E5-2630 v4 processors featuring 10 hyper-threaded cores each (40 logical CPUs in total) and running Fedora 25.
The other two platforms are an Oracle T7-2 server with 2 SPARC M7 processors (512 logical CPUs in total) and an Oracle X5-4 server with 4 Intel Xeon E7-8895 v3 processors (144 logical CPUs in total).
As most high-level conclusions hold across all evaluated platforms,
we focus our presentation on X6-2. 
We include results from the other two systems in the longer version of the paper.
}

\remove{
We run experiments on three different platforms.
For this paper, we focus on a dual-socket x86-based system with 40 logical CPUs in total.
The qualitative results from other two systems (a four-socket x86-based system with 144 logical CPUs in total and 
a dual-socket SPARC-based system with 512 logical CPUs in total) 
were similar, and are included in the longer version of the paper~\cite{gcr-arxiv}.
}

In all experiments, we vary the number of threads up to twice the capacity of each machine.
We do not pin threads to cores, relying on the OS to make its choices.
In all experiments, we employ a scalable memory allocator~\cite{ADM11}.
%which does not use the pthread mutex primitives for synchronization by itself.
We disable the turbo mode on Intel-based platforms to avoid the effect of that mode, which varies with the number of threads, on the results.
Each reported experiment has been run 3 times in exactly the same configuration.
Presented results are the average of results reported by each of those 3 runs.

\subsection{AVL Tree Microbenchmark}
\label{sec:evaluation-avl-tree}

\begin{figure*}[t]
\subfloat[][MCS spin]{\includegraphics[width=0.25\linewidth]{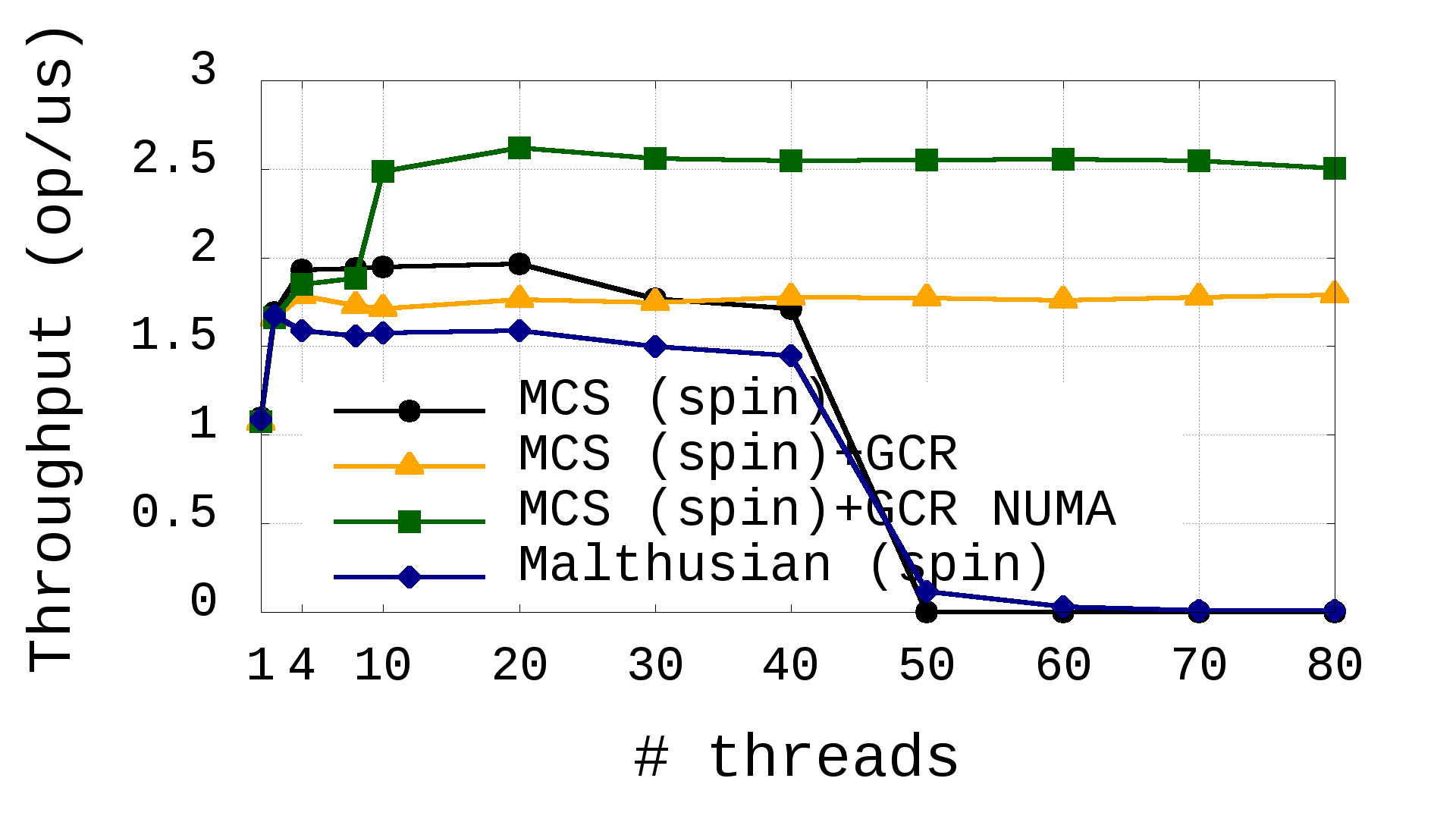}}
\subfloat[][MCS spin-then-park]{\includegraphics[width=0.25\linewidth]{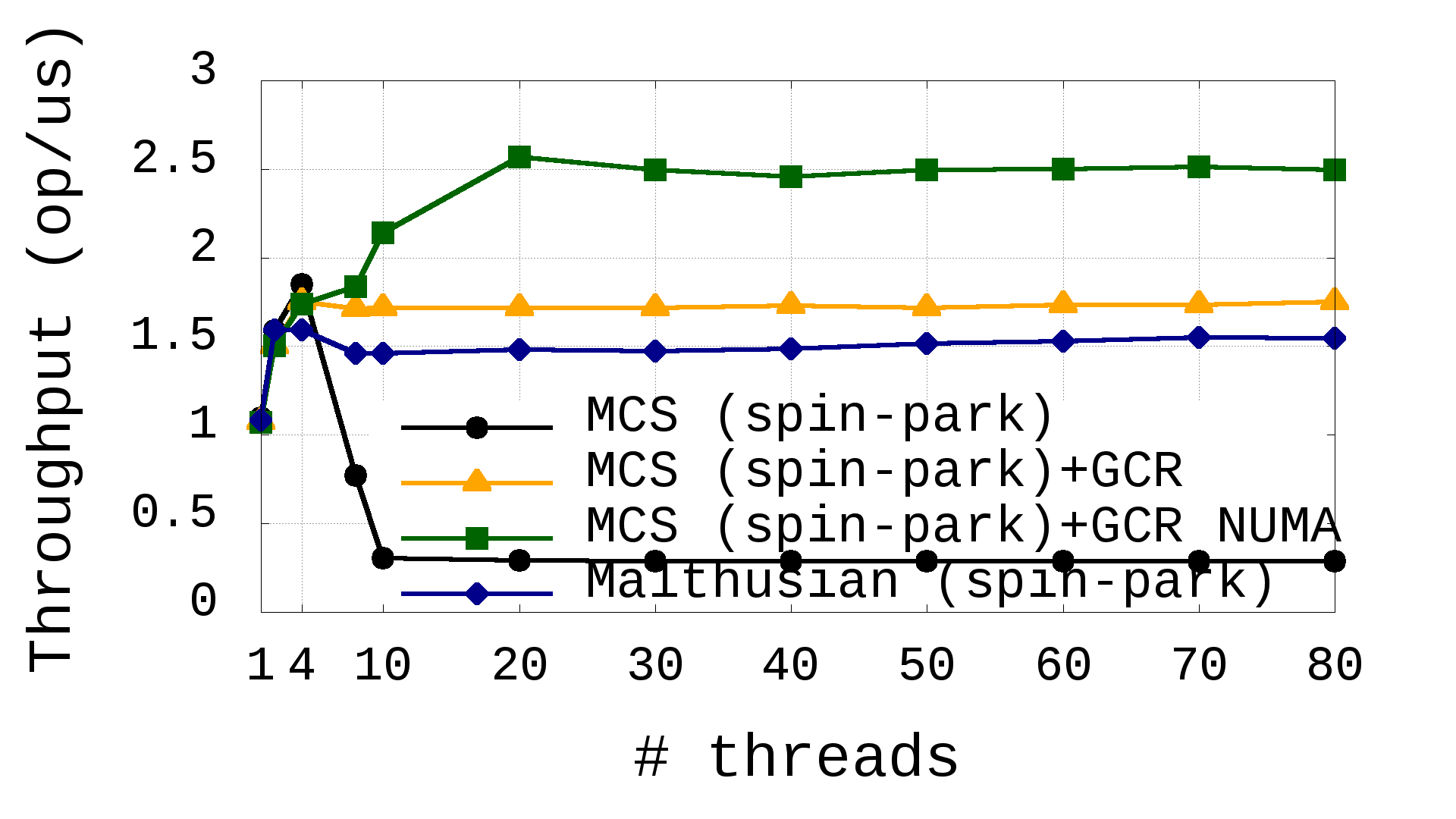}}
\subfloat[][Test-Test-Set]{\includegraphics[width=0.25\linewidth]{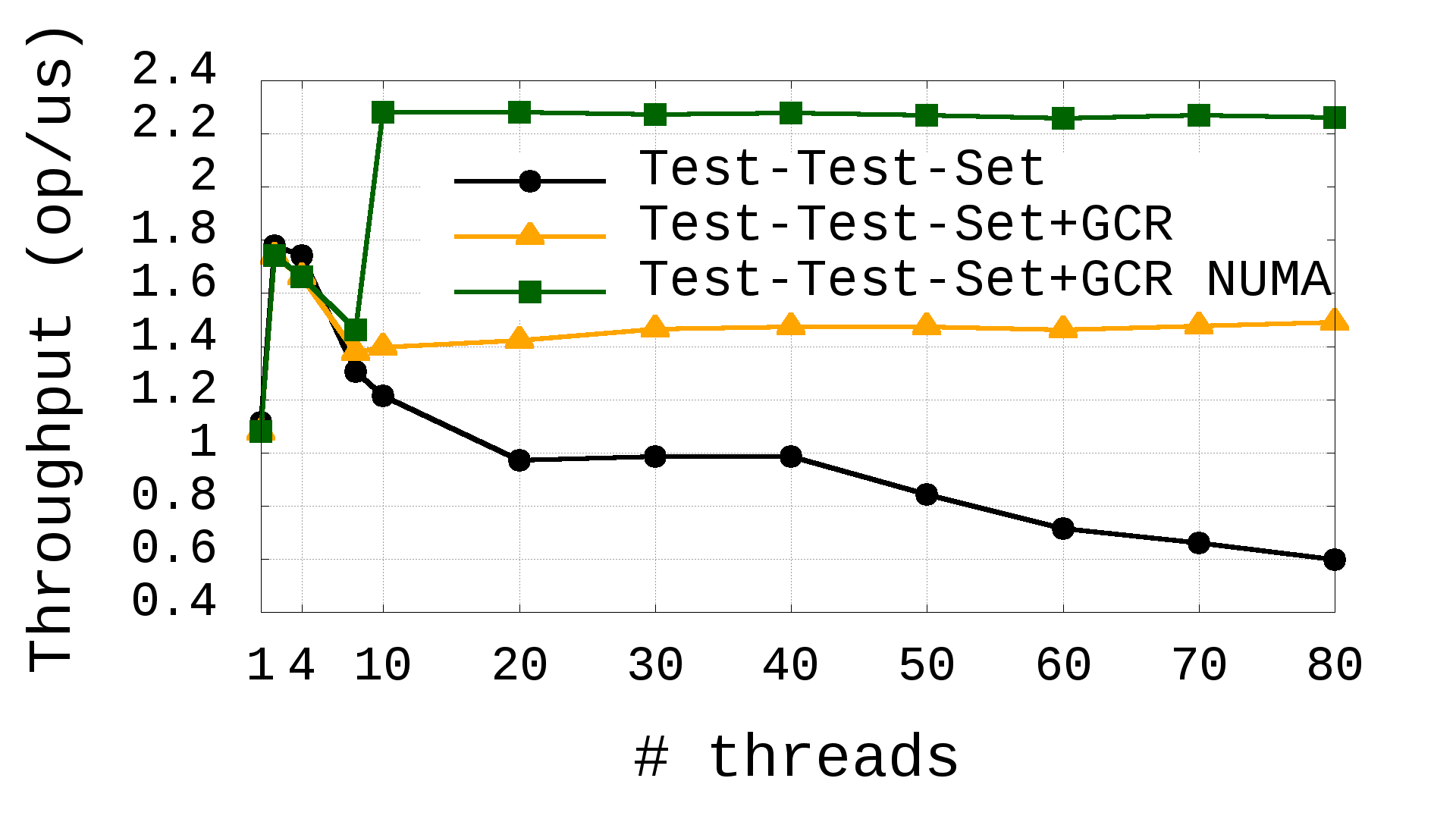}}
\subfloat[][Pthread mutex]{\includegraphics[width=0.25\linewidth]{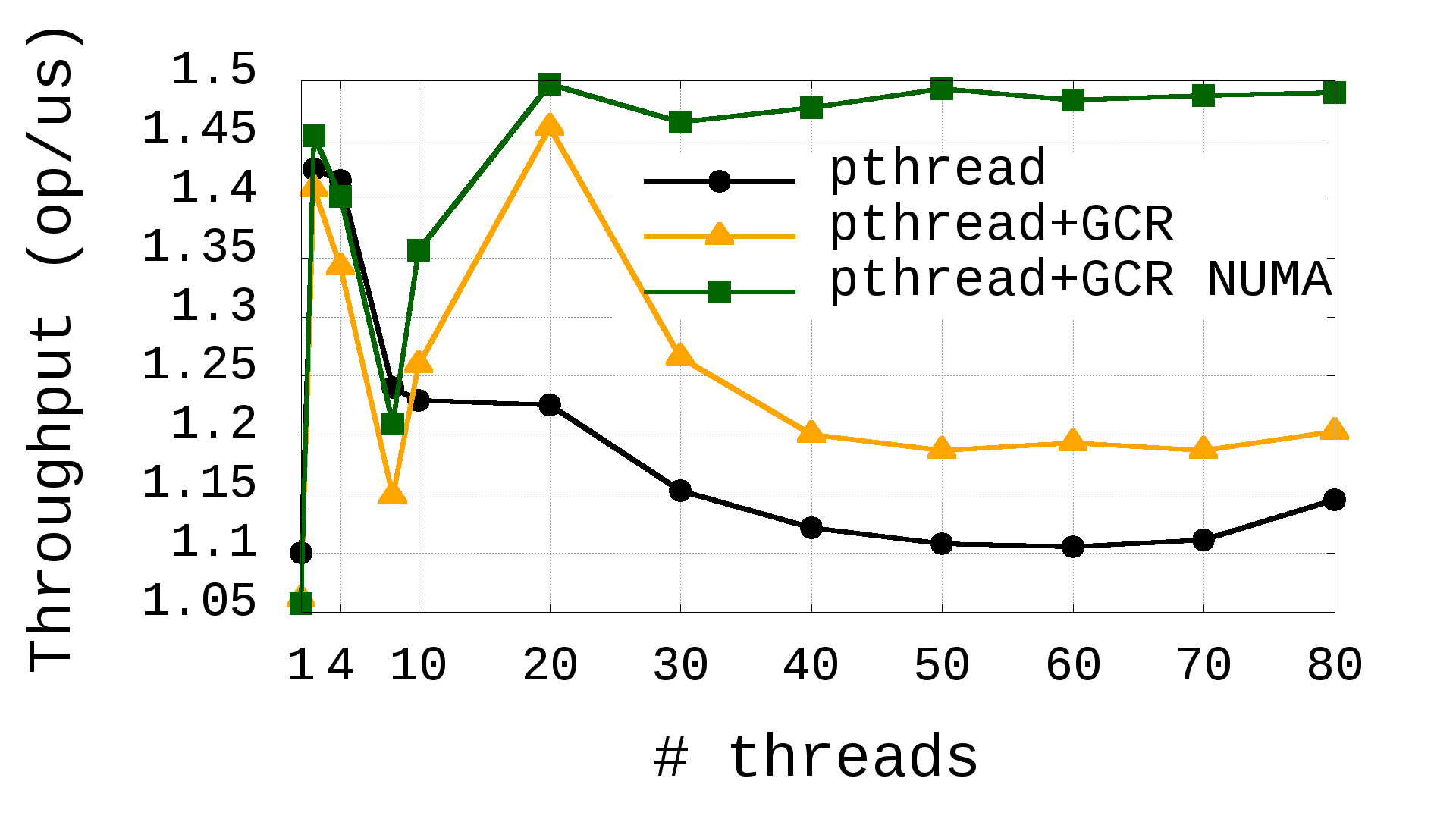}}
\caption{Throughput results for the MCS, Test-Test-Set  and POSIX pthread mutex locks (AVL tree).}
\figlabel{fig:X6-2=avl-tree-with-delay-absolute-perf}
\end{figure*}

\begin{figure*}[t]
\subfloat[][MCS Spin]{\includegraphics[width=0.25\linewidth]{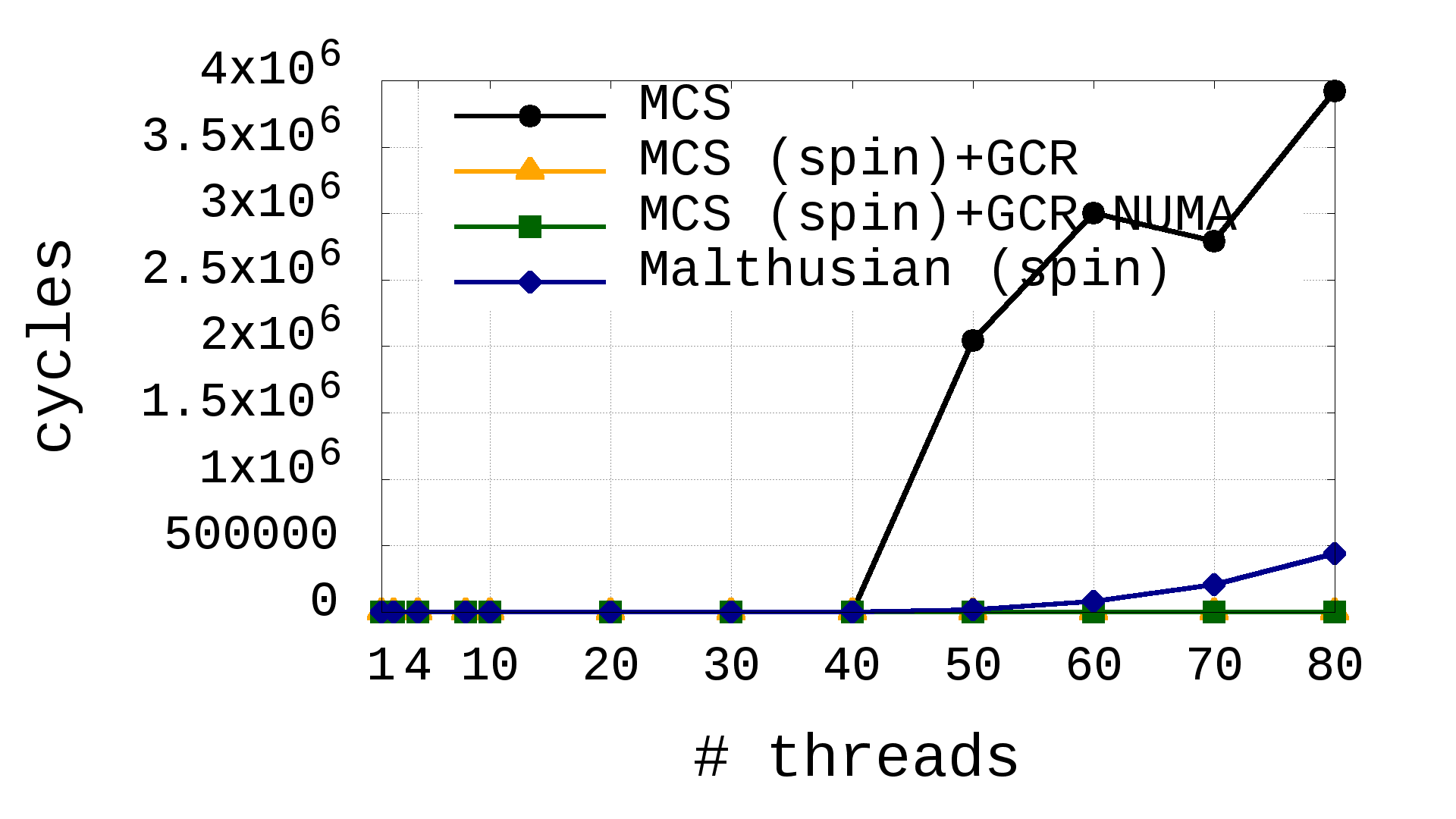}}
\subfloat[][MCS Spin-then-park]{\includegraphics[width=0.25\linewidth]{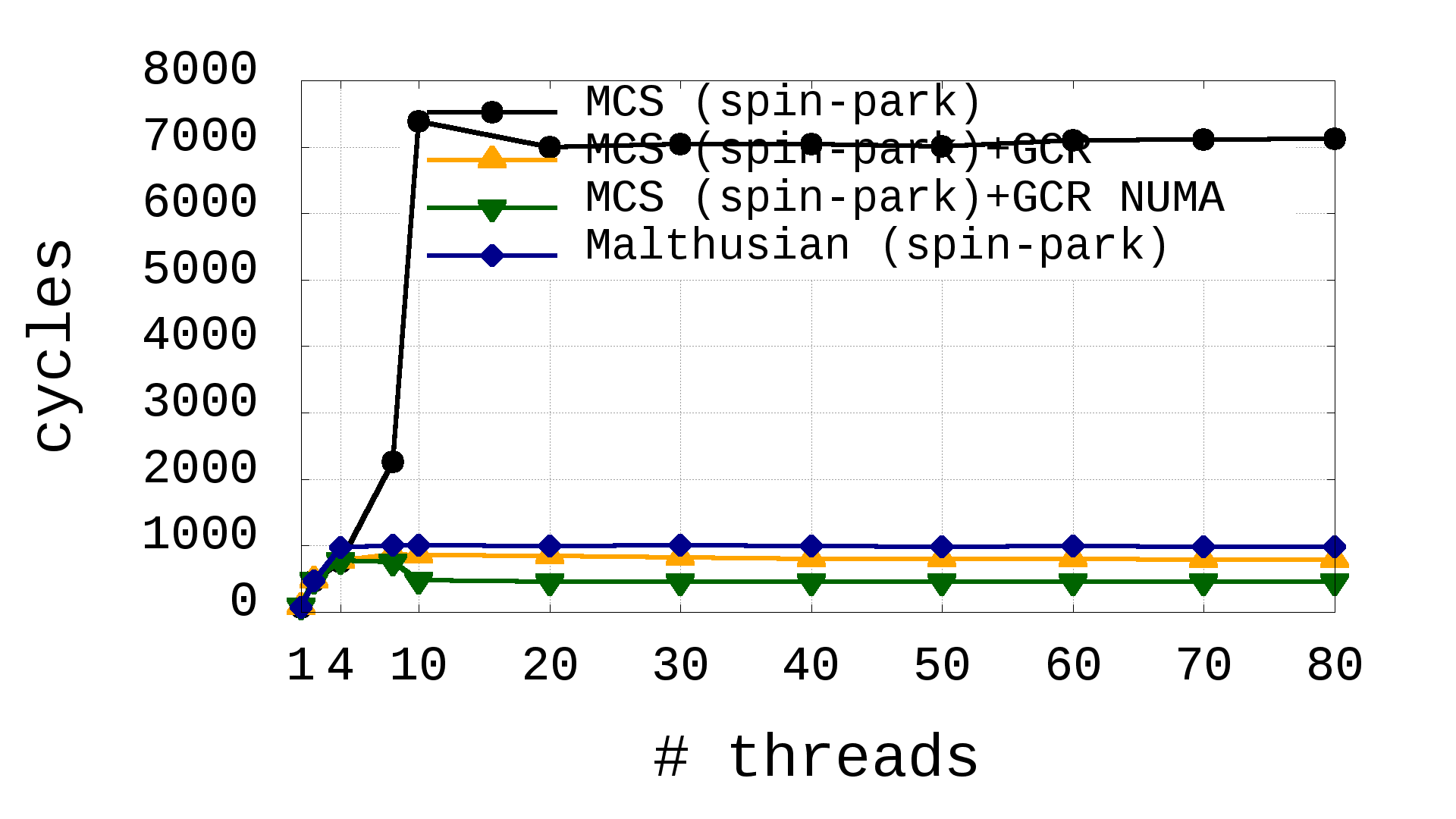}}
\subfloat[][Test-Test-Set]{\includegraphics[width=0.25\linewidth]{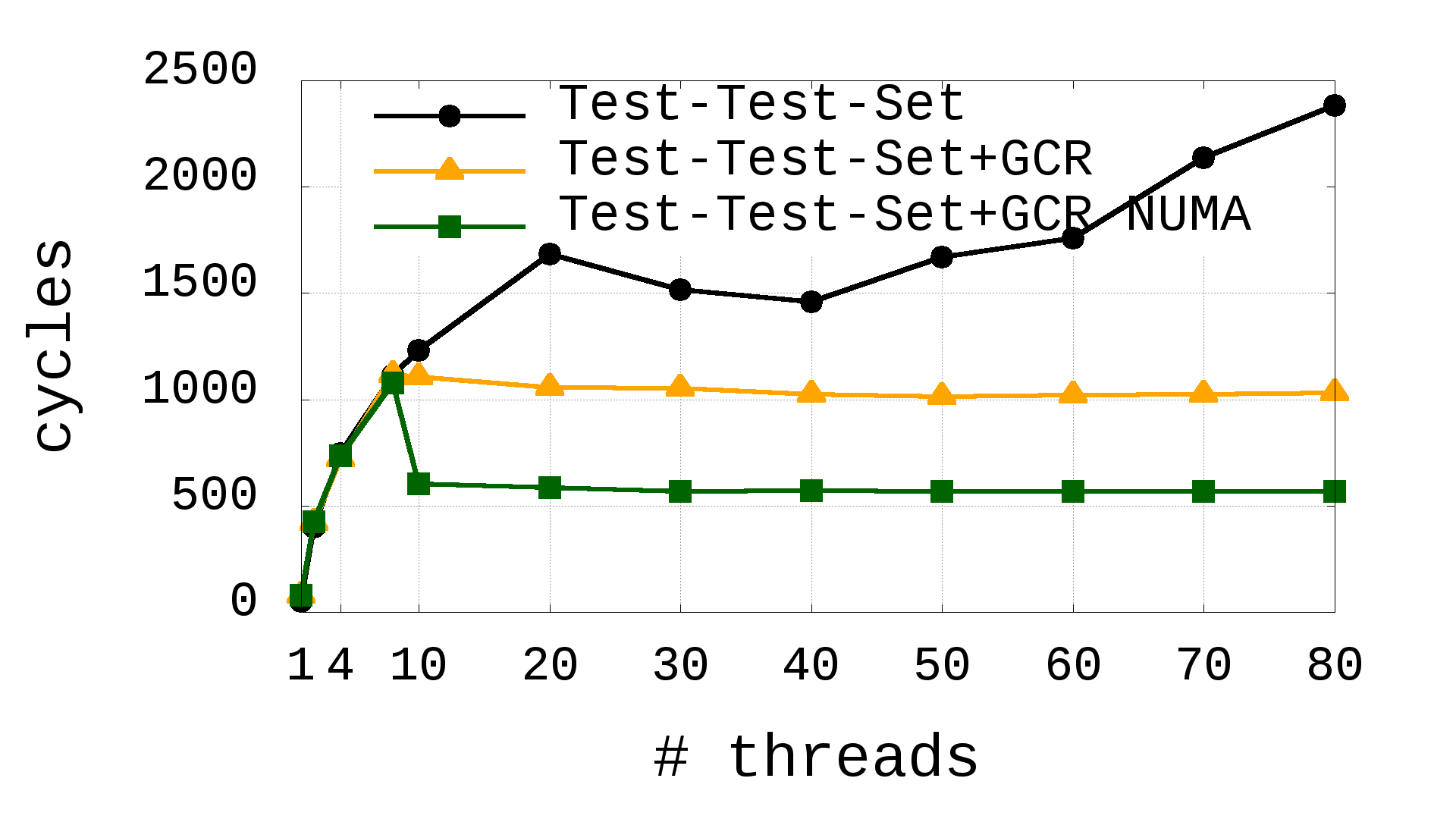}}
\subfloat[][Pthread mutex]{\includegraphics[width=0.25\linewidth]{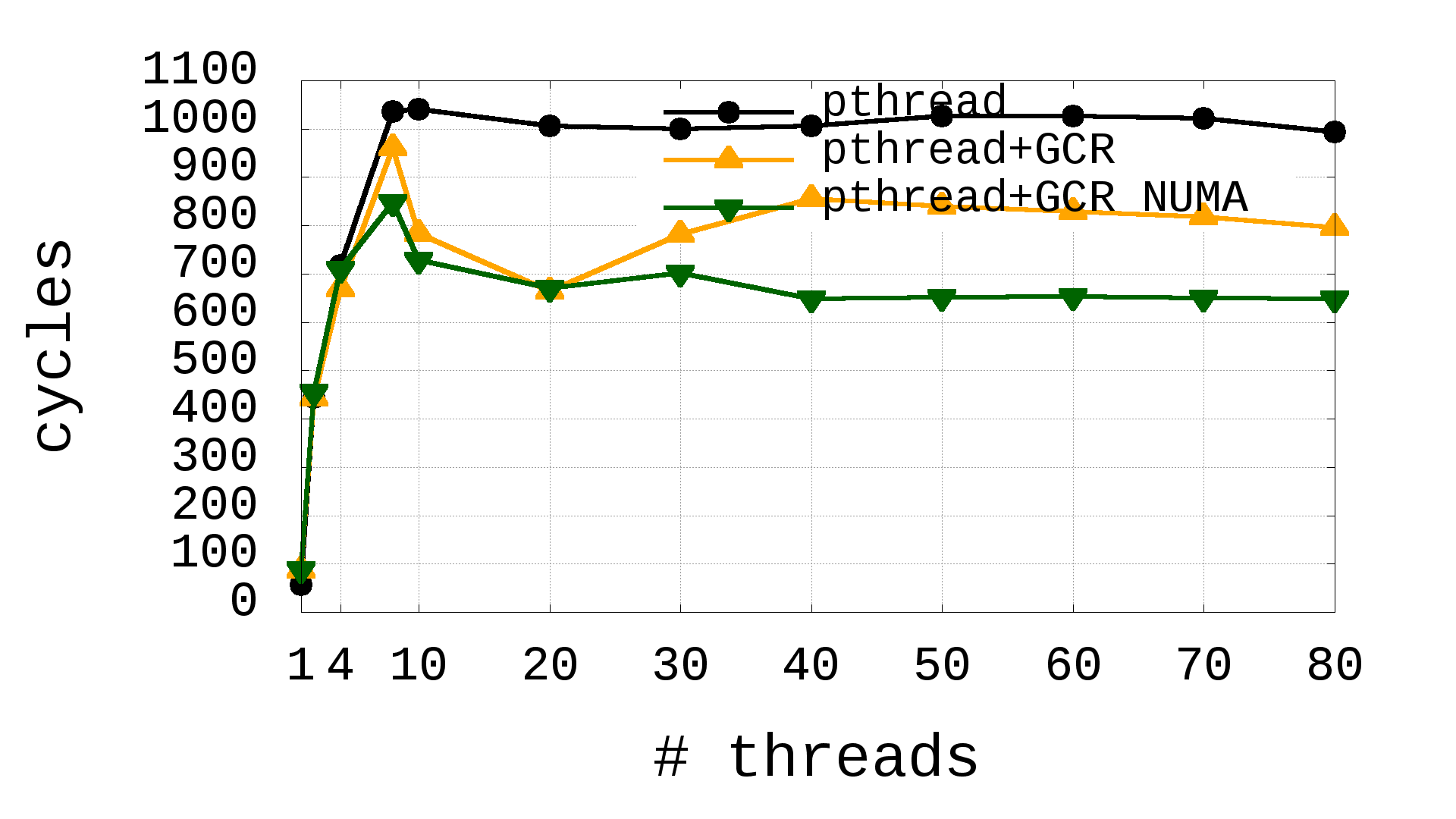}}
\caption{Lock handeoff time for the MCS, Test-Test-Set  and POSIX pthread mutex locks (AVL tree).}
\figlabel{fig:X6-2=avl-tree-with-delay-handoff}
\end{figure*}

\begin{figure*}[t]
\subfloat[][MCS Spin]{\includegraphics[width=0.25\linewidth]{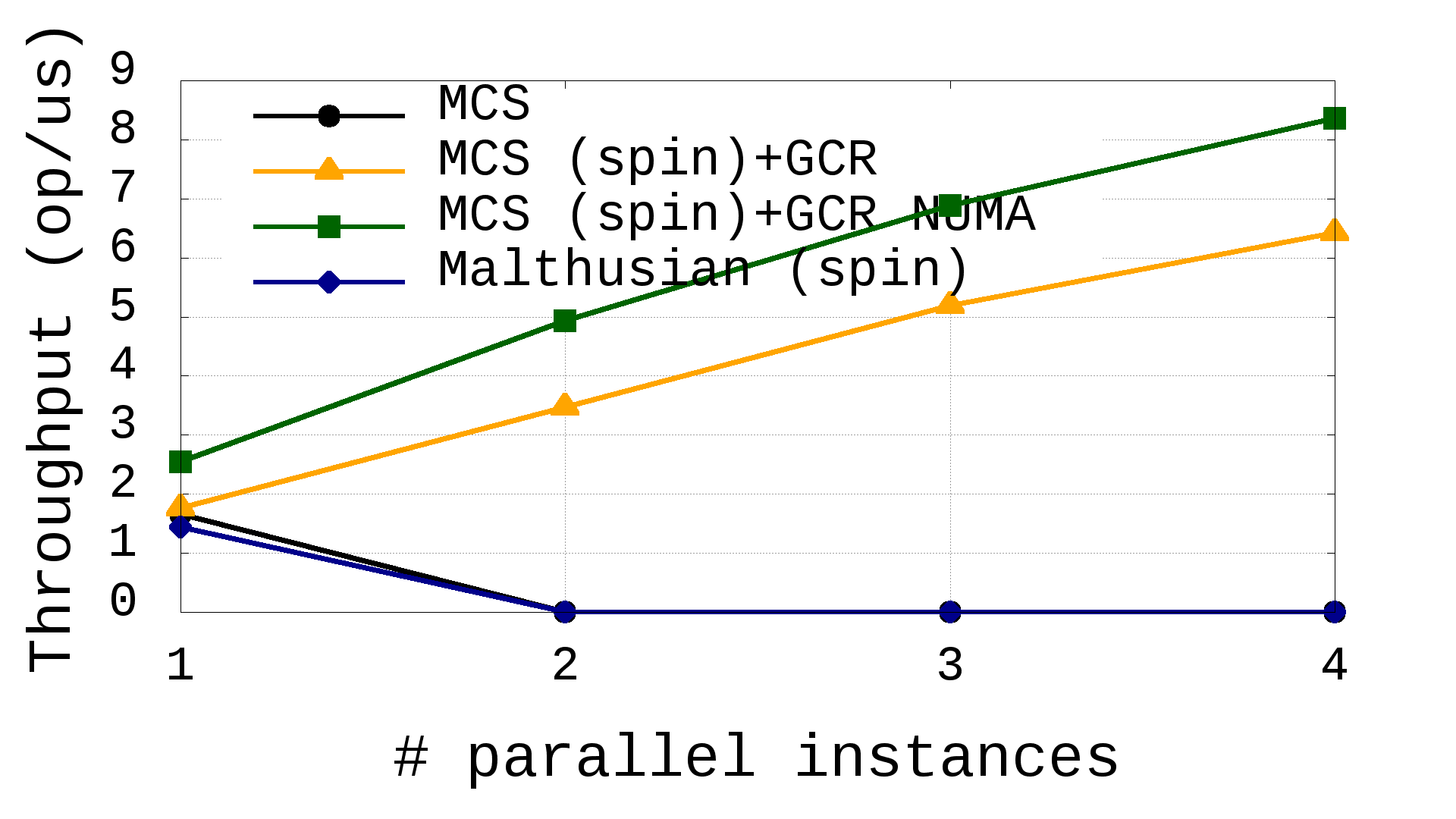}}
\subfloat[][MCS Spin-then-park]{\includegraphics[width=0.25\linewidth]{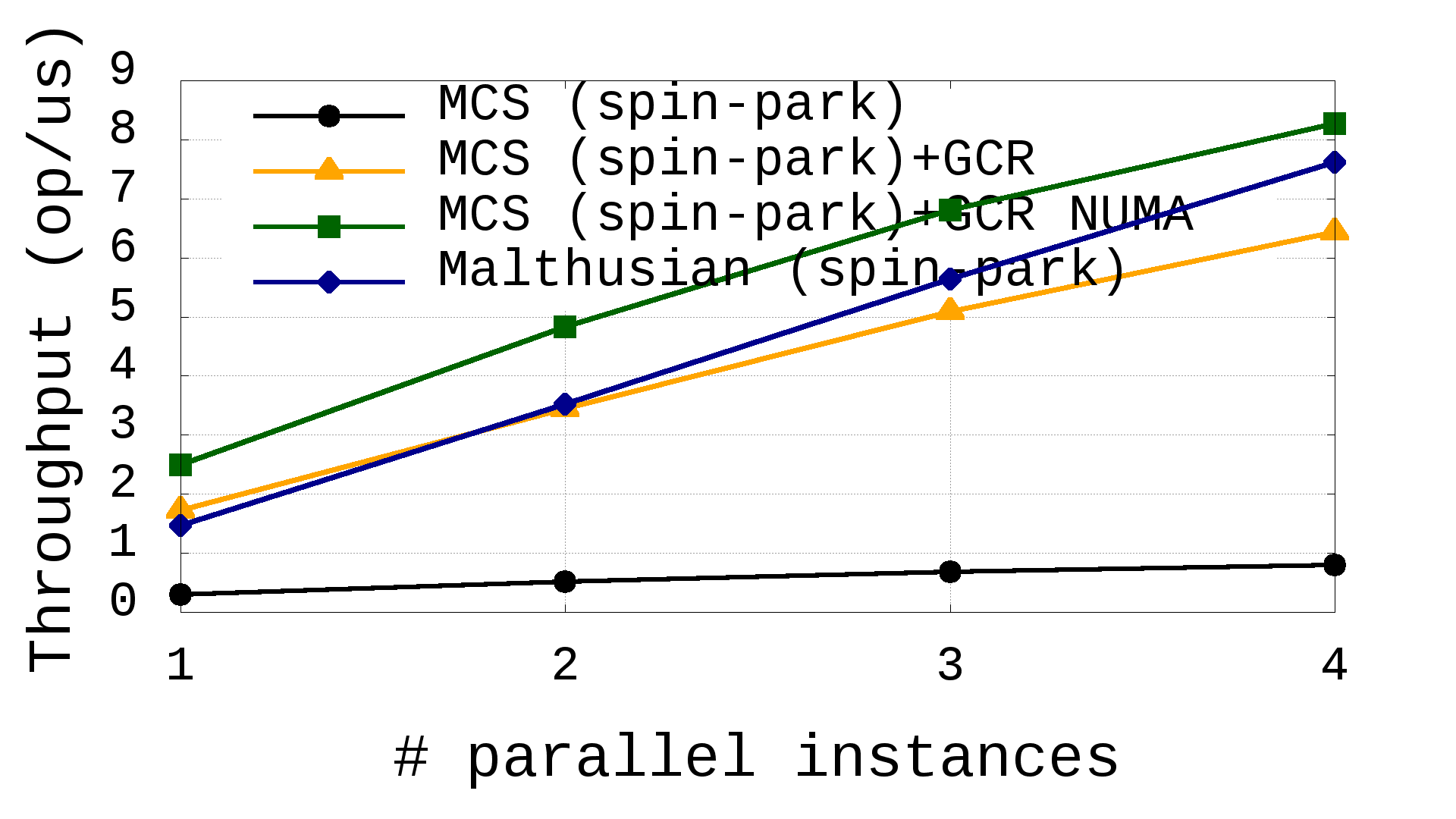}}
\subfloat[][Test-Test-Set]{\includegraphics[width=0.25\linewidth]{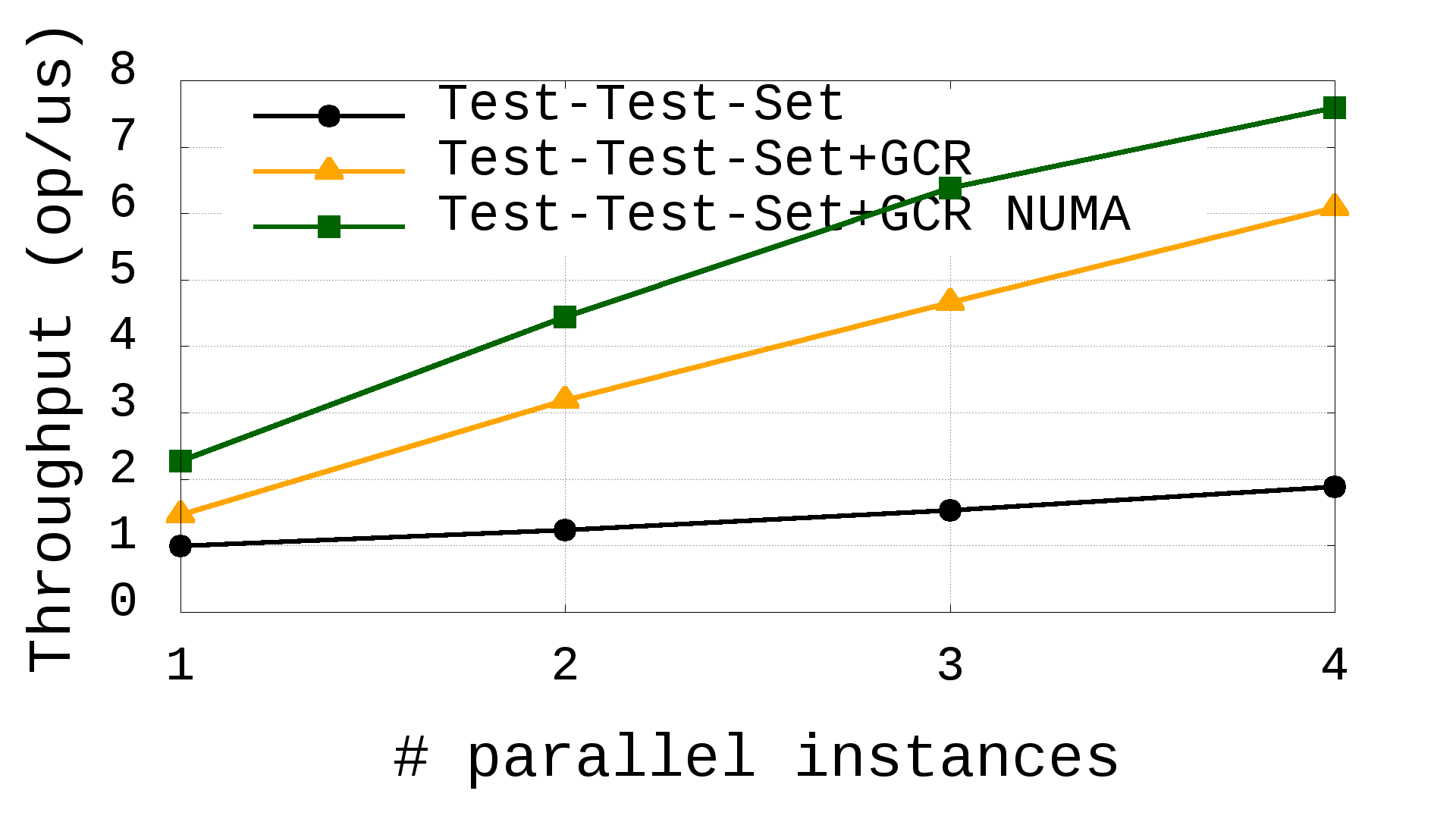}}
\subfloat[][Pthread mutex]{\includegraphics[width=0.25\linewidth]{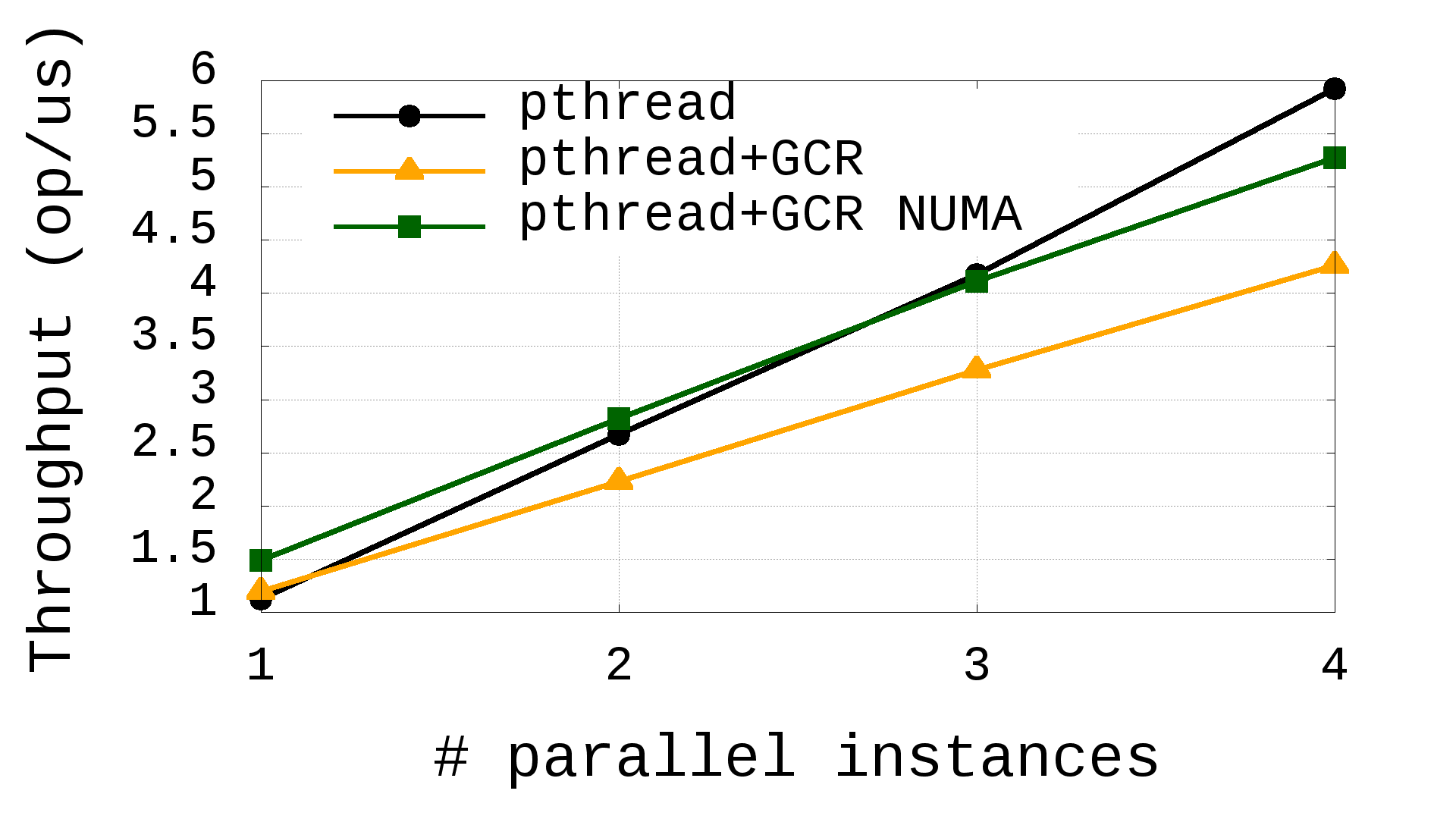}}
\caption{Total throughput measured with multiple instances of the microbenchmark, each run with $40$ threads (AVL tree).}
\figlabel{fig:X6-2=avl-tree-multicopy-absolute-perf}
\end{figure*}

The microbenchmark uses a sequential AVL tree implementation protected by a single lock.
The tree supports the API of a key-value map, including operations for inserting, removing and looking up keys (and associated values) stored in the tree.
After initial warmup, not included in the measurement interval, all threads are synchronized to start running at the same time, 
and apply tree operations chosen uniformly and at random from the given distribution,
with keys chosen uniformly and at random from the given range.
At the end of this time period (lasting $10$ seconds), the total number of operations is calculated, and the throughput is reported.
The reported results are for the key range of $4096$ and
threads performing $80\%$ lookup operations, while the rest is split evenly between inserts and removes.
%(We experimented with a range of operation mixes and various key ranges, 
%and found the relative performance of the locks with and without GCR being similar).
The tree is pre-initialized to contain roughly half of the key range.
Finally, the microbenchmark allows to control the amount of the external work, i.e., the duration of a non-critical section
(simulated by a pseudo-random number calculation loop).
In this experiment, we use a non-critical section duration that allows scalability up to a small number of threads.

\textbf{Detailed performance of GCR on top of several locks:}
The absolute performance of the AVL tree benchmark (in terms of the total throughput) with several locks is shown in \figref{fig:X6-2=avl-tree-with-delay-absolute-perf}.
\figref{fig:X6-2=avl-tree-with-delay-absolute-perf}~(a) and~(b) show how the popular MCS lock~\cite{MS91} performs without GCR, with GCR and with GCR-NUMA, 
and how those locks compare to the recent Malthusian lock~\cite{Dice17}, which implements a concurrency restriction mechanism directly into the MCS lock.
Locks in \figref{fig:X6-2=avl-tree-with-delay-absolute-perf}~(a) employ the spinning waiting policy, while those in \figref{fig:X6-2=avl-tree-with-delay-absolute-perf}~(b) employ the spin-then-park policy.
In addition, \figref{fig:X6-2=avl-tree-with-delay-absolute-perf}~(c) and~(d) compare the performance 
achieved with the simple Test-Test-Set (TTAS) lock and the POSIX pthread mutex lock, respectively, when used without GCR, with GCR and with GCR-NUMA.
The concurrency restriction mechanism of a Malthusian lock cannot be applied directly into the simple TTAS or POSIX pthread mutex locks, 
so we do not include a Malthusian variant in those two cases.

With the spinning policy (\figref{fig:X6-2=avl-tree-with-delay-absolute-perf}~(a)),
GCR has a small detrimental effect ($2\%$ slowdown for a single thread, and in general, at most $12\%$ slowdown) on the performance 
of MCS as long as the machine is not oversubscribed.
This is because all threads remain running on their logical CPUs and the lock handoff is fast at the time that GCR introduces certain (albeit, small) overhead.
The Malthusian lock performs similarly to (but worse than) GCR.
MCS with GCR-NUMA, however, tops the performance chart as it limits the amount of cross-socket communication incurred by other altrenatives when 
the lock is handed off between threads running on different sockets.
%(This communication cannot be avoided with just two threads, each running on a different socket, which is 
%why the performance of MCS with GCR-NUMA dips at two threads, but recovers immediately after that).
The performance of the MCS and Malthusian locks plummets once the number of running threads exceeds the capacity of the machine.
At the same time, GCR (and GCR-NUMA) are not sensitive to that as they park excessive threads, preserving the overall performance.
In case of GCR-NUMA, for instance, this performance is close to the peak achieved with $10$ threads.

The MCS and Malthusian locks with the spin-then-park policy exhibit a different performance pattern (\figref{fig:X6-2=avl-tree-with-delay-absolute-perf}~(b)).
Specifically, the former shows poor performance at the relatively low number of threads.
This is because as the number of threads grows, the waiting threads start quitting spinning and park, adding the overhead of unparking for each lock handoff.
The Malthusian lock with its concurrency restriction mechanism avoids that.
Yet, its performance is slightly worse than that of MCS with GCR.
%because the Malthusian lock manages the queue of passive threads under the lock, while in GCR (passive) threads maintain the queue outside of the critical path.
Once again, MCS with GCR-NUMA easily beats all other contenders.

In summary, the results in \figref{fig:X6-2=avl-tree-with-delay-absolute-perf}~(a) and (b) show that despite being generic, the concurrency restriction 
mechanism of GCR performs superiorly to that of the specialized Malthusian lock.
Besides, unlike the Malthusian lock, the choice of a waiting policy for the underlying lock becomes much less crucial when GCR (or GCR-NUMA) is used.
\extabstract{We expand on this point later.}

The TTAS and pthread mutex locks exhibit yet another performance pattern (\figref{fig:X6-2=avl-tree-with-delay-absolute-perf}~(c) and~(d)).
Similarly to the MCS spin-then-park variant, their performance drops at low thread counts, however they manage to maintain reasonable
throughput even as the number of threads grows.
Along with that, both GCR and GCR-NUMA variants mitigate the drop in the performance.

\remove{
 \figref{fig:X6-2=avl-tree-absolute-perf} shows the results of the same experiment, but with an empty non-critical section.
In this case, there is no any scalability, as the lock becomes saturated with just two threads.
For all cases, GCR maintains the performance achieved at $2$ threads.
This is what can be expected from GCR in case of contention as it keeps only two threads active.
While this is far from being optimal in this case, this is far better than performance achieved by underlying locks without GCR.

Once again, GCR-NUMA achieves better performance by keeping the set of active threads running on the same socket.
Without external work, the set of active threads remains stable as the same thread that released the lock immediately tries to acquire it again.
This in turn improves cache locality even further 
as the data accessed by a thread is cached on the core on which the thread is running, and not just on the socket.
GCR-NUMA is not helpful with just two threads, each running on a different socket, as it does not passivate a thread on another socket 
if it is the only thread to wait for the lock. 
This is why its performance dips at two threads, but recovers immediately after that.
}

We also run experiments in which we measured the handoff time for each of the locks presented in \figref{fig:X6-2=avl-tree-with-delay-absolute-perf}, that is,
the interval between a timestamp taken right before the current lock holder calls \code{Unlock()} and right after the next lock holder returns from \code{Lock()}.
Previous work has shown that the performance of a parallel system is dictated by the length of its critical sections~\cite{EE10}, which is composed
of the time required to acquire and release the lock (captured by the handoff data), and the time a lock holder spends in the critical section.
Indeed, the data in \figref{fig:X6-2=avl-tree-with-delay-handoff} shows correlation between the throughput achieved and the handoff time.
That is, in all cases where the throughput of a lock degraded in \figref{fig:X6-2=avl-tree-with-delay-absolute-perf}, the handoff time has increased.
At the same time, GCR (and GCR-NUMA) manages to maintain a constant handoff time across virtually all thread counts.

In a different experiment, we run multiple instances of the microbenchmark, each configured to use the number of threads equal to
the number of logical CPUs ($40$).
This illustrates the case in which an application with a configurable number of threads chooses to set that number based on the
machine capacity (as it typically happens by default, for instance, in OpenMP framework implementations).
\figref{fig:X6-2=avl-tree-multicopy-absolute-perf} presents the results for the same set of locks 
as \figrangeref{fig:X6-2=avl-tree-with-delay-absolute-perf}{fig:X6-2=avl-tree-with-delay-handoff}.
Both GCR and GCR-NUMA scale well up to $4$ instances for all tested locks.
Except for pthread mutex, all locks without GCR (or GCR-NUMA) perform much worse, especially when the number
of instances is larger than one (which is when the machine is oversubscribed). 
Pthread mutex fares relatively well, although it should be noted that its single instance performance is worse than, e.g., the one achieved by MCS spin.

We note that as the number of instances grows, the Malthusian spin-then-park lock handles contention 
better than GCR (but not GCR-NUMA) on top of MCS spin-then-park (cf.~\figref{fig:X6-2=avl-tree-multicopy-absolute-perf}~(b)).
We attribute that to the fact that the Malthusian lock employs less active threads than GCR (and GCR-NUMA), which in the case of 
a heavily oversubscribed machine plays an important role.
%Indeed, in experiments with a lower threshold for joining the passive set (not shown), we have seen that both GCR and GCR-NUMA narrow the gap 
%from the Malthusian lock.
Note that the Malthusian spin lock does not provide any relief from contention compared to the MCS spin lock (cf.~\figref{fig:X6-2=avl-tree-multicopy-absolute-perf}~(a)),
at the time that both GCR and GCR-NUMA scale linearly with the number of instances.
This is because in this case, threads passivated by the Malthusian lock spin and take resources away 
from the active threads at the time that GCR and GCR-NUMA park those passive threads.

\remove{
The microbenchmark supports variable key range from which keys are drawn for tree operations.
While we experimented with a range of operation mixes and various key ranges, 
the relative performance of the locks with and without GCR was similar. 
Thus, we opted to show results for one setting, in which the key range is set to $4096$ and
threads perform 40\% look up operations, while the rest is split evenly between inserts and removes.
The tree is pre-initialized to contain roughly half of the key range.
}

\begin{figure*}[t]
\subfloat[][GCR, X6-2]{\includegraphics[width=0.58\linewidth, Clip=1.5cm 0 0 0]{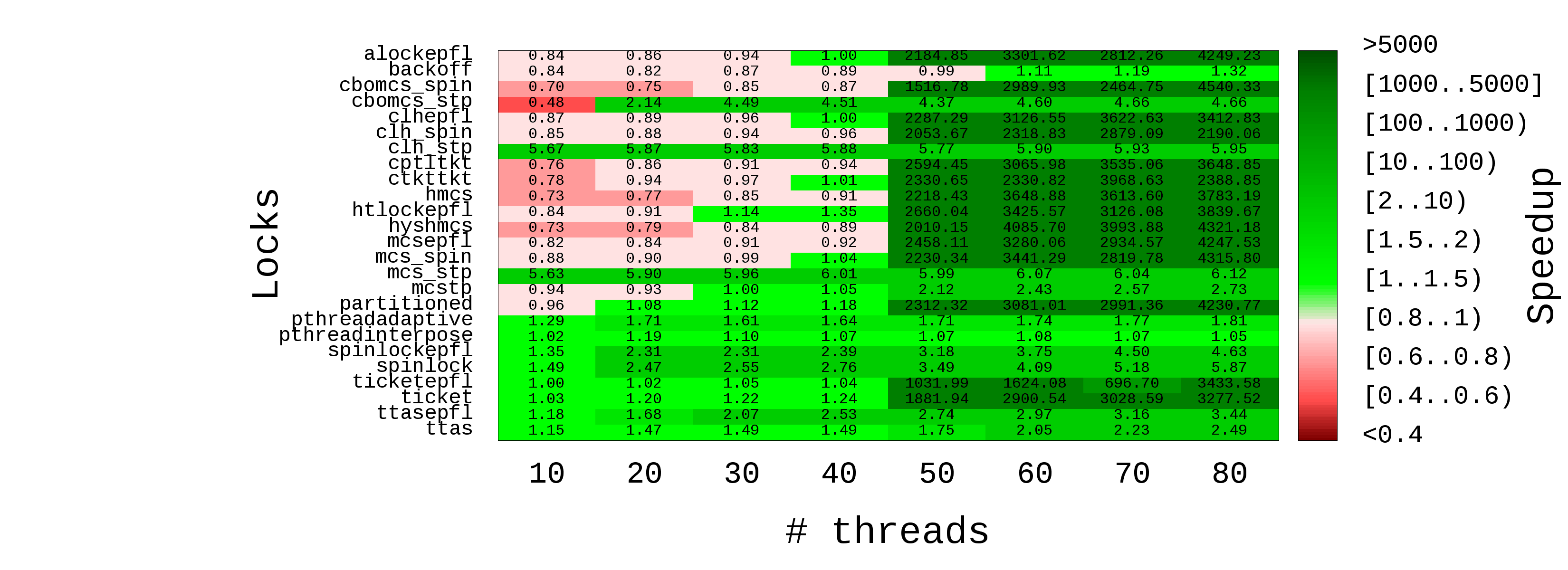}}
\subfloat[][GCR-NUMA, X6-2]{\includegraphics[width=0.58\linewidth, Clip=1.5cm 0 0 0]{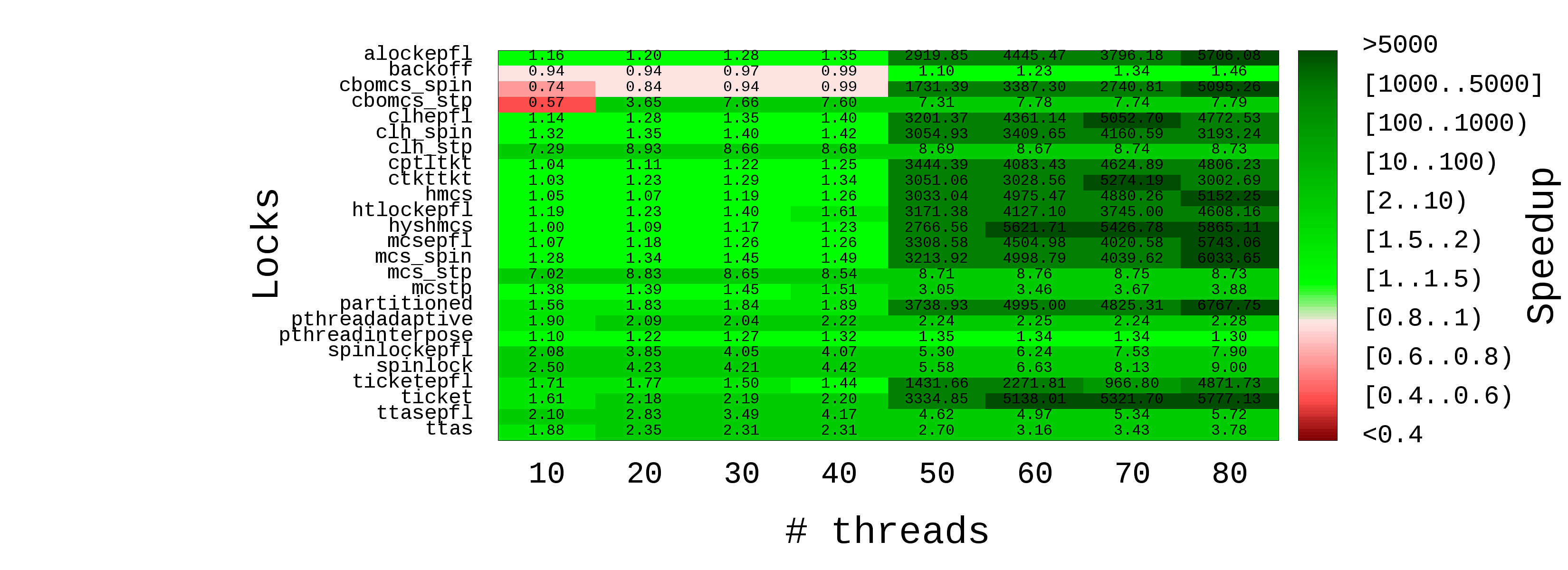}}\\
\subfloat[][GCR, T7-2]{\includegraphics[width=0.58\linewidth, Clip=1.5cm 0 0 0]{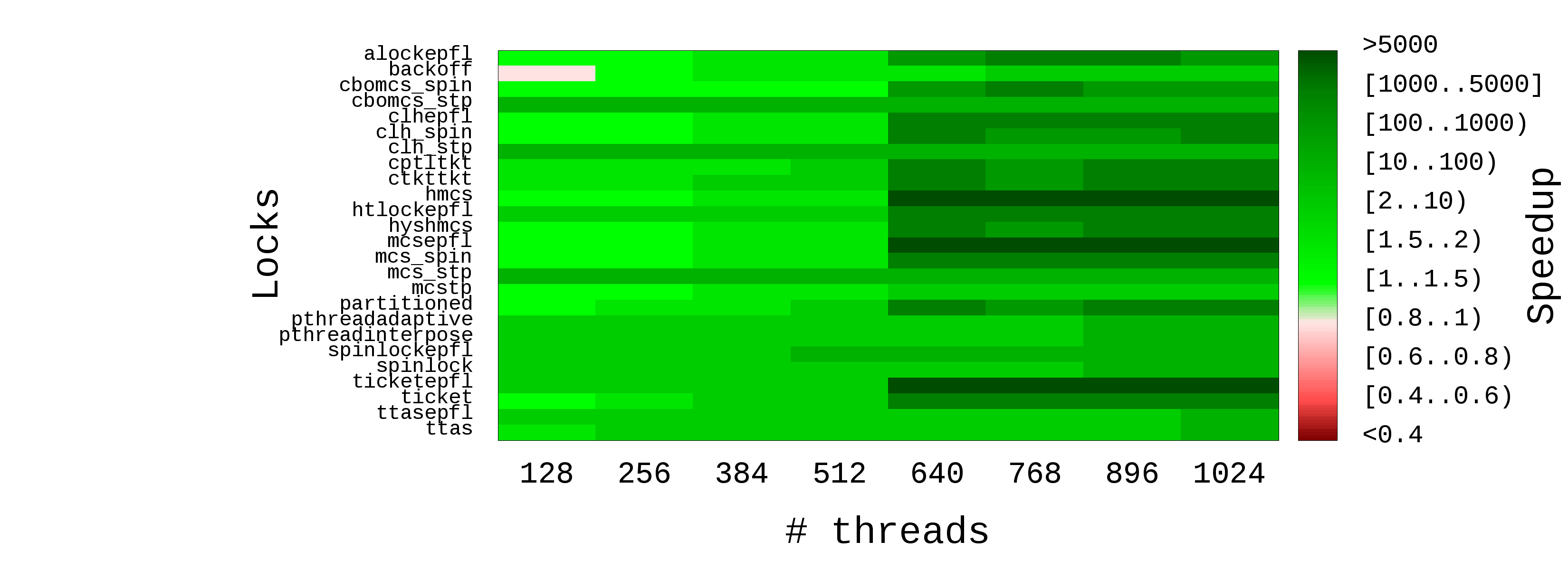}}
\subfloat[][GCR-NUMA, T7-2]{\includegraphics[width=0.58\linewidth, Clip=1.5cm 0 0 0]{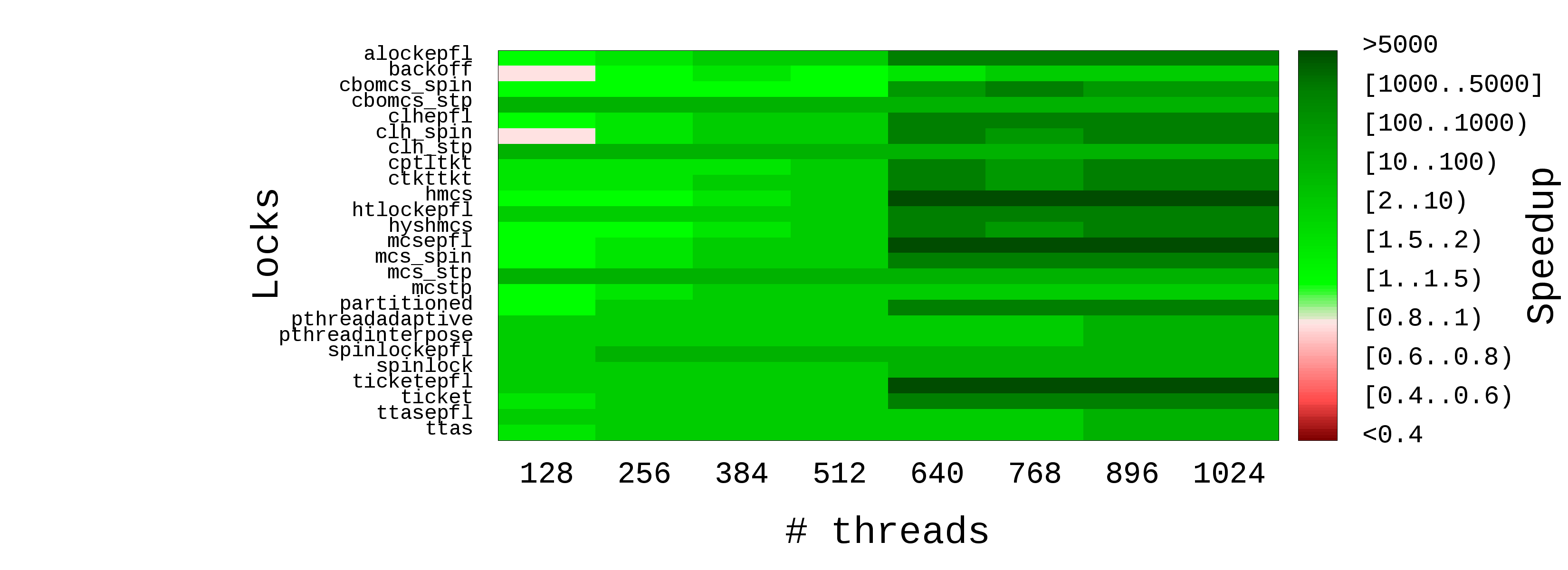}}\\
\subfloat[][GCR, X5-4]{\includegraphics[width=0.58\linewidth, Clip=1.5cm 0 0 0]{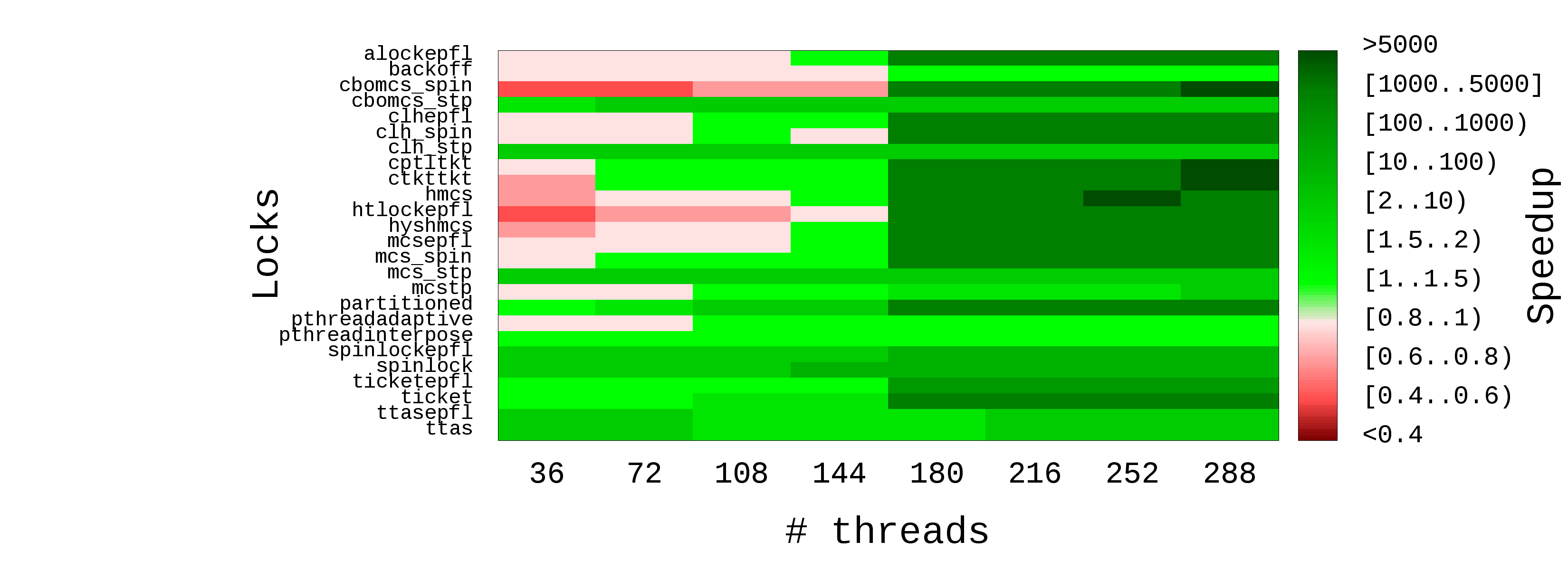}}
\subfloat[][GCR-NUMA, X5-4]{\includegraphics[width=0.58\linewidth, Clip=1.5cm 0 0 0]{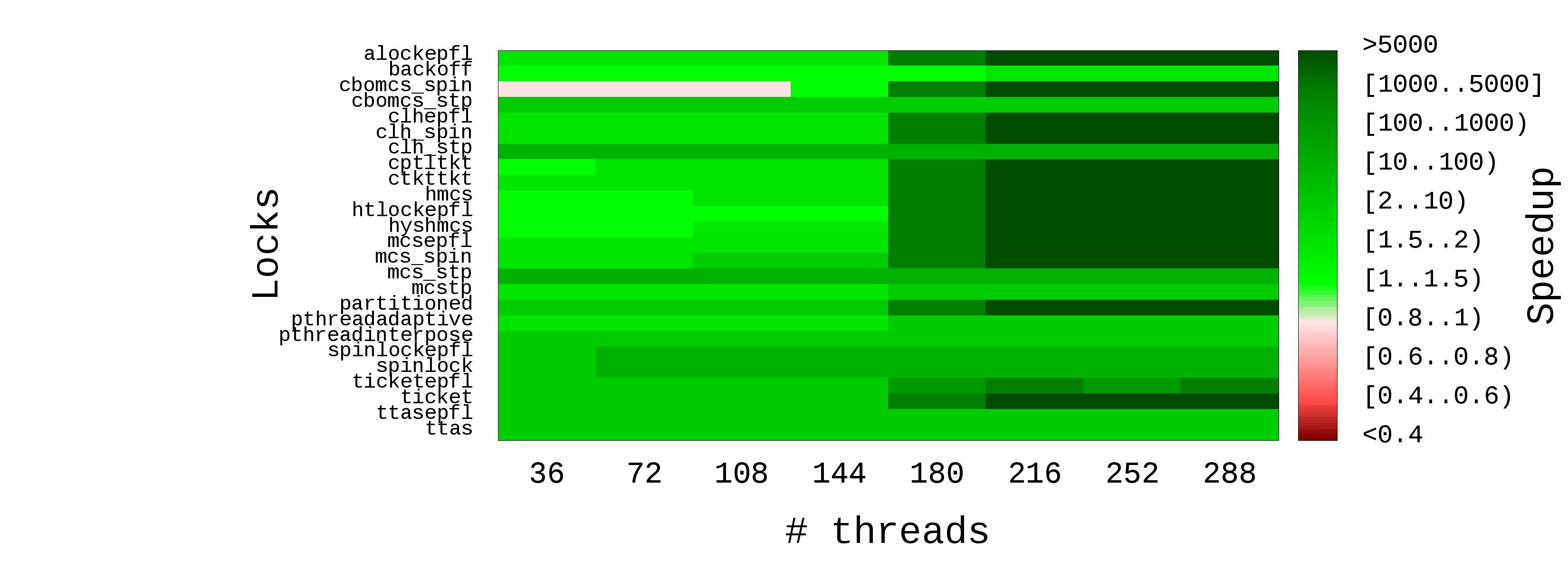}}
\caption{Speedup achieved by GCR and GCR-NUMA over base lock implementations (AVL tree).}
\figlabel{fig:avl-tree-perf-heatmap}
\end{figure*}

\begin{figure*}
\subfloat[][Base]{\includegraphics[width=0.43\linewidth, Clip=1.2cm 0 0 0 ]{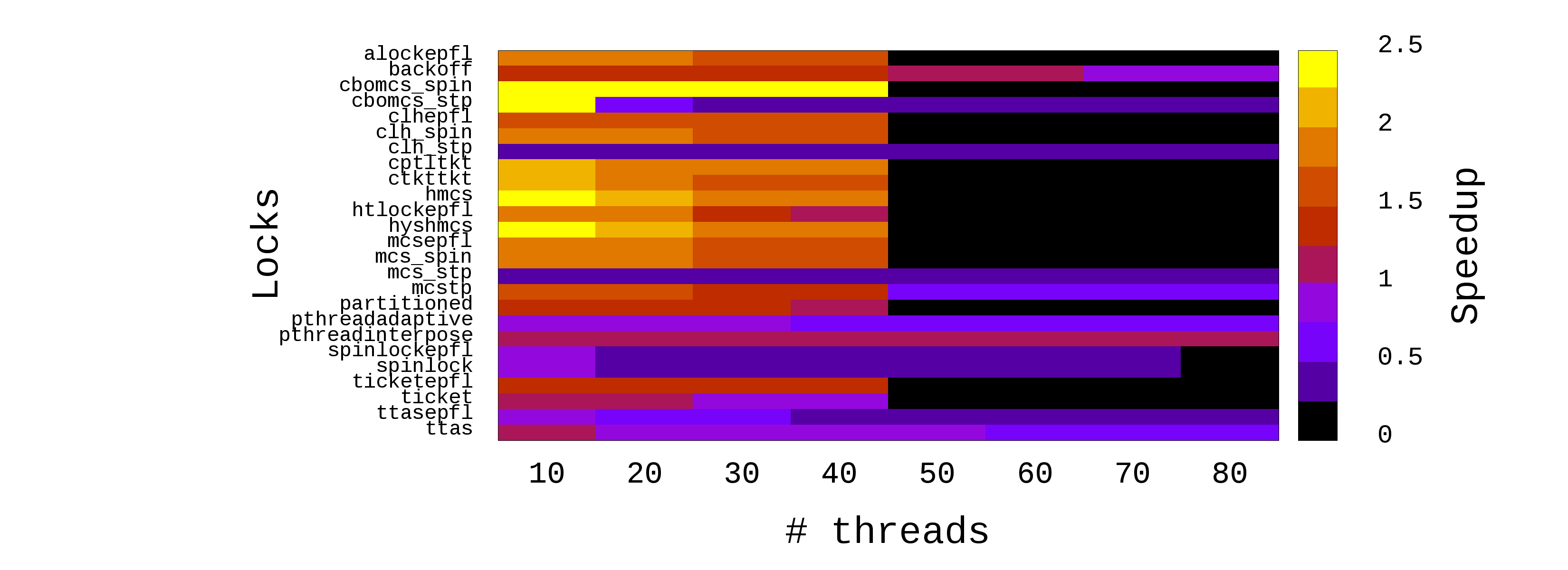}}
\subfloat[][GCR]{\includegraphics[width=0.43\linewidth, Clip=1.2cm 0 0 0]{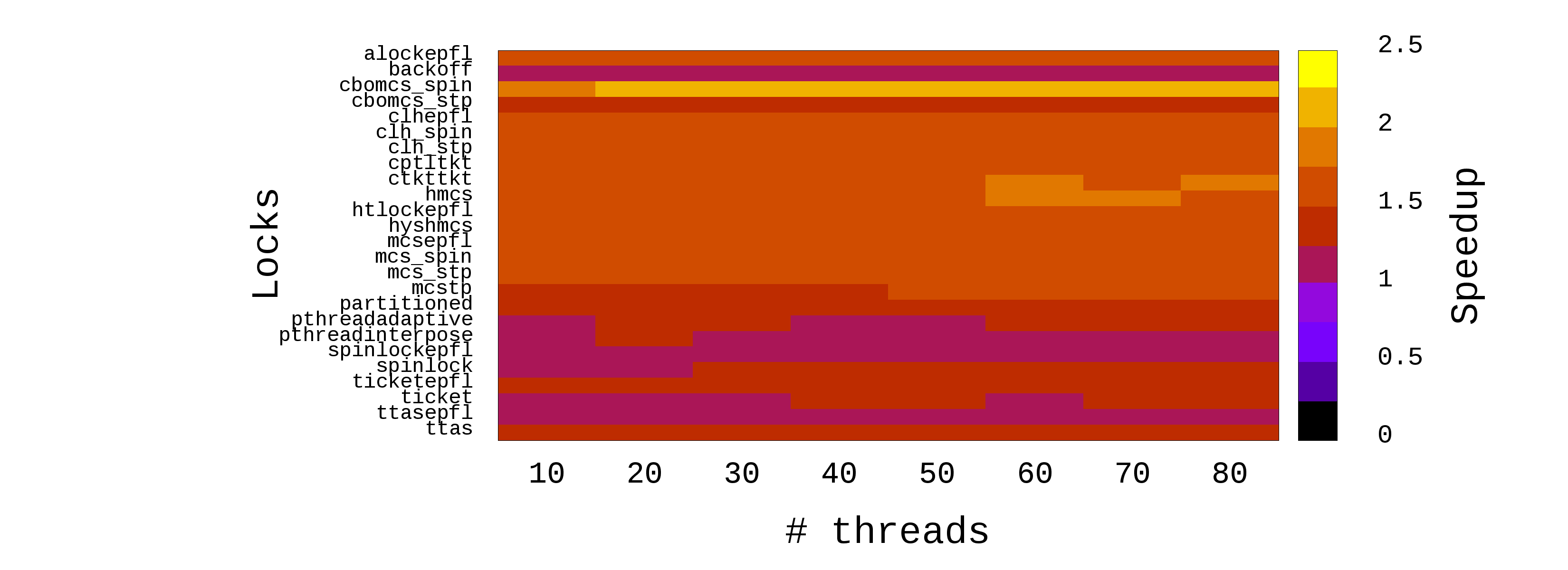}}
\subfloat[][GCR-NUMA]{\includegraphics[width=0.43\linewidth, Clip=1.2cm 0 0 0]{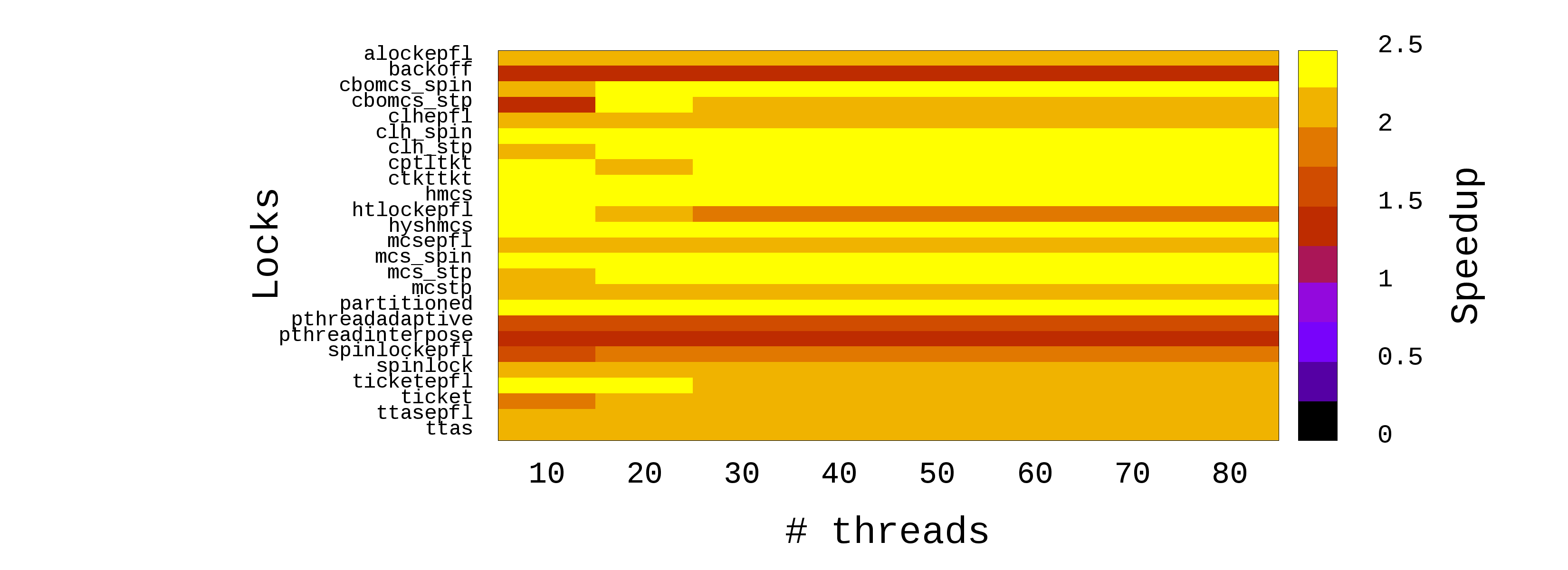}}
\caption{Speedup for the base, GCR and GCR-NUMA locks when normalized to the performance of the MCS (spin-then-park) lock (without GCR, with GCR and with GCR-NUMA, respectively) with a single thread (AVL tree).}
\figlabel{fig:X6-2-avl-tree-normalized-perf-heatmap}
\end{figure*}

\begin{figure*}[t]
\subfloat[][Base]{\includegraphics[width=0.43\linewidth, Clip=1.1cm 0 0 0]{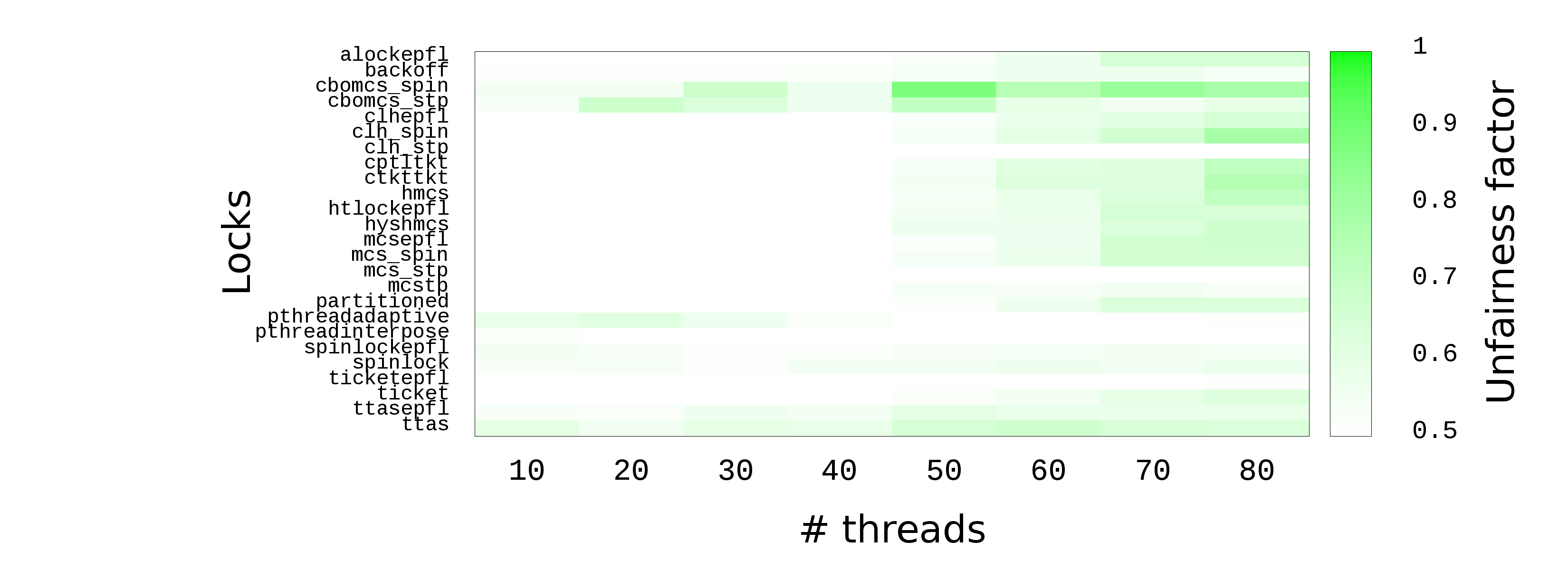}}
\subfloat[][GCR]{\includegraphics[width=0.43\linewidth, Clip=1.1cm 0 0 0]{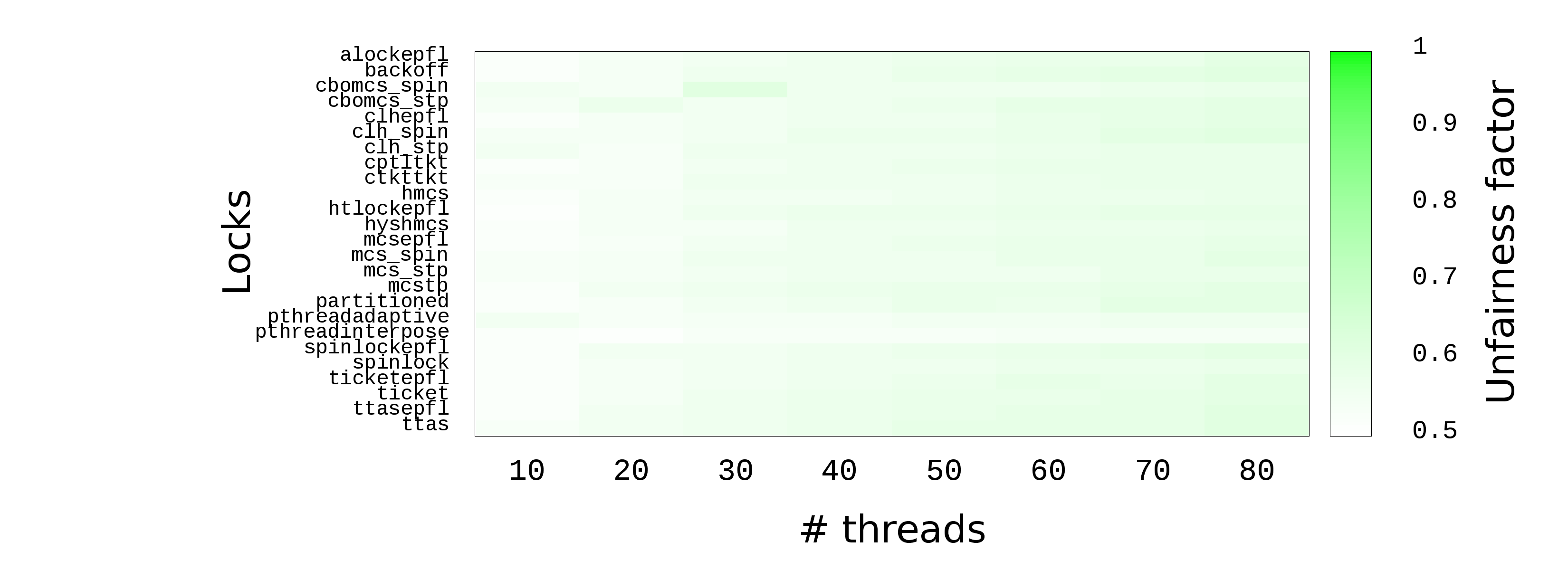}}
\subfloat[][GCR-NUMA]{\includegraphics[width=0.43\linewidth, Clip=1.1cm 0 0 0]{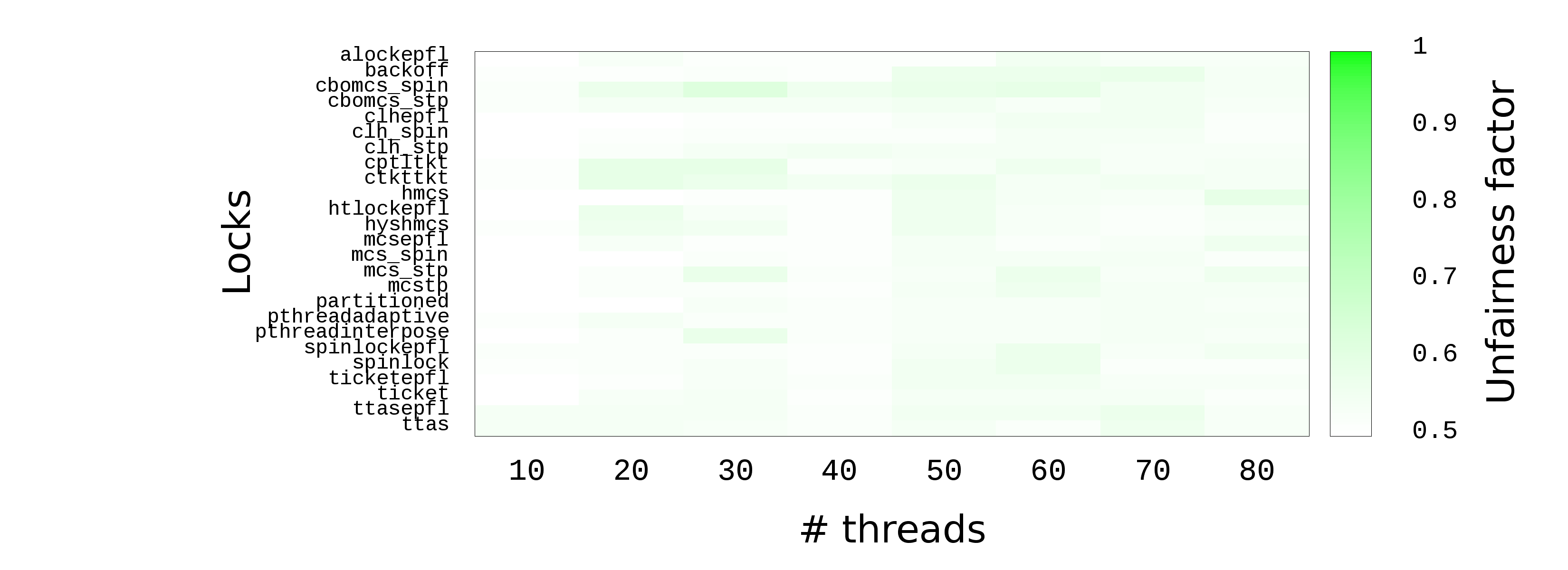}}
\caption{Unfairness factor for the base locks and when GCR is used (AVL tree).}
\label{fig:X6-2-avl-tree-unfairness-heatmap}
\end{figure*}

\textbf{GCR on top of 24 locks:} After presenting results for some concrete locks, 
we show in \figref{fig:avl-tree-perf-heatmap} a heat map encoding the speedup achieved by GCR and GCR-NUMA
when used with each of the $24$ locks provided by LiTL, for all the three machines.
A cell in \figref{fig:avl-tree-perf-heatmap} at row X and column Y 
represents the throughput achieved with Y threads when the GCR (GCR-NUMA) library is used with lock X divided
by throughput achieved when the lock X itself is used (i.e., without GCR or GCR-NUMA).
The shades of red colors represent slowdown (speedup below $1$, which in most cases falls in the range of [0.8..1), i.e., 
less than 20\% slowdown), while the shades of green colors represent positive speedup;
the intensity of the color represents how substantial the slowdown/speedup is .
In other words, ideally, we want to see heat maps of green colors, as dark as possible.
\extabstract{(The raw speedup numbers for X6-2 can be found in tables in the Appendix.)}
%For this experiment, as well as for the rest of the figures in this section, we use the same amount of external work as in \figref{fig:X6-2=avl-tree-with-delay-absolute-perf}.
We also provide raw speedup numbers for X6-2, but remove them for other machines for better readability.
 
\remove{
First, we note that for a single-thread performance, the GCR variant incurs between $12\%$--$16\%$ overhead on Intel machines and 
a slightly higher overhead of about $20\%$ on the SPARC machine.
Some of this overhead, however, can be attributed to the mechanism of the dynamic linkage.
We confirmed that by statically linking the GCR library with a few lock libraries and observing a single-thread overhead dropping to $10\%$.
(We note that a noticeable difference between dynamically linked and statically linked GCR libraries appeared only with a single thread).
However, for some of the locks, starting with as little as $4$ threads, the overhead of the GCR library gets overweighted by the benefits it offers.
}

Until the machines become oversubscribed, the locks that do not gain from GCR are mostly NUMA-aware locks, such as \code{cbomcs\_spin},
\code{cbomcs\_stp}, \code{ctcktck} (which are the variants of Cohort locks~\cite{topc15-dice}) 
and such as \code{hyshmcs} and \code{hmcs} (which are NUMA-aware hierarchical MCS locks~\cite{ppopp15-chabbi}).
This means that, unsurprisingly, putting a (non-NUMA-aware) GCR mechanism in front of a NUMA-aware lock is not a good idea.

%This is perhaps surprising, as the major motivation for GCR was restricting concurrency on systems with more threads than computational resources.
%We believe this is a result of improved cache locality achieved by GCR as it allows fewer threads to circulate through the contended 
%lock at any given time interval.

When machines are oversubscribed, however, GCR achieves gains for all locks, often resulting in more than 4000x throughout 
increase compared to the base lock. \extabstract{ (cf.\ Table~\ref{table:X6-2-avl-tree-GCR} in the Appendix).}
Those are the cases when the performance with the base lock (without GCR) plummets, while GCR manages to avoid the drop.
In general, locks that use spin-then-park policy tend to experience a relatively smaller drop in performance 
(and thus relatively smaller benefit for GCR -- up to 6x).

\remove{
The performance of the \code{backoff} lock (which is a simple test-and-set spinlock with exponential backoff) is noteworthy.
It is, perhaps, the least sensitive lock to the concurrency restriction even when the machine is oversubscribed.
This is because this lock ensures (by letting all threads spin on one global location)
that it can be acquired only by a thread running on a CPU, unlike, e.g., 
locks that form a queue of waiting threads, where a lock can be handed off to a thread waiting for CPU time.
Notice that a simple \code{TTAS} (Test-Test-Set) lock also performs relatively well, but lacking the backoff mechanism, 
this lock is not equipped to deal with increased contention.
In practice, however, it is extremely challenging to tune the backoff mechanism~\cite{JSA10}.
}

%Tables~\ref{table:X6-2-avl-tree-GCR-NUMA} presents the same data, but with the NUMA-aware variant of GCR (denoted as GCR-NUMA thereafter).
When considering the speedup achieved with GCR-NUMA\extabstract{(right column of \figref{fig:avl-tree-perf-heatmap})}, 
we see the benefit of the concurrency restriction increasing, in some case dramatically,
while in most cases, the benefit shows up at much lower thread counts.
In particular, it is worth noting that above half the capacity of machines, 
the performance of virtually all locks is improved with GCR-NUMA.
When machines are oversubscribed, GCR-NUMA achieves over 6000x performance improvement for some of the locks.

A different angle on the same data is given by \figref{fig:X6-2-avl-tree-normalized-perf-heatmap}.
%We show the plots for the base locks and for those employed by GCR and GCR-NUMA.
Here we normalize the performance of all locks to that achieved with \code{mcs\_stp} (MCS spin-the-park) with a single thread. % (with our without the respective GCR variant).
The data in \figref{fig:X6-2-avl-tree-normalized-perf-heatmap}(a) for locks without GCR 
echoes results of several studies comparing between different locks
and concluding that the best performing lock varies across thread counts~\cite{TGT13, GLQ16}.
Notably, the normalized performance with GCR and GCR-NUMA is more homogeneous (\figref{fig:X6-2-avl-tree-normalized-perf-heatmap}(b) and~(c)).
With a few exceptions, all locks deliver similar performance with GCR (and GCR-NUMA) and across virtually all thread counts. % at and above certain point.
%For GCR, this point is four threads, while for GCR-NUMA, this point is ten, which means the latter maintains the peak performance even when the machine
%is overloaded.
The conclusion from these data is that GCR and GCR-NUMA deliver much more stable performance, mostly independent
of the type of the underlying lock or the waiting policy it uses.

\textbf{Is better performance with GCR traded for fairness?}
It is natural to ask how the fairness of each lock is affected once the GCR mechanism is used.
There are many ways to assess fairness; we show one that we call the \emph{unfairness factor}.
To calculate this factor, we sort the number of operations reported by each thread at the end of the run,
and calculate the portion of operations completed by the upper half of threads.
Note that the unfairness factor is a value between $0.5$ and $1$, where a strictly fair lock would produce a value of $0.5$ 
(since all threads would apply the same number of operations) and an unfair lock would produce a value close to $1$.

The unfairness factors of all the locks, without GCR, with GCR and with GCR-NUMA, are shown in 
the corresponding heat maps in Figure~\ref{fig:X6-2-avl-tree-unfairness-heatmap}.
While in some cases GCR produces slightly higher unfairness factor, it appears to smooth high unfairness 
factors of some of the locks, generating a much more homogenous heat map.
This is because some of the locks, e.g., Test-Test-Set, can be grossly unfair mainly due to caching effects.
That is, if multiple threads attempt to acquire the lock at the same time,
the thread on the same core or socket as a previous lock holder is likely to win as 
it has the lock word in its cache.
GCR restricts the number of threads competing for the lock, and shuffles those threads periodically,
achieving long-term fairness.
Interestingly, GCR-NUMA achieves even greater fairness, as it picks active threads from the same socket.
Thus, it reduces the chance that the same thread(s) will acquire the lock repeatedly while another thread
on a different socket fails to do that due to expensive remote cache misses.

\begin{figure*}[!t]
\subfloat[][GCR]{\adjincludegraphics[width=0.5\linewidth,Clip=1.2cm 0 1.6cm 0]{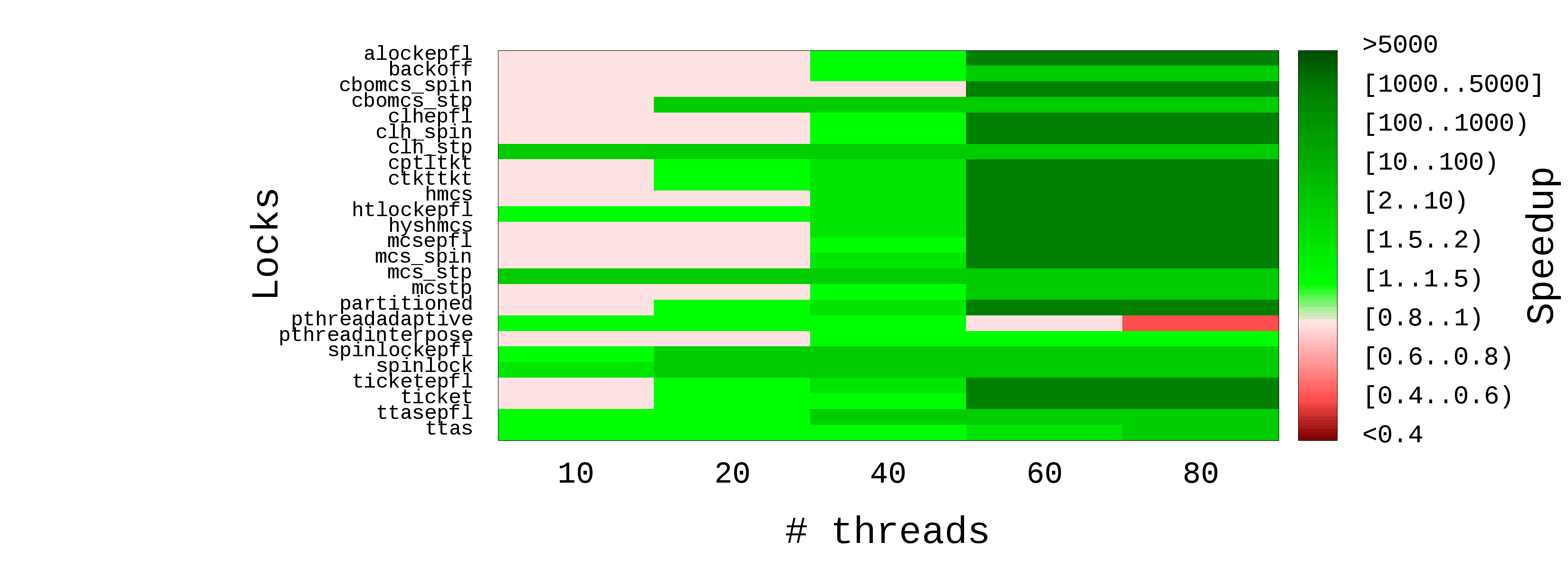}}
\subfloat[][GCR-NUMA]{\adjincludegraphics[width=0.5\linewidth,Clip=2.75cm 0 0 0]{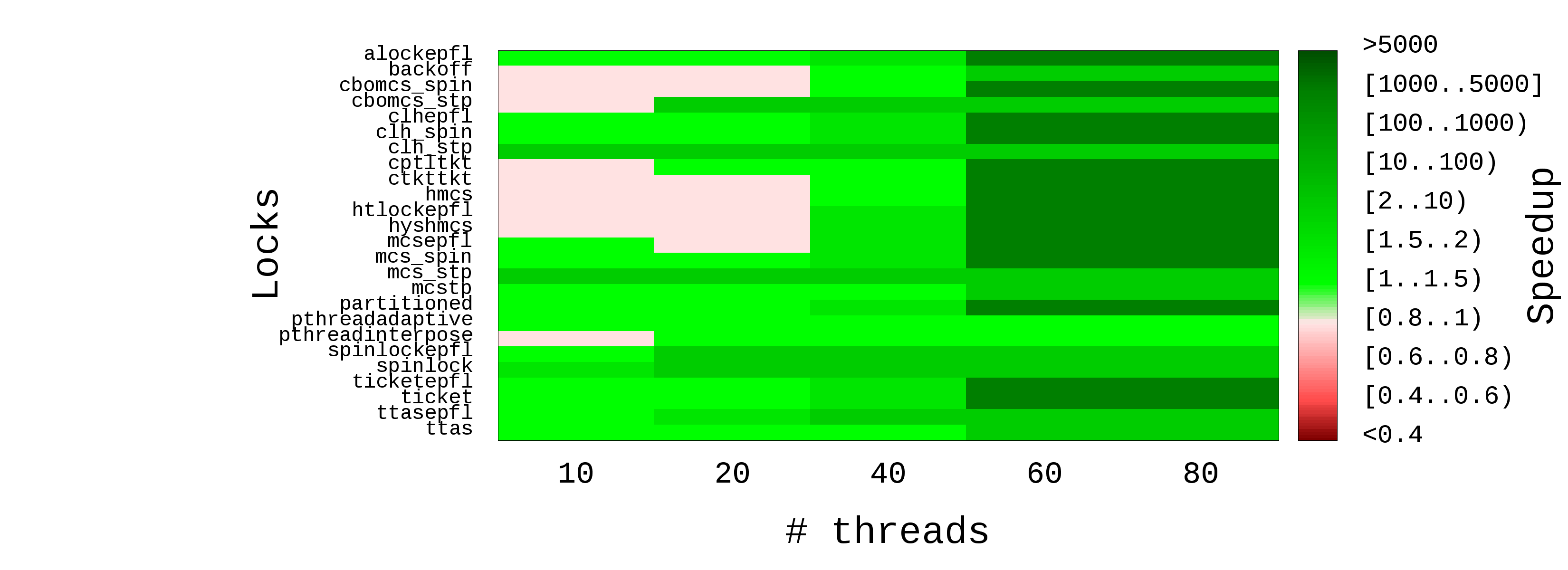}}\\
%\subfloat[][GCR, T7-2]{\includegraphics[width=0.5\linewidth]{figures/ol-bur-t7-2/kyoto/kyoto-heatmap-perf-base_v_gcr.png}}
%\subfloat[][GCR-NUMA, T7-2]{\includegraphics[width=0.5\linewidth]{figures/ol-bur-t7-2/kyoto/kyoto-heatmap-perf-base_v_gcrnumaaware.png}}\\
%\subfloat[][GCR, X5-4]{\includegraphics[width=0.5\linewidth]{figures/ol-bur-x5-4/kyoto/kyoto-heatmap-perf-base_v_gcr.png}}
%\subfloat[][GCR-NUMA, X5-4]{\includegraphics[width=0.5\linewidth]{figures/ol-bur-x5-4/kyoto/kyoto-heatmap-perf-base_v_gcrnumaaware.png}}
\caption{Speedup achieved by GCR and GCR-NUMA over base lock implementations (Kyoto).}
\figlabel{fig:kyoto-perf-heatmap}
\end{figure*}

\subsection{Kyoto Cabinet}
We report on our experiments with the Kyoto Cabinet~\cite{kyotocabinet} \code{kccachetest} benchmark run in a 
\code{wicked} mode, which exercises an in-memory database.
Similarly to~\cite{Dice17}, we modified the benchmark to use the standard POSIX pthread mutex locks, which
we interpose with locks from the LiTL library.
We also modified the benchmark to run for a fixed time and report the aggregated work completed. 
Finally, we fixed the key range at a constant (10M) elements.
% to allow more fair comparison of the results achieved with a varying number of threads.
(Originally, the benchmark set the key range dependent on the number of threads).
All those changes were also applied to Kyoto in~\cite{Dice17} to allow fair comparison of performance across different thread counts.
The length of each run was $60$ seconds. %, and we report the mean results out of $3$ runs made in the same configuration.

Kyoto employs multiple locks, each protecting a slot comprising of a number of buckets in a hash table; the latter
is used to implement a database~\cite{kyotocabinet}.
Given that the \code{wicked} mode exercises a database with random operations and random keys, one should expect a lower 
load on each of the multiple slot locks compared to the load on the central lock used to protect the access to the AVL tree in the microbenchmark above.
%Thus, the benefit of CR is expected to be lower in this case.
Yet, Kyoto provides a good example of how GCR behaves in a real application setting.

The results are presented in \figref{fig:kyoto-perf-heatmap}, where % (for space considerations, we include only results from X6-2).
we run GCR and GCR-NUMA on top of 24 locks provided by LiTL.
Similarly to \figref{fig:avl-tree-perf-heatmap}, each cell decodes the slowdown/speedup achieved by GCR or GCR-NUMA, respectively, compared to the base lock.
As \figref{fig:kyoto-perf-heatmap} shows, both GCR and GCR-NUMA deliver robust gains (at times, over x1000), and those gains start for virtually all locks 
even before the machine becomes oversubscribed.

\remove{
\begin{table*}
\begin{center}
\scalebox{0.8}{
\input{figures/avidac.us.oracle.com/kyoto/kyoto-perf-base_v_gcr_v_gcrnumaaware.tbl}
}
\end{center}
\caption{Relative performance of lock implementations with and without GCR (Kyoto, X6-2).}
\label{table:X6-2-kyoto}
\end{table*}
}

\subsection{LevelDB}
LevelDB is an open-source key-value storage library~\cite{leveldb}. % originally written by Google.
We experimented with the release 1.20 of the library, which included the \code{db\_bench} benchmark.
We used \code{db\_bench} to create a database with the default 1M key-value pairs.
This database was used subsequently in the \code{readrandom} mode of \code{db\_bench},
in which threads read random keys from the database.
(The same database was used for all runs, since \code{readrandom} does not modify it).
Following the example of Kyoto, we modified the \code{readrandom} mode to run for a fixed time 
(rather than run a certain number of operations, which caused the runtime to grow disproportionally for some locks under contention).
The reported numbers are the aggregated throughput in the \code{readrandom} mode.
The length of each run was $10$ seconds.% and we report the mean results out of $3$ runs made in the same configuration.

\begin{figure*}[!t]
\subfloat[][GCR (10M keys)]{\adjincludegraphics[width=0.5\linewidth,Clip=0.5cm 0 1.65cm 0]{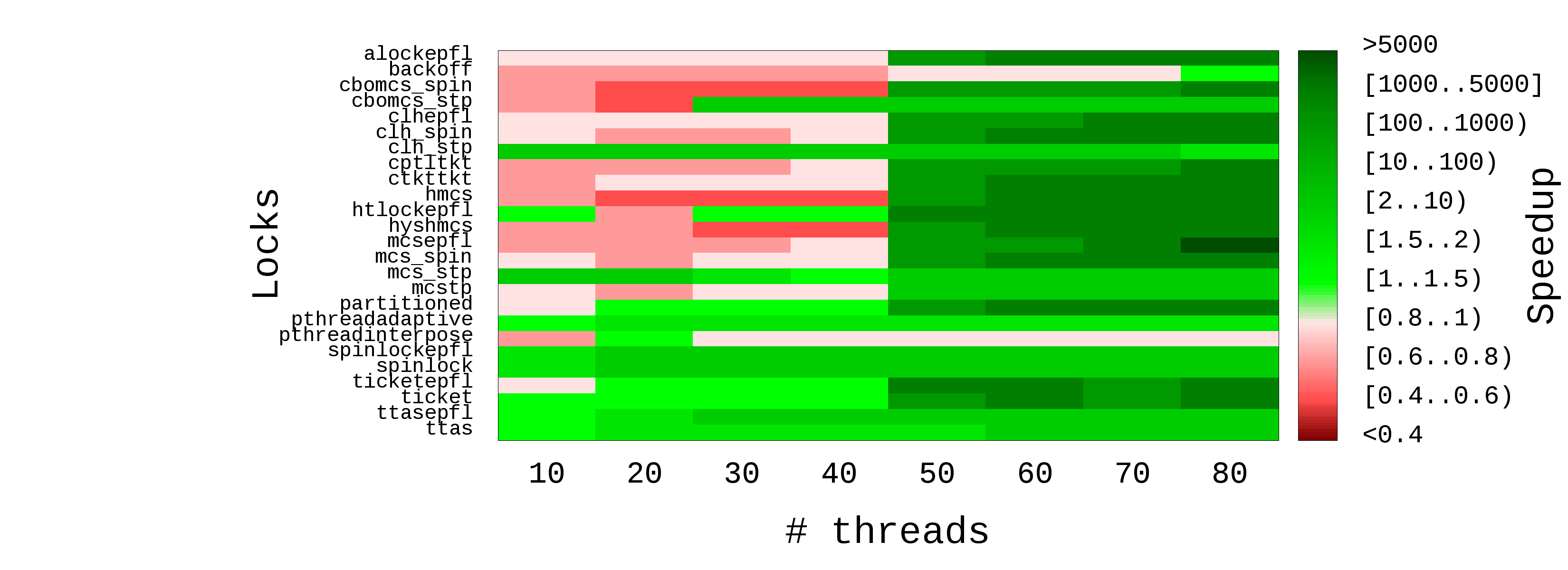}}
\subfloat[][GCR-NUMA (10M keys)]{\adjincludegraphics[width=0.5\linewidth,Clip=2.75cm 0 0 0]{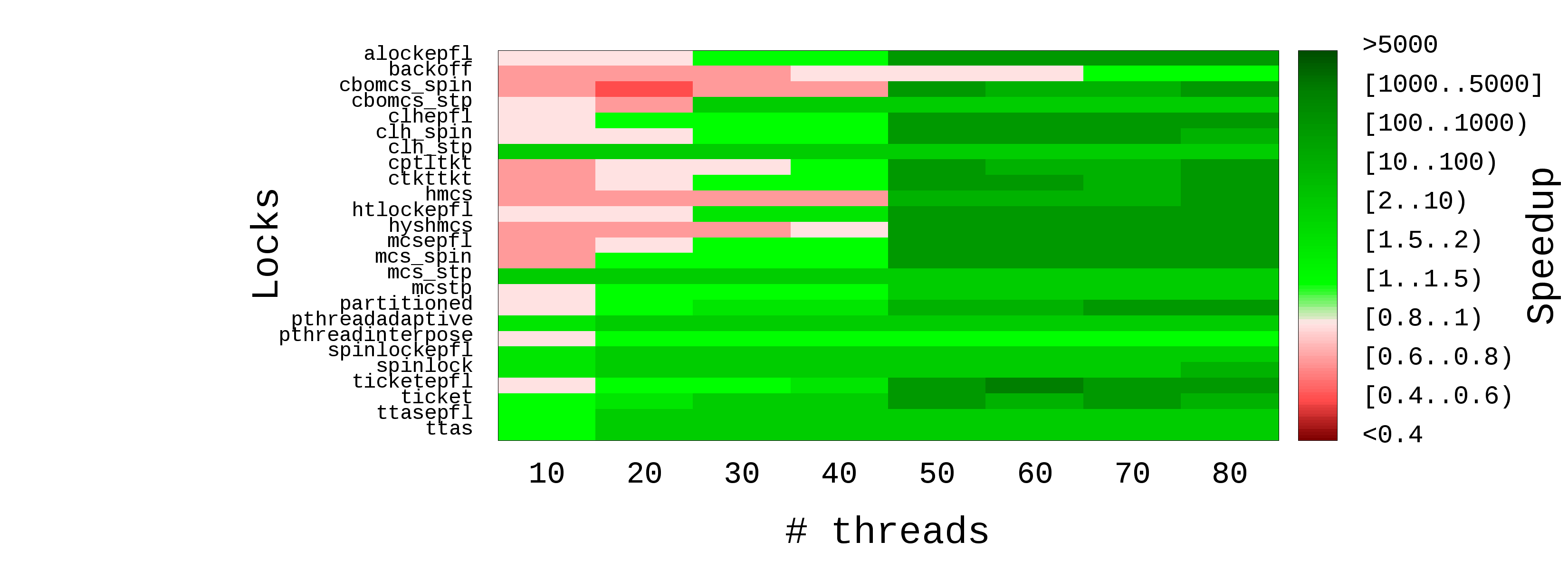}}\\
\subfloat[][GCR (empty DB)]{\adjincludegraphics[width=0.5\linewidth,Clip=0.5cm 0 1.65cm 0]{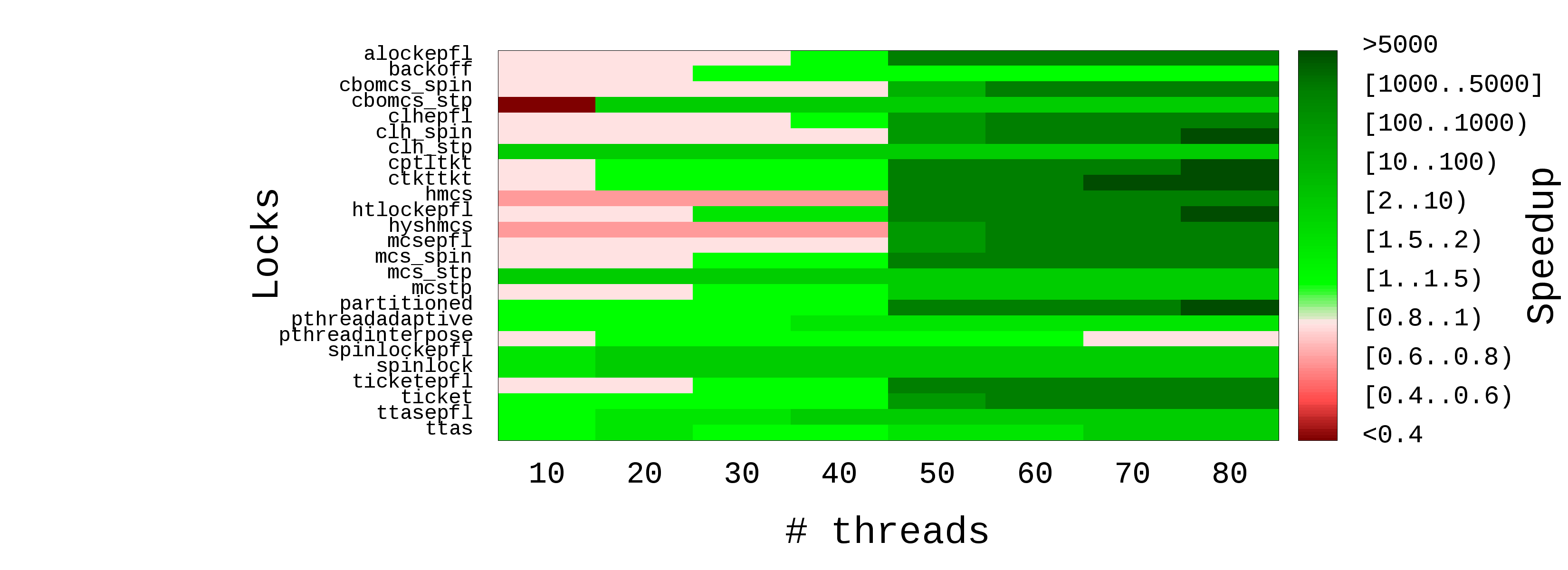}}
\subfloat[][GCR-NUMA (empty DB)]{\adjincludegraphics[width=0.5\linewidth,Clip=2.75cm 0 0 0]{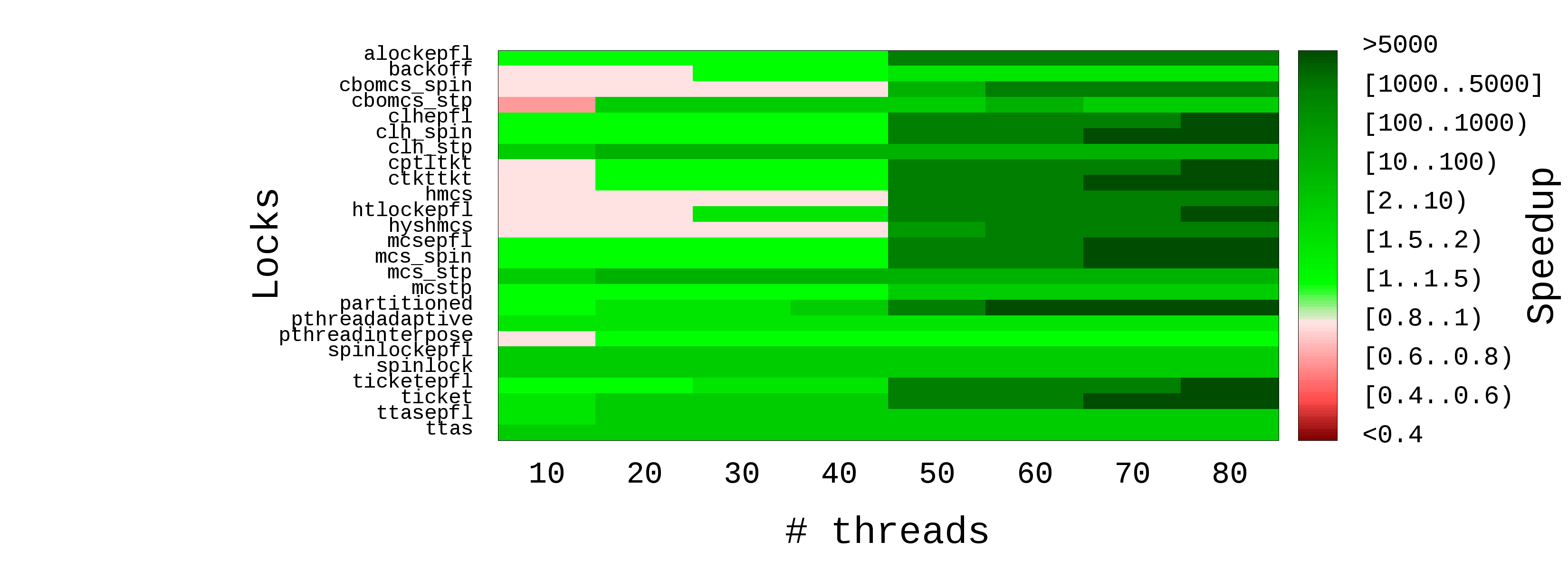}}\\
\caption{Speedup achieved by GCR and GCR-NUMA over base lock implementations (LevelDB).}
\figlabel{fig:leveldb-perf-heatmap}
\end{figure*}

As its name suggests, the \code{readrandom} mode of \code{db\_bench} is composed of 
\code{Get} operations on the database with random keys.
Each \code{Get} operation acquires a global (per-database) lock in order to take a consistent snapshot of pointers to internal database structures 
(and increment reference counters to prevent deleting those structures while \code{Get} is running).
The search operation itself, however, executes without holding the database lock, but acquires locks protecting (sharded) LRU cache as it seeks to 
update the cache structure with keys it has accessed.
Thus, like in the case of Kyoto, the contention is spread over multiple locks.

The results are presented in \figref{fig:leveldb-perf-heatmap}~(a) and~(b).
%Due to lack of space, we show only the results from the X6-2 machine; other platforms exhibit similar behavior.
As expected, GCR gains are relatively modest, yet when the machine is oversubscribed, 
it achieves positive speedups for all locks but two (the \code{backoff} and the \code{pthreadinterpose} locks. 
The latter is simply a wrapper around the standard POSIX pthread mutex.).
GCR-NUMA helps to extract more performance, but only slightly as low lock contention limits its benefit as well.

In order to explore how increased contention on the database lock would affect the speedups achieved by GCR and GCR-NUMA, 
we run the same experiment with an empty database.
In this case, the work outside of the critical sections (searching a key) is minimal and does not involve acquiring any other lock.
The results are presented in \figref{fig:leveldb-perf-heatmap}~(c) and~(d).
Overall, %echoing the AVL tree microbenchmark results, 
the increased contention leads to increased speedups achieved by GCR and GCR-NUMA. 
%for all thread counts above $1$.
In particular, when the machine is oversubscribed, all locks but one benefit from GCR (and all locks benefit from GCR-NUMA).

\section{Conclusion}
We have presented GCR, a generic concurrency restriction mechanism, and GCR-NUMA, the extension of GCR to the NUMA settings.
GCR wraps any underlying lock and controls which threads are allowed to compete for its acquisition.
The idea is to keep the lock saturated by as few threads as possible, while parking all other excessive threads
that would otherwise compete for the lock, create contention and consume valuable system resources.
Extensive evaluation with more than two dozen locks shows substantial speedup achieved by GCR on various systems and benchmarks;
the speedup grows even larger when GCR-NUMA is used.

\extabstract{
As a future direction, we would like to make our approach adaptive, disabling the concurrency 
restriction mechanism when the underlying lock is not or lightly contended.
This would help reduce the overhead of GCR in those cases.
Extending GCR for reader-writer locks is another direction for future research.
}

\bibliography{refs}

%%% -*-BibTeX-*-
%%% Do NOT edit. File created by BibTeX with style
%%% ACM-Reference-Format-Journals [18-Jan-2012].

\begin{thebibliography}{24}

%%% ====================================================================
%%% NOTE TO THE USER: you can override these defaults by providing
%%% customized versions of any of these macros before the \bibliography
%%% command.  Each of them MUST provide its own final punctuation,
%%% except for \shownote{}, \showDOI{}, and \showURL{}.  The latter two
%%% do not use final punctuation, in order to avoid confusing it with
%%% the Web address.
%%%
%%% To suppress output of a particular field, define its macro to expand
%%% to an empty string, or better, \unskip, like this:
%%%
%%% \newcommand{\showDOI}[1]{\unskip}   % LaTeX syntax
%%%
%%% \def \showDOI #1{\unskip}           % plain TeX syntax
%%%
%%% ====================================================================

\ifx \showCODEN    \undefined \def \showCODEN     #1{\unskip}     \fi
\ifx \showDOI      \undefined \def \showDOI       #1{#1}\fi
\ifx \showISBNx    \undefined \def \showISBNx     #1{\unskip}     \fi
\ifx \showISBNxiii \undefined \def \showISBNxiii  #1{\unskip}     \fi
\ifx \showISSN     \undefined \def \showISSN      #1{\unskip}     \fi
\ifx \showLCCN     \undefined \def \showLCCN      #1{\unskip}     \fi
\ifx \shownote     \undefined \def \shownote      #1{#1}          \fi
\ifx \showarticletitle \undefined \def \showarticletitle #1{#1}   \fi
\ifx \showURL      \undefined \def \showURL       {\relax}        \fi
% The following commands are used for tagged output and should be
% invisible to TeX
\providecommand\bibfield[2]{#2}
\providecommand\bibinfo[2]{#2}
\providecommand\natexlab[1]{#1}
\providecommand\showeprint[2][]{arXiv:#2}

\bibitem[\protect\citeauthoryear{Afek, Dice, and Morrison}{Afek
  et~al\mbox{.}}{2011}]%
        {ADM11}
\bibfield{author}{\bibinfo{person}{Yehuda Afek}, \bibinfo{person}{Dave Dice},
  {and} \bibinfo{person}{Adam Morrison}.} \bibinfo{year}{2011}\natexlab{}.
\newblock \showarticletitle{Cache Index-aware Memory Allocation}. In
  \bibinfo{booktitle}{\emph{Proceedings of ACM ISMM}}. \bibinfo{pages}{55--64}.
\newblock


\bibitem[\protect\citeauthoryear{Boyd-Wickizer, Kaashoek, Morris, and
  Zeldovich}{Boyd-Wickizer et~al\mbox{.}}{2012}]%
        {BKM12}
\bibfield{author}{\bibinfo{person}{S. Boyd-Wickizer}, \bibinfo{person}{M.
  Kaashoek}, \bibinfo{person}{R. Morris}, {and} \bibinfo{person}{N.
  Zeldovich}.} \bibinfo{year}{2012}\natexlab{}.
\newblock \showarticletitle{Non-scalable locks are dangerous}. In
  \bibinfo{booktitle}{\emph{Proceedings of the Linux Symposium}}.
\newblock


\bibitem[\protect\citeauthoryear{Chabbi, Fagan, and Mellor-Crummey}{Chabbi
  et~al\mbox{.}}{2015}]%
        {ppopp15-chabbi}
\bibfield{author}{\bibinfo{person}{Milind Chabbi}, \bibinfo{person}{Michael
  Fagan}, {and} \bibinfo{person}{John Mellor-Crummey}.}
  \bibinfo{year}{2015}\natexlab{}.
\newblock \showarticletitle{{High Performance Locks for Multi-level NUMA
  Systems}}. In \bibinfo{booktitle}{\emph{Proceedings of the ACM PPoPP}}.
\newblock


\bibitem[\protect\citeauthoryear{Chadha, Mahlke, and Narayanasamy}{Chadha
  et~al\mbox{.}}{2012}]%
        {cases12-chadha}
\bibfield{author}{\bibinfo{person}{Gaurav Chadha}, \bibinfo{person}{Scott
  Mahlke}, {and} \bibinfo{person}{Satish Narayanasamy}.}
  \bibinfo{year}{2012}\natexlab{}.
\newblock \showarticletitle{{When Less is More (LIMO):Controlled Parallelism
  For Improved Efficiency}}. In \bibinfo{booktitle}{\emph{Conference on
  Compilers, Architectures and Synthesis for Embedded Systems (CASES)}}.
\newblock


\bibitem[\protect\citeauthoryear{Craig}{Craig}{1993}]%
        {CLH}
\bibfield{author}{\bibinfo{person}{Travis Craig}.}
  \bibinfo{year}{1993}\natexlab{}.
\newblock \bibinfo{booktitle}{\emph{Building {FIFO} and priority-queueing spin
  locks from atomic swap}}.
\newblock \bibinfo{type}{Technical Report} TR 93-02-02.
  \bibinfo{institution}{U. of Washington, Dept. of Computer Science}.
\newblock


\bibitem[\protect\citeauthoryear{David, Guerraoui, and Trigonakis}{David
  et~al\mbox{.}}{2013}]%
        {TGT13}
\bibfield{author}{\bibinfo{person}{Tudor David}, \bibinfo{person}{Rachid
  Guerraoui}, {and} \bibinfo{person}{Vasileios Trigonakis}.}
  \bibinfo{year}{2013}\natexlab{}.
\newblock \showarticletitle{Everything You Always Wanted to Know About
  Synchronization but Were Afraid to Ask}. In
  \bibinfo{booktitle}{\emph{Proceedings of the ACM Symposium on Operating
  Systems Principles (SOSP)}}. \bibinfo{pages}{33--48}.
\newblock


\bibitem[\protect\citeauthoryear{Dice}{Dice}{2017}]%
        {Dice17}
\bibfield{author}{\bibinfo{person}{Dave Dice}.}
  \bibinfo{year}{2017}\natexlab{}.
\newblock \showarticletitle{Malthusian Locks}. In
  \bibinfo{booktitle}{\emph{Proc. of ACM EuroSys}}. \bibinfo{pages}{314--327}.
\newblock


\bibitem[\protect\citeauthoryear{Dice and Kogan}{Dice and Kogan}{2019}]%
        {DK19}
\bibfield{author}{\bibinfo{person}{Dave Dice} {and} \bibinfo{person}{Alex
  Kogan}.} \bibinfo{year}{2019}\natexlab{}.
\newblock \showarticletitle{Compact {NUMA}-aware Locks}. In
  \bibinfo{booktitle}{\emph{Proc. of ACM EuroSys}}.
\newblock


\bibitem[\protect\citeauthoryear{Dice, Marathe, and Shavit}{Dice
  et~al\mbox{.}}{2015}]%
        {topc15-dice}
\bibfield{author}{\bibinfo{person}{David Dice}, \bibinfo{person}{Virendra~J.
  Marathe}, {and} \bibinfo{person}{Nir Shavit}.}
  \bibinfo{year}{2015}\natexlab{}.
\newblock \showarticletitle{{Lock Cohorting: A General Technique for Designing
  NUMA Locks}}.
\newblock \bibinfo{journal}{\emph{ACM TOPC}} \bibinfo{volume}{1},
  \bibinfo{number}{2}, Article \bibinfo{articleno}{13} (\bibinfo{date}{Feb}
  \bibinfo{year}{2015}).
\newblock


\bibitem[\protect\citeauthoryear{Eyerman and Eeckhout}{Eyerman and
  Eeckhout}{2010}]%
        {EE10}
\bibfield{author}{\bibinfo{person}{Stijn Eyerman} {and} \bibinfo{person}{Lieven
  Eeckhout}.} \bibinfo{year}{2010}\natexlab{}.
\newblock \showarticletitle{Modeling Critical Sections in {Amdahl's} Law and
  Its Implications for Multicore Design}. In
  \bibinfo{booktitle}{\emph{Proceedings of ACM ISCA}}.
  \bibinfo{pages}{362--370}.
\newblock
\showISBNx{978-1-4503-0053-7}


\bibitem[\protect\citeauthoryear{FAL Labs}{FAL Labs}{[n. d.]}]%
        {kyotocabinet}
FAL Labs \bibinfo{year}{[n. d.]}\natexlab{}.
\newblock \bibinfo{title}{{Kyoto Cabinet}}.
\newblock
\newblock
\newblock
\shownote{http://fallabs.com/kyotocabinet.}


\bibitem[\protect\citeauthoryear{Guiroux, Lachaize, and Qu{\'{e}}ma}{Guiroux
  et~al\mbox{.}}{2016}]%
        {GLQ16}
\bibfield{author}{\bibinfo{person}{Hugo Guiroux}, \bibinfo{person}{Renaud
  Lachaize}, {and} \bibinfo{person}{Vivien Qu{\'{e}}ma}.}
  \bibinfo{year}{2016}\natexlab{}.
\newblock \showarticletitle{Multicore Locks: The Case Is Not Closed Yet}. In
  \bibinfo{booktitle}{\emph{Proceedings of USENIX ATC}}.
  \bibinfo{pages}{649--662}.
\newblock


\bibitem[\protect\citeauthoryear{He, Scherer, and Scott}{He
  et~al\mbox{.}}{2005}]%
        {hipc05-he}
\bibfield{author}{\bibinfo{person}{Bijun He}, \bibinfo{person}{William~N.
  Scherer}, {and} \bibinfo{person}{Michael~L. Scott}.}
  \bibinfo{year}{2005}\natexlab{}.
\newblock \showarticletitle{{Preemption Adaptivity in Time-published
  Queue-based Spin Locks}}. In \bibinfo{booktitle}{\emph{Proc. of High
  Performance Computing (HiPC)}}. \bibinfo{pages}{7--18}.
\newblock


\bibitem[\protect\citeauthoryear{Heirman, Carlson, Van~Craeynest, Hur, Jaleel,
  and Eeckhout}{Heirman et~al\mbox{.}}{2014}]%
        {hpca14-heirman}
\bibfield{author}{\bibinfo{person}{W. Heirman}, \bibinfo{person}{T.E. Carlson},
  \bibinfo{person}{K. Van~Craeynest}, \bibinfo{person}{I. Hur},
  \bibinfo{person}{A. Jaleel}, {and} \bibinfo{person}{L. Eeckhout}.}
  \bibinfo{year}{2014}\natexlab{}.
\newblock \showarticletitle{{Undersubscribed Threading on Clustered Cache
  Architectures}}. In \bibinfo{booktitle}{\emph{Proceedings of IEEE HPCA}}.
\newblock


\bibitem[\protect\citeauthoryear{Johnson, Athanassoulis, Stoica, and
  Ailamaki}{Johnson et~al\mbox{.}}{2009}]%
        {damon09-johnson}
\bibfield{author}{\bibinfo{person}{Ryan Johnson}, \bibinfo{person}{Manos
  Athanassoulis}, \bibinfo{person}{Radu Stoica}, {and}
  \bibinfo{person}{Anastasia Ailamaki}.} \bibinfo{year}{2009}\natexlab{}.
\newblock \showarticletitle{{A New Look at the Roles of Spinning and
  Blocking}}. In \bibinfo{booktitle}{\emph{Proceedings of the International
  Workshop on Data Management on New Hardware (DaMoN)}}.
  \bibinfo{publisher}{ACM}.
\newblock


\bibitem[\protect\citeauthoryear{Johnson, Stoica, Ailamaki, and Mowry}{Johnson
  et~al\mbox{.}}{2010}]%
        {JSA10}
\bibfield{author}{\bibinfo{person}{Ryan Johnson}, \bibinfo{person}{Radu
  Stoica}, \bibinfo{person}{Anastasia Ailamaki}, {and} \bibinfo{person}{Todd~C.
  Mowry}.} \bibinfo{year}{2010}\natexlab{}.
\newblock \showarticletitle{Decoupling Contention Management from Scheduling}.
  In \bibinfo{booktitle}{\emph{Proceedings of ACM ASPLOS}}.
  \bibinfo{pages}{117--128}.
\newblock


\bibitem[\protect\citeauthoryear{LevelDB}{LevelDB}{[n. d.]}]%
        {leveldb}
LevelDB \bibinfo{year}{[n. d.]}\natexlab{}.
\newblock \bibinfo{title}{LevelDB}.
\newblock
\newblock
\newblock
\shownote{https://github.com/google/leveldb.}


\bibitem[\protect\citeauthoryear{Lim and Agarwal}{Lim and Agarwal}{1993}]%
        {tocs93-lim}
\bibfield{author}{\bibinfo{person}{Beng-Hong Lim} {and} \bibinfo{person}{Anant
  Agarwal}.} \bibinfo{year}{1993}\natexlab{}.
\newblock \showarticletitle{{Waiting Algorithms for Synchronization in
  Large-scale Multiprocessors}}.
\newblock \bibinfo{journal}{\emph{ACM Transactions on Computing Systems}}
  (\bibinfo{year}{1993}).
\newblock


\bibitem[\protect\citeauthoryear{LiTL}{LiTL}{[n. d.]}]%
        {LiTL}
LiTL \bibinfo{year}{[n. d.]}\natexlab{}.
\newblock \bibinfo{title}{{LiTL: Library for Transparent Lock interposition}}.
\newblock
\newblock
\newblock
\shownote{https://github.com/multicore-locks/litl.}


\bibitem[\protect\citeauthoryear{Mellor-Crummey and Scott}{Mellor-Crummey and
  Scott}{1991}]%
        {MS91}
\bibfield{author}{\bibinfo{person}{John~M. Mellor-Crummey} {and}
  \bibinfo{person}{Michael~L. Scott}.} \bibinfo{year}{1991}\natexlab{}.
\newblock \showarticletitle{Algorithms for Scalable Synchronization on
  Shared-memory Multiprocessors}.
\newblock \bibinfo{journal}{\emph{ACM Trans. Comp. Syst.}} \bibinfo{volume}{9},
  \bibinfo{number}{1} (\bibinfo{year}{1991}), \bibinfo{pages}{21--65}.
\newblock


\bibitem[\protect\citeauthoryear{Pusukuri, Gupta, and Bhuyan}{Pusukuri
  et~al\mbox{.}}{2011}]%
        {iiswc11-pusukuri}
\bibfield{author}{\bibinfo{person}{Kishore~Kumar Pusukuri},
  \bibinfo{person}{Rajiv Gupta}, {and} \bibinfo{person}{Laxmi~N. Bhuyan}.}
  \bibinfo{year}{2011}\natexlab{}.
\newblock \showarticletitle{{Thread Reinforcer: Dynamically Determining Number
  of Threads via OS Level Monitoring}}. In \bibinfo{booktitle}{\emph{Proc. of
  IEEE IISWC}}.
\newblock


\bibitem[\protect\citeauthoryear{Radovic and Hagersten}{Radovic and
  Hagersten}{2003}]%
        {RH03}
\bibfield{author}{\bibinfo{person}{Zoran Radovic} {and} \bibinfo{person}{Erik
  Hagersten}.} \bibinfo{year}{2003}\natexlab{}.
\newblock \showarticletitle{Hierarchical Backoff Locks for Nonuniform
  Communication Architectures}. In \bibinfo{booktitle}{\emph{Proceedings of EEE
  HPCA}}. \bibinfo{pages}{241--252}.
\newblock


\bibitem[\protect\citeauthoryear{Raman, Kim, Oh, Lee, and August}{Raman
  et~al\mbox{.}}{2011}]%
        {pldi11-raman}
\bibfield{author}{\bibinfo{person}{Arun Raman}, \bibinfo{person}{Hanjun Kim},
  \bibinfo{person}{Taewook Oh}, \bibinfo{person}{Jae~W. Lee}, {and}
  \bibinfo{person}{David~I. August}.} \bibinfo{year}{2011}\natexlab{}.
\newblock \showarticletitle{{Parallelism Orchestration Using DoPE: The Degree
  of Parallelism Executive}}. In \bibinfo{booktitle}{\emph{Proc. of ACM PLDI}}.
\newblock


\bibitem[\protect\citeauthoryear{Yoo and Lee}{Yoo and Lee}{2008}]%
        {spaa08-yoo}
\bibfield{author}{\bibinfo{person}{Richard~M. Yoo} {and}
  \bibinfo{person}{Hsien-Hsin~S. Lee}.} \bibinfo{year}{2008}\natexlab{}.
\newblock \showarticletitle{{Adaptive Transaction Scheduling for Transactional
  Memory Systems}}. In \bibinfo{booktitle}{\emph{Proc. of ACM SPAA}}.
\newblock


\end{thebibliography}

\extabstract{
\appendix

\section{Correctness proofs}
In this section, we argue that the GCR algorithm does not introduce starvation as long as the 
underlying lock is starvation-free, the OS scheduler does not starve any thread 
and the underling architecture supports starvation-free atomic increment and swap operations.
On a high level, our argument is built on top of two observations, namely that once a thread
enters the queue of waiting threads, it eventually reaches the top of the queue,
and that a thread at the top of the queue eventually calls the \code{Lock} function of the underlying lock.

\begin{lemma}
\label{lemma:1}
The \code{tail} pointer always either holds a \code{NULL} value or points to a node whose \code{next} pointer is \code{NULL}.
\end{lemma}

\begin{proofsketch}
The \code{tail} pointer initially holds \code{NULL}.
From inspecting the code, the \code{tail} pointer may change only in \lineref{push:5} or \lineref{pop:5}.
Consider the change in \lineref{push:5}.
The value of the \code{tail} pointer is set to a node whose \code{next} field was initialized to NULL (cf.~\lineref{push:3}).
Thus, the lemma holds when the change in \lineref{push:5} takes place.

The \code{next} pointer of a node gets modified only in \lineref{push:7}.
The node whose \code{next} pointer gets modified is the one pointed by \code{tail} before 
the change of \code{tail} in \lineref{push:3} took place.
As a result, when \lineref{push:7} is executed, the \code{tail} pointer does not point anymore to the node whose
\code{next} pointer gets updated, and the lemma holds.

Next, consider the change to \code{tail} in \lineref{pop:5} done with a CAS instruction.
If CAS is successful, the \code{tail} pointer gets a \code{NULL} value.
Otherwise, the \code{tail} pointer is not updated (and thus 
remains pointing to a node whose \code{next} pointer is \code{NULL}).
Thus, the lemma holds in both cases.
\end{proofsketch}

We say that a passive thread enters the queue if it finishes executing \lineref{push:5} 
and leaves the queue when it finishes executing a successful CAS in \lineref{pop:6} or an assignment in \lineref{pop:17}.

\begin{lemma}
\label{lemma:2}
The \code{event} field of any node in the queue, except for, perhaps, the node pointed by \code{top}, is $0$.
\end{lemma}

\begin{proofsketch}
First, we observe that only a thread whose node has a non-zero \code{event} value 
can call \code{popSelfFromQueue} (cf.~\lineref{lock:12}).

Next, we show that at most one node in the queue has a non-zero \code{event} value.
The \code{event} field is initialized to $0$ (\lineref{push:4}), and is set to $1$ either in \lineref{push:10}
or in \lineref{pop:18}.
In the former case, this happens when the corresponding thread finds the queue empty (\lineref{push:5}).
Thus, when it sets the \code{event} field of its node to $1$, the claim holds.
In the latter case, a thread $t$ sets the \code{event} field in the node of its successor in the queue.
However, it does so after removing its node from the queue (by updating the \code{top} pointer to its successor in \lineref{pop:17}).
Based on the observation above, $t$'s node contains a non-zero \code{event} field.
By removing its node from the queue and setting the \code{event} field in the successor node, $t$ maintains the claim.

Finally, we argue that the node with a non-zero \code{event} value is the one pointed by \code{top}.
Consider, again, the two cases where the \code{event} field gets set.
In the first case, it is set by a thread that just entered the queue and found the queue empty (and thus updated \code{top} to point to its node).
In the second case, it is set by a thread $t$ that just updated \code{top} to point to the node of its successor in the queue.
At this point (i.e., after executing \lineref{pop:17} and before executing \lineref{pop:18}), no node in the queue contains a non-zero \code{event} value.
Thus, based on the observation above, no thread can call \code{popSelfFromQueue}.
At the same time, any new thread entering the queue will find at least $t$'s successor there and thus, will not change the \code{top} pointer.
Hence, when $t$ executes \lineref{pop:18}, it sets the \code{event} field of the node pointed by \code{top}.
\end{proofsketch}

We refer to a thread whose node is pointed by \code{top} as a \emph{thread at the top of the queue}.
\begin{lemma}
\label{lemma:3}
Only a thread at the top of the queue can call \code{popSelfFromQueue}.
\end{lemma}

\begin{proofsketch}
After entering the queue, a thread may leave it (by calling \code{popSelfFromQueue}) 
only after it finds the \code{event} field in its node holding a non-zero value (cf.~\lineref{lock:12}).
According to Lemma~\ref{lemma:2}, this can only be the thread at the top of the queue.
\end{proofsketch}

We define the state of the queue at time $T$ to be all nodes that are reachable from \code{top} at time $T$.
We order the nodes in the queue according to their rank, which is the number of links (\code{next} pointers) to be traversed
from $top$ to reach the node.
The rank of the node pointed by \code{top} is $0$, the rank of the next node is $1$ and so on.
The rank of a (passive) thread is simply a rank of its corresponding node in the queue.

\begin{lemma}
\label{lemma:4}
The queue preserves the FIFO order, that is, a thread $t$ that enters the queue will leave the queue after 
all threads that entered the queue before $t$ and before all threads that enter the queue after $t$.
\end{lemma}

\begin{proofsketch}
By inspecting the code, the only place where the \code{next} pointer of a node may change is in \lineref{push:7} 
(apart from the initialization in~\lineref{push:3}).
If a thread executes \lineref{push:7}, then according to Lemma~\ref{lemma:1}, it changes the \code{next} pointer
of a node of its predecessor in the queue from \code{NULL} to the thread's node.
Thus, once a \code{next} pointer is set to a non-\code{NULL} value, it would never change again
(until the node is deleted, which will happen only after the respective thread leaves the queue).
Hence, threads that enter the queue after $t$ will have a higher rank than $t$.

Lemma~\ref{lemma:3} implies that only a thread with rank $0$ can leave the queue.
Thus, any thread that joins the queue after $t$ will leave the queue after $t$, as $t$'s rank will reach $0$ first.
Also, any thread that enters the queue before $t$ will have a lower rank than $t$.
Therefore, $t$ cannot leave the queue before all those threads do.
\end{proofsketch}

\begin{lemma}
\label{lemma:5}
A thread at the top of the queue eventually calls the \code{Lock} function of the underlying lock.
\end{lemma}

\begin{proofsketch}
By inspecting the code, when a thread $t$ reaches the top of the queue, its \code{event} field changes to $1$ (\lineref{push:10} and \lineref{pop:18}).
Thus, it reaches the \code{while} loop in \lineref{lock:14}.
It would stay in this loop as long as the \code{topApproved} field does not change or when all active threads have left.
If the latter does not happen, however, assuming that the lock is starvation-free and no thread holds it indefinitely long, active
threads will circulate through the lock, incrementing the \code{numAcqs} counter.
Once this counter reaches the threshold, the \code{topApproved} field will be set, releasing $t$ from the \code{while} loop.

Following that, the only other place where $t$ may spin before calling the \code{Lock} function of the underlying lock
is in the \code{for} loop in \lineref{push:11}.
This would occur in a rare case where $t$ may not find a successor in \lineref{pop:2}, but the successor will appear and update
\code{tail} (in \lineref{push:5}) right before $t$ does (in \lineref{pop:5}).
However, assuming a scheduler that does not starve any thread, the successor will eventually update the \code{next}
pointer in $t$'s node, allowing $t$ to break from the \code{for} loop.
\end{proofsketch}

\begin{lemma}
\label{lemma:6}
A thread that enters the queue eventually becomes a thread at the top of the queue.
\end{lemma}

\begin{proofsketch}
Consider a thread $t$ entering the queue by executing \lineref{push:5}.
If it finds no nodes in the queue (i.e., \code{tail} holds a \code{NULL} value), $t$ sets \code{top}
to point to its node (\lineref{push:9}) and the Lemma trivially holds.
Otherwise, some other thread $t_1$ is at the top of the queue.
By Lemma~\ref{lemma:5}, that thread will eventually call the \code{Lock} function of the underlying lock.
However, before doing so, it will remove itself from the queue (\lineref{lock:21}).
According to Lemma~\ref{lemma:4}, the queue preserves the FIFO order.
Thus, the next thread after $t_1$ will become the thread at the top of the queue, and 
eventually call the \code{Lock} function of the underlying lock.
By applying the same argument recursively, we can deduct that $t$ will reach the top of the queue (after all threads
with a lower rank do so).
\end{proofsketch}

\begin{theorem}
When GCR is applied to a starvation-free lock L, the resulting lock is starvation-free.
\end{theorem}

\begin{proofsketch}
Given that the underlying lock L is starvation-free, we need to show that every thread calling GCR's \code{Lock} function would
eventually call L's \code{Lock} function.
In case a thread finds at most one active thread in \lineref{lock:3}, it proceeds by calling to the L's \code{Lock} function
(after executing an atomic FAA instruction, which we assume is starvation-free).
%\footnote{We note that this assumption 
%does not hold on platforms that do not support atomic FAA and require emulation via CAS or Load-Linked/Store-Conditional instructions.}).
Otherwise, it proceeds on the slow path by entering the queue (\lineref{lock:10}).
Assuming the atomic swap operation is starvation-free, the thread would eventually execute \lineref{push:5} and enter the queue.
By Lemma~\ref{lemma:6}, it will eventually reach the top of the queue, and by Lemma~\ref{lemma:5}, 
it will eventually call L's \code{Lock} function. 
\end{proofsketch}

\section{Raw Speedup Data}
In this section, we include the raw speedup data for heat map plots generated for the AVL tree microbenchmark experiment 
run on X6-2 and presented in \figref{fig:avl-tree-perf-heatmap}~(a) and (b).
As mentioned in Section~\ref{sec:evaluation-avl-tree}, a cell at row X and column Y in Table~\ref{table:X6-2-avl-tree-GCR} 
(and the corresponding color-coded cell in \figref{fig:avl-tree-perf-heatmap}~(a))
represents the throughput achieved with Y threads when the GCR library is used with lock X divided
by throughput achieved when the GCR library is not used (i.e., with the lock X itself).
Table~\ref{table:X6-2-avl-tree-GCR-NUMA} presents the raw speedup data for GCR-NUMA.

\begin{table*}
\begin{center}
\scalebox{0.8}{
\begin{tabular}{l||cccccccccccc}
Lock & 1 & 2 & 4 & 8 & 10 & 20 & 30 & 40 & 50 & 60 & 70 & 80\\
\hline
alockepfl &  0.98 &  0.98 &  0.96 &  0.84 &  0.84 &  0.86 &  0.94 &  1.00 & 2184.85 & 3301.62 & 2812.26 & 4249.23\\
backoff &  0.97 &  0.96 &  0.96 &  0.86 &  0.84 &  0.82 &  0.87 &  0.89 &  0.99 &  1.11 &  1.19 &  1.32\\
cbomcs\_spin &  0.98 &  0.97 &  0.97 &  0.71 &  0.70 &  0.75 &  0.85 &  0.87 & 1516.78 & 2989.93 & 2464.75 & 4540.33\\
cbomcs\_stp &  0.98 &  0.98 &  1.00 &  0.60 &  0.48 &  2.14 &  4.49 &  4.51 &  4.37 &  4.60 &  4.66 &  4.66\\
clhepfl &  0.97 &  0.98 &  0.98 &  0.86 &  0.87 &  0.89 &  0.96 &  1.00 & 2287.29 & 3126.55 & 3622.63 & 3412.83\\
clh\_spin &  0.97 &  0.98 &  0.95 &  0.86 &  0.85 &  0.88 &  0.94 &  0.96 & 2053.67 & 2318.83 & 2879.09 & 2190.06\\
clh\_stp &  0.98 &  0.99 &  1.00 &  2.60 &  5.67 &  5.87 &  5.83 &  5.88 &  5.77 &  5.90 &  5.93 &  5.95\\
cptltkt &  0.96 &  0.97 &  0.98 &  0.74 &  0.76 &  0.86 &  0.91 &  0.94 & 2594.45 & 3065.98 & 3535.06 & 3648.85\\
ctkttkt &  0.97 &  0.99 &  1.01 &  0.77 &  0.78 &  0.94 &  0.97 &  1.01 & 2330.65 & 2330.82 & 3968.63 & 2388.85\\
hmcs &  0.98 &  0.99 &  0.98 &  0.74 &  0.73 &  0.77 &  0.85 &  0.91 & 2218.43 & 3648.88 & 3613.60 & 3783.19\\
htlockepfl &  0.98 &  0.94 &  0.99 &  0.80 &  0.84 &  0.91 &  1.14 &  1.35 & 2660.04 & 3425.57 & 3126.08 & 3839.67\\
hyshmcs &  0.98 &  0.99 &  0.96 &  0.77 &  0.73 &  0.79 &  0.84 &  0.89 & 2010.15 & 4085.70 & 3993.88 & 4321.18\\
mcsepfl &  0.98 &  0.98 &  0.98 &  0.81 &  0.82 &  0.84 &  0.91 &  0.92 & 2458.11 & 3280.06 & 2934.57 & 4247.53\\
mcs\_spin &  0.98 &  0.98 &  0.93 &  0.89 &  0.88 &  0.90 &  0.99 &  1.04 & 2230.34 & 3441.29 & 2819.78 & 4315.80\\
mcs\_stp &  0.98 &  0.95 &  0.95 &  2.22 &  5.63 &  5.90 &  5.96 &  6.01 &  5.99 &  6.07 &  6.04 &  6.12\\
mcstp &  0.98 &  0.98 &  0.98 &  0.94 &  0.94 &  0.93 &  1.00 &  1.05 &  2.12 &  2.43 &  2.57 &  2.73\\
partitioned &  0.97 &  0.98 &  0.97 &  0.92 &  0.96 &  1.08 &  1.12 &  1.18 & 2312.32 & 3081.01 & 2991.36 & 4230.77\\
pthreadadaptive &  0.97 &  0.99 &  0.96 &  1.17 &  1.29 &  1.71 &  1.61 &  1.64 &  1.71 &  1.74 &  1.77 &  1.81\\
pthreadinterpose &  0.96 &  0.99 &  0.95 &  0.93 &  1.02 &  1.19 &  1.10 &  1.07 &  1.07 &  1.08 &  1.07 &  1.05\\
spinlockepfl &  0.97 &  0.97 &  0.96 &  1.16 &  1.35 &  2.31 &  2.31 &  2.39 &  3.18 &  3.75 &  4.50 &  4.63\\
spinlock &  0.97 &  0.98 &  1.02 &  1.31 &  1.49 &  2.47 &  2.55 &  2.76 &  3.49 &  4.09 &  5.18 &  5.87\\
ticketepfl &  0.97 &  0.97 &  1.05 &  0.95 &  1.00 &  1.02 &  1.05 &  1.04 & 1031.99 & 1624.08 & 696.70 & 3433.58\\
ticket &  0.97 &  0.97 &  1.00 &  0.99 &  1.03 &  1.20 &  1.22 &  1.24 & 1881.94 & 2900.54 & 3028.59 & 3277.52\\
ttasepfl &  0.97 &  0.98 &  0.97 &  1.09 &  1.18 &  1.68 &  2.07 &  2.53 &  2.74 &  2.97 &  3.16 &  3.44\\
ttas &  0.97 &  0.98 &  0.95 &  1.05 &  1.15 &  1.47 &  1.49 &  1.49 &  1.75 &  2.05 &  2.23 &  2.49\\
\end{tabular}
}
\end{center}
\caption{Relative performance of lock implementations with and without GCR (AVL tree, X6-2).}
\label{table:X6-2-avl-tree-GCR}
\end{table*}

\begin{table*}
\begin{center}
\scalebox{0.8}{
\begin{tabular}{l||cccccccccccc}
Lock & 1 & 2 & 4 & 8 & 10 & 20 & 30 & 40 & 50 & 60 & 70 & 80\\
\hline
alockepfl &  0.98 &  0.99 &  0.98 &  0.90 &  1.16 &  1.20 &  1.28 &  1.35 & 2919.85 & 4445.47 & 3796.18 & 5706.08\\
backoff &  0.97 &  0.96 &  0.98 &  0.87 &  0.94 &  0.94 &  0.97 &  0.99 &  1.10 &  1.23 &  1.34 &  1.46\\
cbomcs\_spin &  0.98 &  0.98 &  1.00 &  0.64 &  0.74 &  0.84 &  0.94 &  0.99 & 1731.39 & 3387.30 & 2740.81 & 5095.26\\
cbomcs\_stp &  0.98 &  0.98 &  1.00 &  0.65 &  0.57 &  3.65 &  7.66 &  7.60 &  7.31 &  7.78 &  7.74 &  7.79\\
clhepfl &  0.97 &  0.98 &  0.96 &  0.95 &  1.14 &  1.28 &  1.35 &  1.40 & 3201.37 & 4361.14 & 5052.70 & 4772.53\\
clh\_spin &  0.97 &  0.92 &  0.98 &  0.95 &  1.32 &  1.35 &  1.40 &  1.42 & 3054.93 & 3409.65 & 4160.59 & 3193.24\\
clh\_stp &  0.98 &  0.99 &  0.99 &  2.87 &  7.29 &  8.93 &  8.66 &  8.68 &  8.69 &  8.67 &  8.74 &  8.73\\
cptltkt &  0.97 &  0.98 &  0.95 &  0.97 &  1.04 &  1.11 &  1.22 &  1.25 & 3444.39 & 4083.43 & 4624.89 & 4806.23\\
ctkttkt &  0.97 &  0.99 &  1.01 &  1.01 &  1.03 &  1.23 &  1.29 &  1.34 & 3051.06 & 3028.56 & 5274.19 & 3002.69\\
hmcs &  0.98 &  0.98 &  0.95 &  0.99 &  1.05 &  1.07 &  1.19 &  1.26 & 3033.04 & 4975.47 & 4880.26 & 5152.25\\
htlockepfl &  0.98 &  0.98 &  0.95 &  1.08 &  1.19 &  1.23 &  1.40 &  1.61 & 3171.38 & 4127.10 & 3745.00 & 4608.16\\
hyshmcs &  0.97 &  0.99 &  0.98 &  0.95 &  1.00 &  1.09 &  1.17 &  1.23 & 2766.56 & 5621.71 & 5426.78 & 5865.11\\
mcsepfl &  0.98 &  0.89 &  0.95 &  0.87 &  1.07 &  1.18 &  1.26 &  1.26 & 3308.58 & 4504.98 & 4020.58 & 5743.06\\
mcs\_spin &  0.98 &  0.98 &  0.96 &  0.97 &  1.28 &  1.34 &  1.45 &  1.49 & 3213.92 & 4998.79 & 4039.62 & 6033.65\\
mcs\_stp &  0.98 &  0.95 &  0.94 &  2.38 &  7.02 &  8.83 &  8.65 &  8.54 &  8.71 &  8.76 &  8.75 &  8.73\\
mcstp &  0.98 &  0.98 &  0.98 &  1.04 &  1.38 &  1.39 &  1.45 &  1.51 &  3.05 &  3.46 &  3.67 &  3.88\\
partitioned &  0.97 &  0.98 &  0.99 &  1.03 &  1.56 &  1.83 &  1.84 &  1.89 & 3738.93 & 4995.00 & 4825.31 & 6767.75\\
pthreadadaptive &  0.97 &  0.98 &  0.94 &  1.42 &  1.90 &  2.09 &  2.04 &  2.22 &  2.24 &  2.25 &  2.24 &  2.28\\
pthreadinterpose &  0.96 &  1.02 &  0.99 &  0.97 &  1.10 &  1.22 &  1.27 &  1.32 &  1.35 &  1.34 &  1.34 &  1.30\\
spinlockepfl &  0.97 &  0.97 &  0.96 &  1.26 &  2.08 &  3.85 &  4.05 &  4.07 &  5.30 &  6.24 &  7.53 &  7.90\\
spinlock &  0.97 &  0.98 &  1.03 &  1.45 &  2.50 &  4.23 &  4.21 &  4.42 &  5.58 &  6.63 &  8.13 &  9.00\\
ticketepfl &  0.97 &  0.97 &  1.08 &  1.13 &  1.71 &  1.77 &  1.50 &  1.44 & 1431.66 & 2271.81 & 966.80 & 4871.73\\
ticket &  0.97 &  0.97 &  0.98 &  1.09 &  1.61 &  2.18 &  2.19 &  2.20 & 3334.85 & 5138.01 & 5321.70 & 5777.13\\
ttasepfl &  0.97 &  0.98 &  0.98 &  1.17 &  2.10 &  2.83 &  3.49 &  4.17 &  4.62 &  4.97 &  5.34 &  5.72\\
ttas &  0.97 &  0.98 &  0.95 &  1.12 &  1.88 &  2.35 &  2.31 &  2.31 &  2.70 &  3.16 &  3.43 &  3.78\\
\end{tabular}
}
\end{center}
\caption{Relative performance of lock implementations with and without GCR-NUMA (AVL tree, X6-2).}
\label{table:X6-2-avl-tree-GCR-NUMA}
\end{table*}
}

\end{document}